\documentclass[letterpaper,11pt]{article}
\usepackage{jheppub}
\pdfoutput=1
\usepackage{amsmath}
\usepackage{graphicx}

\usepackage[small]{subfigure}
\usepackage{placeins}
\usepackage{xcolor}
\usepackage{slashed}

\usepackage{hyperref}
\newcommand{\refcite}[1]{ref.~\cite{#1}}
\newcommand{\refscite}[1]{refs.~\cite{#1}}
\newcommand{\eq}[1]{eq.~\eqref{eq:#1}}
\newcommand{\eqs}[2]{eqs.~\eqref{eq:#1} and \eqref{eq:#2}}

\renewcommand{\sec}[1]{sec.~\ref{sec:#1}}

\newcommand{\fig}[1]{fig.~\ref{fig:#1}}
\newcommand{\figs}[2]{figs.~\ref{fig:#1} and \ref{fig:#2}}
\newcommand{\app}[1]{appendix~\ref{app:#1}}

\newcommand{\abs}[1]{\lvert#1\rvert}
\newcommand{\ord}[1]{\mathcal{O}(#1)}
\newcommand{\df}{\mathrm{d}}
\newcommand{\eps}{\epsilon}

\newcommand{\cO}{{\mathcal O}}

\newcommand{\cI}{{\mathcal I}}
\newcommand{\cL}{{\mathcal L}}
\newcommand{\cM}{{\mathcal M}}

\newcommand{\bn}{\bar{n}}

\newcommand{\GeV}{\,\mathrm{GeV}}
\newcommand{\TeV}{\,\mathrm{TeV}}

\newcommand{\nn}{\nonumber}

\newcommand{\Msquared}{A}                  
\newcommand{\as}{\alpha_s}
\newcommand{\muMS}{\mu_{\rm MS}}

\newcommand{\Ecm}{E_\mathrm{cm}}

\newcommand{\sub}{\mathrm{sub}}
\newcommand{\LO}{\mathrm{LO}}
\newcommand{\NLO}{\mathrm{NLO}}
\newcommand{\hadcm}{\mathrm{cm}}
\newcommand{\lep}{\mathrm{lep}}

\newcommand{\Tau}{\mathcal{T}}

\newcommand{\cut}{{\mathrm{cut}}}

\newcommand{\altM}{\overline{\Msquared}}
\def\SCETI{$\text{SCET}_\text{I}$ }

\allowdisplaybreaks[4]

\begin{document}


\preprint{\vbox{\hbox{MIT-CTP 5007}\hbox{DESY 18-120}}}

\title{\boldmath Power Corrections for N-Jettiness Subtractions at ${\cal O}(\alpha_s)\!\!$}

\author[1]{Markus A.~Ebert,}
\emailAdd{ebert@mit.edu}

\author[2,3]{Ian Moult,}
\emailAdd{ianmoult@lbl.gov}

\author[1]{Iain W.~Stewart,}
\emailAdd{iains@mit.edu}

\author[4]{Frank J.~Tackmann,}
\emailAdd{frank.tackmann@desy.de}

\author[1]{Gherardo Vita,}
\emailAdd{vita@mit.edu}

\author[5]{and Hua Xing Zhu}
\emailAdd{zhuhx@zju.edu.cn}

\affiliation[1]{Center for Theoretical Physics, Massachusetts Institute of Technology, Cambridge, MA 02139, USA\vspace{0.5ex}}
\affiliation[2]{Berkeley Center for Theoretical Physics, University of California, Berkeley, CA 94720, USA\vspace{0.5ex}}
\affiliation[3]{Theoretical Physics Group, Lawrence Berkeley National Laboratory, Berkeley, CA 94720, USA\vspace{0.5ex}}
\affiliation[4]{Theory Group, Deutsches Elektronen-Synchrotron (DESY), D-22607 Hamburg, Germany\vspace{0.5ex}}
\affiliation[5]{Department of Physics, Zhejiang University, Hangzhou, Zhejiang 310027, China\vspace{0.5ex}}

\abstract{
We continue the study of power corrections for $N$-jettiness
subtractions by analytically computing the complete next-to-leading power corrections at $\cO(\alpha_s)$ for color-singlet production. This includes
all nonlogarithmic terms and all partonic channels for Drell-Yan and gluon-fusion Higgs production.
These terms are important to further improve the numerical performance of the
subtractions, and to better understand the structure of power corrections
beyond their leading logarithms, in particular their universality.  
We emphasize the importance of computing the power corrections
differential in both the invariant mass, $Q$, and rapidity, $Y$, of the color-singlet system,
which is necessary to account for the rapidity dependence in the subtractions.
This also clarifies apparent disagreements in the literature. 
Performing a detailed numerical study, we find excellent agreement of our analytic results with a previous numerical extraction.
}

\maketitle

\section{Introduction}
\label{sec:intro}

The precision study of the Standard Model at the LHC, as well as increasingly sophisticated searches for physics beyond the Standard Model, require precision predictions for processes in a complicated hadron collider environment.
When calculating higher-order QCD corrections, the presence of infrared divergences require techniques to isolate and cancel all divergences. Completely analytic calculations are only possible for some of the simplest cases, e.g.~\cite{Anastasiou:2003ds, Dulat:2017brz, Mistlberger:2018etf}, while for more complicated processes, in particular those involving jets in the final state, numerical techniques are typically required.

At next-to-leading order (NLO) the FKS~\cite{Frixione:1995ms,Frixione:1997np} and CS~\cite{Catani:1996jh, Catani:1996vz, Catani:2002hc} subtraction schemes provide generic subtractions for arbitrary processes, and have been used to great success. At next-to-next-to-leading order (NNLO), due to the more complicated structure of infrared singularities, the development of general subtraction schemes has proven more difficult. While subtraction schemes have been demonstrated both for color-singlet production \cite{Catani:2007vq, Caola:2017dug, Herzog:2018ily, Magnea:2018hab, DelDuca:2016csb, DelDuca:2016ily, Kardos:2018kqd}, as well as for several processes involving jets in the final state \cite{GehrmannDeRidder:2005cm,Czakon:2010td,Boughezal:2011jf,Czakon:2014oma,Boughezal:2015dva,Boughezal:2015aha,Gaunt:2015pea}, significant work is still required before efficient NNLO subtractions can be achieved for arbitrary colored final states.

$N$-jettiness subtractions \cite{Boughezal:2015dva, Gaunt:2015pea} are based on the $N$-jettiness resolution variable $\Tau_N$~\cite{Stewart:2009yx, Stewart:2010tn}, and are applicable to generic $N$-jet final states. They have successfully been applied to NNLO calculations for a variety of color-singlet final states \cite{Campbell:2016jau, Campbell:2016yrh, Boughezal:2016wmq, Campbell:2017aul,Heinrich:2017bvg}, as well as final states involving a single jet \cite{Boughezal:2015dva, Boughezal:2015aha, Boughezal:2015ded, Boughezal:2016dtm,Campbell:2017dqk,Abelof:2016pby}. They are also a key ingredient in one of the first methods for combining NNLO calculations with parton showers~\cite{Alioli:2012fc, Alioli:2015toa}. 
The leading-power subtraction terms are given by an all-orders factorization formula derived in \refscite{Stewart:2009yx, Stewart:2010tn} using soft-collinear effective theory (SCET)~\cite{Bauer:2000ew, Bauer:2000yr, Bauer:2001ct, Bauer:2001yt, Bauer:2002nz}. Required ingredients are explicitly known to NNLO with up to a single jet in the final state  \cite{Becher:2006qw, Stewart:2010qs, Becher:2010pd, Berger:2010xi, Jouttenus:2011wh, Kelley:2011ng, Monni:2011gb, Hornig:2011iu, Kang:2015moa, Gaunt:2014xga, Gaunt:2014cfa, Boughezal:2015eha, Li:2016tvb, Boughezal:2017tdd, Campbell:2017hsw}.

An important feature of $N$-jettiness subtractions is that power corrections in the resolution variable can be calculated in an expansion about the soft and collinear limits, allowing the numerical performance of the subtractions to be systematically improved. Recently there has been significant interest in understanding subleading power corrections to collider cross sections~\cite{Laenen:2008ux,Laenen:2010uz,Bonocore:2014wua,Bonocore:2015esa,Kolodrubetz:2016uim,Bonocore:2016awd,Moult:2016fqy,Boughezal:2016zws,Moult:2017jsg,Moult:2017rpl,Feige:2017zci,DelDuca:2017twk,Chang:2017atu,Beneke:2017ztn,Boughezal:2018mvf,Bahjat-Abbas:2018hpv}.
Advances in the understanding of subleading power limits using SCET~\cite{Moult:2017rpl,Feige:2017zci} have allowed the leading logarithmic (LL) next-to-leading power (NLP) corrections to be computed at NLO and NNLO \cite{Moult:2016fqy,Moult:2017jsg}, with independent calculations of the same terms done by a second group in \refscite{Boughezal:2016zws,Boughezal:2018mvf}. The leading logarithms have also been resummed to all orders for pure glue QCD for 2-jettiness in $H\to gg$ \cite{Moult:2018jjd}.

\pagebreak[2]

The inclusion of the leading logarithms was found to improve the numerical accuracy, and thereby the computational efficiency, of $N$-jettiness subtractions for color-singlet production by up to an order of magnitude~\cite{Moult:2016fqy, Moult:2017jsg}. The analytic calculation of the next-to-leading logarithmic (NLL) power corrections is important for several reasons. Theoretically, it greatly furthers our understanding of the power corrections, since the perturbative structure becomes significantly more nontrivial at NLL as compared to at LL. From a practical perspective, they provide further substantial improvements in the numerical performance of the subtractions. In particular, they make the subtractions much more robust in cases where there are accidental cancellations between different channels or where the NLL terms are numerically enhanced relative to the LL terms.

In this paper, we compute the full NLP corrections at $\cO(\alpha_s)$ for both Drell-Yan and gluon-fusion Higgs production in all partonic channels, including the nonlogarithmic terms (which are the NLL terms at $\cO(\alpha_s)$).
One focus of this paper is to derive a master formula for the NLP corrections to $0$-jettiness at $\cO(\alpha_s)$ and to discuss in detail the subleading-power calculation, in particular the treatment of measurements at subleading power, which in our case are the invariant mass $Q$ and rapidity $Y$ of the color-singlet system.
More generally, one can consider measuring any observable that does not vanish for the partonic process at lowest order in perturbation theory,
which we refer to as Born measurements.
Our analysis lays the ground
for extending the calculation of the NLP corrections to higher powers, higher orders in $\alpha_s$,
and to more complicated processes.

We also perform a detailed numerical study comparing our analytic results with the full nonsingular result extracted numerically from MCFM8~\cite{Campbell:1999ah, Campbell:2010ff, Campbell:2015qma, Boughezal:2016wmq}, which allows us both to verify our analytic calculation and to probe the typical size of the higher-power corrections in various different partonic channels.
We find that the NLL power corrections can exhibit a much more pronounced $Y$ dependence from PDF effects than is the case at LL, which demonstrates the importance of calculating the power corrections fully differential in the Born phase space.

Our discussion of the Born measurements also allows us to clarify an apparent disagreement in the recent literature regarding the LL power corrections. As we discuss in more detail in \sec{compare_lit},
since the calculations in \refscite{Boughezal:2016zws, Boughezal:2018mvf} are not differential in the color-singlet rapidity $Y$, their results can only be used fully integrated over $Y$.%
\footnote{We thank the authors of \refscite{Boughezal:2016zws, Boughezal:2018mvf} for discussions
and confirmation of this point.}
In contrast, the results computed here, which agree with those previously derived by a subset of the present authors in \refscite{Moult:2016fqy, Moult:2017jsg}, are differential in $Y$.
When integrating over all $Y$, integration by parts is used to explicitly show that these LL results are equivalent to those of \refscite{Boughezal:2016zws, Boughezal:2018mvf}. In \sec{compare_lit} we also compare our new differential NLL results with the integrated results of \refcite{Boughezal:2018mvf}.

The outline of this paper is as follows: In \sec{Njettiness}, we briefly review $N$-jettiness subtractions for color-singlet production and define our notation. In \sec{mom_route}, we discuss in some detail the treatment of Born measurements at subleading power. In \sec{master_formula}, we derive a formula to NLP for the soft and collinear power corrections for $0$-jettiness. Although our primary focus is on NLO, the general strategy is valid to higher orders as well. In \sec{results_singlet}, we use our master formula to derive explicit results for the NLP power corrections at NLO for both Drell-Yan and gluon-fusion Higgs production. In \sec{compare_lit}, we provide a detailed comparison with the literature for those partonic channels where results are available. In \sec{results}, we present a detailed numerical study, and compare our analytic results with a previous numerical extraction. We conclude in \sec{conclusions}.

\section{\boldmath N-Jettiness Subtractions, Definitions and Notation}
\label{sec:Njettiness}

In this section we briefly review $N$-jettiness subtractions~\cite{Boughezal:2015dva, Gaunt:2015pea} in the context of color-singlet production, and discuss the structure of the power corrections to the subtraction scheme. This also allows us to define the notation that will be used in the rest of this paper. For a detailed discussion, we refer the reader to \refcite{Gaunt:2015pea}.

To compute a cross section for color-singlet production $\sigma(X)$, where $X$ denotes some set of cuts on the Born phase space, we write the cross section as an integral over the differential cross section in the resolution variable $\Tau_0$
\begin{align}
\sigma(X)
&= \sigma(X, \Tau_\cut) + \int_{\Tau_\cut}\! \df\Tau_0\, \frac{\df\sigma(X)}{\df\Tau_0}
\,,\end{align}
where
\begin{align}
\sigma(X, \Tau_\cut) &\equiv \int^{\Tau_\cut}\!\df\Tau_0\, \frac{\df\sigma(X)}{\df\Tau_0}
\,.
\end{align}

For a general measure, the 0-jettiness $\Tau_0$ can be defined as~\cite{Stewart:2010pd, Jouttenus:2011wh}
\begin{align} \label{eq:Tau0_0}
 \Tau_0 = \sum_i \min \biggl\{ \frac{2 q_a \cdot k_i}{Q_a} \,, \frac{2 q_b \cdot k_i}{Q_b} \biggr\}
\,,\end{align}
where the sum runs over all hadronic momenta $k_i$ in the final state.
Here, $q_{a,b}$ are projected Born momenta (referred to as label momenta in SCET), which are given in terms of the
total leptonic invariant mass $Q$ and rapidity $Y$ as
\begin{align} \label{eq:Born}
q_a^\mu = x_a \Ecm \frac{n^\mu}{2} = Q e^Y \frac{n^\mu}{2}
\,,\qquad
q_b^\mu = x_b \Ecm \frac{\bn^\mu}{2} = Q e^{-Y} \frac{\bn^\mu}{2}
\,,\end{align}
where
\begin{equation}
 n^\mu = (1,0,0,1) \,,\quad \bn^\mu = (1,0,0,-1)
\end{equation}
are lightlike vectors along the beam directions.
The choice in \eq{Born} corresponds to parameterizing the Born phase space in terms of $Q$ and $Y$,
and this choice already enters in the leading-power factorization theorem, where the beam functions
are evaluated at $x_{a,b} = Q e^{\pm Y}$.

The $Q_{a,b}$ measures in \eq{Tau0_0} determine the different definitions of 0-jettiness.
Two different definitions, originally introduced in \refscite{Stewart:2009yx, Berger:2010xi}
as beam thrust, are the leptonic and hadronic definitions given by
\begin{alignat}{4} \label{eq:Tau0_2}
 &\text{leptonic:}\quad & Q_a &= Q_b = Q \,,\qquad
  & \Tau_0^\lep
  &= \sum_i \min \biggl\{ \frac{x_a \Ecm}{Q} n \cdot k_i \,,\, \frac{x_b \Ecm}{Q} \bn \cdot k_i \biggr\}
\nn\\* &&&&
  &= \sum_i \min \biggl\{ e^Y n \cdot k_i \,,\, e^{-Y} \bn \cdot k_i \biggr\}
\nn\\
 &\text{hadronic:}\qquad & Q_{a,b} &= x_{a,b} \Ecm \,,
 & \Tau_0^\hadcm &= \sum_i \min \Bigl\{ n \cdot k_i \,,\, \bn \cdot k_i \Bigr\}
\,.\end{alignat}
It has been shown \cite{Moult:2016fqy} that the power corrections for the hadronic definition are poorly behaved, and grow exponentially with rapidity, while the $e^{\pm Y}$ factor in the measure for the leptonic definition exactly avoids this effect.

For later convenience, we write the dimensionful and dimensionless $0$-jettiness resolution variables
in terms of a $Y$-dependent parameter $\rho(Y)$ as
\begin{align}\label{eq:Tau0}
\Tau_0^x
&= \sum_i \min \Bigl\{ \rho_x \,k_i^+, \rho_x^{-1}\,k_i^- \Bigr\}
\,, \qquad
\tau^x \equiv \frac{\Tau_0^x}{Q}
\,,\end{align}
with
\begin{align}\label{eq:Njet_def}
\text{leptonic:} && &\rho_\lep  = e^Y\,, \qquad \tau^\lep=\frac{\Tau_0^\lep}{Q}
\,, \nn \\
\text{hadronic:} && & \rho_\hadcm = 1\,, \qquad \tau^{\hadcm}=\frac{\Tau_0^\hadcm}{Q}
\,.\end{align}
In the following, we will mostly drop the subscript $0$ on $\Tau_0$, since there should be no cause for confusion that our results are for 0-jettiness.
For generic results that apply to both the leptonic and hadronic definitions we also drop the superscript and simply use $\Tau$ and $\tau$, keeping a generic parameter $\rho$ when necessary.

To implement the $N$-jettiness subtractions, we now add and subtract a subtraction term to the cross section (suppressing the dependence on the Born measurements $X$ for simplicity)
\begin{align} \label{eq:nsub_master}
\sigma
&=
\sigma^\sub(\Tau_{\cut})
+ \int_{\Tau_\cut}\! \df\Tau_0\, \frac{\df\sigma}{\df\Tau_0}
+ \bigl[\sigma(\Tau_\cut) - \sigma^\sub(\Tau_\cut) \bigr]
\nn \\
&\equiv \sigma^\sub(\Tau_\cut) + \int_{\Tau_\cut}\! \df\Tau_0\, \frac{\df\sigma}{\df\Tau_0}
+ \Delta \sigma(\Tau_\cut)
\,.\end{align}
Since $\Tau_0$ is a zero jet resolution variable, for $\tau = \Tau_0/Q \to 0$ we can expand the differential cross section $\df\sigma/\df\tau$ and its cumulative $\sigma(\tau_\cut)$ about the soft and collinear limits from $\tau\to 0$ and $\tau_\cut\to 0$ as
\begin{align}\label{eq:xsec_expand}
\frac{\df \sigma}{\df\tau}
&= \frac{\df\sigma^{(0)}}{\df\tau} + \frac{\df\sigma^{(2)}}{\df\tau}+ \frac{\df\sigma^{(4)}}{\df\tau} + \dotsb
\,, \\ \nn
\sigma(\tau_\cut)
&= \sigma^{(0)}(\tau_\cut) + \sigma^{(2)}(\tau_\cut) + \sigma^{(4)}(\tau_\cut) + \dotsb
\,.\end{align}
Here $\df\sigma^{(0)}/\df\tau$ and $\sigma^{(0)}(\tau_\cut)$ contain all leading-power terms,
\begin{align}
\frac{\df\sigma^{(0)}}{\df\tau}
\sim \delta(\tau)+ \biggl[\frac{ \ln^j\tau  }{\tau}\biggr]_+
\,, \qquad
\sigma^{(0)}(\tau_\cut) \sim \ln^{j}\tau_\cut
\,.\end{align}
These terms must be included in the subtraction term to obtain a finite result, namely
\begin{equation}
\sigma^\sub(\Tau_\cut) = \sigma^{(0)}(\tau_\cut = \Tau_\cut/Q)\, [1 + \ord{\tau_\cut}]
\,.\end{equation}
The further terms in the series expansion in \eq{xsec_expand} are suppressed by powers of $\tau$
\begin{align}\label{eq:scaling_lam2}
\tau \frac{\df\sigma^{(2k)}}{\df\tau} &\sim \ord{\tau^k \ln^j\tau}
\,, \qquad
\sigma^{(2k)}(\tau_\cut) \sim \ord{\tau_\cut^k \ln^j\tau_\cut}
\,.\end{align}
While these terms with $k\ge 1$ do not need to be included in the subtraction term, the size of the neglected term, $\Delta \sigma(\tau_\cut)$ is determined by the leading-power corrections that are left out of $\sigma^{\rm sub}$. Therefore, including additional power corrections in $\sigma^{\rm sub}$ can significantly improve the performance of the subtraction. Indeed, general scaling arguments imply that up to an order of magnitude in performance can be gained for each subleading power logarithm that is included in the subtractions~\cite{Gaunt:2015pea}. For the leading logarithms, this was explicitly confirmed for most partonic channels in the numerical studies in \refscite{Moult:2016fqy, Moult:2017jsg}. Here, we extend the calculation to the NLL terms at $\cO(\alpha_s)$, which yields the nonlogarithmic terms, hence giving the complete NLP result. The remaining NLO power corrections then scale at worst as $\alpha_s \tau_{\cut}^2 \log \tau_{\cut}$, and will be very small, as we will see in our numerical studies.

\section{Born Measurements at Subleading Power}
\label{sec:mom_route}

We begin by discussing in some detail the treatment of the Born measurements, $Q^2$ and $Y$, which plays an important role at subleading powers. We will use the soft and collinear expansions from SCET, which  provide a convenient language when discussing the power expansion of QCD amplitudes at fixed order. We will not need to employ any of the field theory technology from SCET for our analysis here.

\subsection{General Setup and Notation}

Consider the production of a color-singlet final state $L$ of fixed invariant mass $Q$ and rapidity $Y$,
together with an arbitrary measurement $\Tau$
that only acts on hadronic radiation and gives $\Tau=0$ at Born level.
Since the observable $\Tau$ resolves soft and collinear emissions
it will induce large logarithms $\ln(\Tau/Q)$.
Our goal is to expand the cross section in $\Tau$ (or $\tau = \Tau/Q$)
in order to systematically understand its logarithmic structure.

Consider proton-proton scattering with the underlying partonic process
\begin{align}
 a(p_a) + b(p_b) \to L(p_1, \cdots) + X(k_1, \cdots)
\,,\end{align}
where $L$ is the leptonic (color-singlet) final state and $X$ denotes additional QCD radiation.
Its cross section reads
\begin{align} \label{eq:sigma1}
 \frac{\df\sigma}{\df Q^2 \df Y \df\Tau} &
 = \int_{0}^{1}\!\! \df \zeta_a \df \zeta_b\, \frac{f_a(\zeta_a)\, f_b(\zeta_b)}{2 \zeta_a \zeta_b \Ecm^2}
   \int\!\biggl(\prod_i \frac{\df^d k_i}{(2\pi)^d} (2\pi) \delta_+(k_i^2) \biggr)
   \int\!\!\frac{\df^d q}{(2\pi)^d} \, |\cM(p_a, p_b; \{k_i\}, q)|^2
   \nn\\* &\quad\times
   (2\pi)^d \delta^{(d)}(p_a + p_b - k - q) \, \delta(Q^2 - q^2)
   \, \delta\biggl(Y - \frac{1}{2}\ln\frac{q^-}{q^+}\biggr)
   \, \delta\bigl[\Tau - \hat\Tau(\{k_i\})\bigr]
\,.\end{align}
Here, the incoming momenta are given by
\begin{align} \label{eq:p_ab}
 p_a^\mu = \zeta_a \Ecm \frac{n^\mu}{2}
\,,\qquad
 p_b^\mu = \zeta_b \Ecm\frac{\bn^\mu}{2}
\,,\end{align}
$k = \sum_i k_i$ is the total outgoing hadronic momentum, and $q$ is the total leptonic momentum.
Since our measurements are not sensitive to the details of the leptonic final state,
we have absorbed the leptonic phase space integral into the matrix element,
\begin{align} \label{eq:M2_1}
 |\cM(p_a, p_b; \{k_i\}, q)|^2 &= \int\df\Phi_L(q) \, |\cM(p_a, p_b; \{k_i\}, \{p_j\})|^2
\,,\nn\\
  \df\Phi_L(q) &= \prod_j \frac{\df^d p_{j}}{(2\pi)^d} (2\pi) \delta_+(p_{j}^2 - m_j^2)
\, (2\pi)^d \delta^{(d)}\Bigl(q - \sum_j p_j\Bigr)
\,.\end{align}
The matrix element $\cM$ contains the renormalization scale $\mu^{2\eps}$,
which as always is associated with the renormalized coupling $\alpha_s(\mu)$, and may also contain virtual corrections.
For now the measurement function $\hat\Tau(\{k_i\})$ is kept arbitrary.

We can now solve the $Q^2$ and $Y$ measurements to fix the incoming momenta as
\begin{align} \label{eq:zeta_ab}
 \zeta_a(k) &= \frac{1}{\Ecm} \biggl(k^- +  e^{+Y} \sqrt{Q^2 + k_T^2} \biggr)
\,,\nn\\
 \zeta_b(k) &= \frac{1}{\Ecm} \biggl(k^+ +  e^{-Y} \sqrt{Q^2 + k_T^2} \biggr)
\,.\end{align}
Taking the Jacobian factors from solving the $\delta$ functions into account, \eq{sigma1} becomes
\begin{align} \label{eq:sigma2}
 \frac{\df\sigma}{\df Q^2 \df Y \df\Tau} &
 = \int\!\biggl(\prod_i \frac{\df^d k_i}{(2\pi)^d} (2\pi) \delta_+(k_i^2) \biggr)
   \frac{f_a(\zeta_a)\, f_b(\zeta_b)}{2 \zeta_a \zeta_b \Ecm^4}
   \Msquared(Q, Y; \{k_i\}) \delta\bigl[\Tau - \hat\Tau(\{k_i\})\bigr]
\,,\end{align}
where we defined
\begin{align} \label{eq:Msquared}
 \Msquared(Q, Y; \{k_i\}) \equiv |\cM(p_a, p_b, \{k_i\}, q=p_a+p_b - k)|^2
\end{align}
to stress that the squared matrix element only depends on the Born measurements $Q$ and $Y$,
which fix the incoming momenta through \eqs{p_ab}{zeta_ab},
and the emission momenta $k_i$.
Note that we have left implicit in our notation in \eq{sigma2} the dependence of $\zeta_{a,b}$ on $k$ through \eq{zeta_ab}.
They are restricted to $\zeta_{a,b} \in [0,1]$, which is implicit in the support of the proton PDFs.

\subsection{Power Expansion in Soft and Collinear Limits}

Instead of solving the $\Tau$ measurement function to express (some of the) $k_i$ in terms of $\Tau$, we find that a convenient strategy to organize the expansion in $\Tau$ is to multipole expand the final state momenta.
At this stage we need only assume that $\Tau$ is a \SCETI observable,
which is true for many definitions of $N$-jettiness. For such observables, it is known from \SCETI that
we can organize the cross section in terms of a power counting parameter $\lambda \sim \sqrt{\tau}$.
All momenta $k_i$ can then be categorized as either collinear or soft modes
(since we work in \SCETI these are often called ultrasoft, although we will not make this distinction),
whose momenta scale as
\begin{align} \label{eq:modes}
 &n{-}\text{collinear}: \quad k_n \sim Q \, (\lambda^2, 1, \lambda)
\,,\\
 \nn&\bn{-}\text{collinear}: \quad k_{\bn} \sim Q \, (1, \lambda^2, \lambda)
\,,\\
 \nn&\text{soft}: \hspace{1.8cm} k_s \sim Q \, (\lambda^2, \lambda^2, \lambda^2)
\,,\end{align}
where we decomposed each momentum into lightcone coordinates
\begin{equation}
 k^\mu = k^- \frac{n^\mu}{2} + k^+ \frac{\bn^\mu}{2} + k_\perp^\mu \equiv (k^+, k^-, k_\perp)
\,.\end{equation}
Here $n$ and $\bar n$ are lightlike vectors satisfying $n\cdot \bn=2$. The components of the momenta that scale like $\lambda^2$ are referred to as residual momenta.
The soft momenta are homogeneous, and have purely residual scaling. Overlap between the soft and collinear modes occurring in integrals over final state momenta is removed by the zero-bin subtraction procedure \cite{Manohar:2006nz}.

The benefit of this decomposition is that it allows one to expand \eq{sigma2} in $\lambda$,
agnostic of the actual measurement $\Tau$.
The LP result is then simply obtained by expanding the cross section through $\lambda^0$,
the NLP result by expanding through $\lambda^2$, etc.
Note that when performing this expansion, all other factors, such as $Q, E_{\rm cm} \sim \lambda^0$.

While the expansion of the matrix element is of course process dependent,
we can give general expressions for the incoming momentum fractions \eq{zeta_ab},
independent of the process and observable $\Tau$.
If $k$ is a soft momentum, then the expansion required at NLP is given by
\begin{align} \label{eq:zeta_ab_soft}
 \zeta_a(k) &= x_a \biggl(1 +  \frac{k^- e^{-Y}}{Q} \biggr) + \cO(\lambda^4)
\,,\nn\\
 \zeta_b(k) &= x_b \biggl(1 +  \frac{k^+ e^{+Y}}{Q} \biggr) + \cO(\lambda^4)
\,,\end{align}
where we factored out the Born momentum fractions
\begin{align}
 x_a = \frac{Q e^{+Y}}{\Ecm}
\,,\qquad
 x_b = \frac{Q e^{-Y}}{\Ecm}
\,.\end{align}
In the $n$-collinear limit, we obtain
\begin{align} \label{eq:zeta_ab_coll}
 \zeta_a(k) &= x_a \biggl[ \biggl(1 + \frac{k^- e^{-Y}}{Q}\biggr) + \frac{k_T^2}{2 Q^2} \biggr] + \cO(\lambda^4)
\,,\nn\\
 \zeta_b(k) &= x_b \biggl[ 1 + \biggl(\frac{k^+ e^{+Y}}{Q} + \frac{k_T^2}{2 Q^2} \biggr) \biggr] + \cO(\lambda^4)
\,.\end{align}
For clarity, we have grouped terms of the same power counting in round brackets.
Similarly, one can obtain the $\bn$-collinear limit,
or any combination as might appear when combining multiple emissions.

\section{Master Formula for Power Corrections to Next-to-Leading Power}
\label{sec:master_formula}

In this section we derive a master formula for the NLP corrections.
This formula applies to any \SCETI observable in color-singlet production.
In \sec{results_singlet}, we will apply it to derive explicit results for Drell-Yan and gluon-fusion Higgs production.

\subsection{General Setup for Color-Singlet \texorpdfstring{\SCETI}{SCET I} Observables}

For reference, we start with the LO cross section for the production
of a color-singlet final state $L$ of invariant mass $Q^2$ and rapidity $Y$,
together with an (up to now arbitrary) measurement $\Tau$ acting only on hadronic radiation,
\begin{align} \label{eq:sigmaLO}
 \frac{\df\sigma^\LO}{\df Q^2 \df Y \df\Tau} &
 = \frac{f_a(x_a)\, f_b(x_b)}{2 x_a x_b \Ecm^4}
   \Msquared^\LO(Q, Y)\, \delta(\Tau)
\,,\end{align}
where $x_{a,b} = \frac{Q}{\Ecm} e^{\pm Y}$ and $\Msquared^\LO$ is the squared matrix element
in the Born kinematics, see \eq{Msquared}.
For future reference, we also define the LO partonic cross section, $\hat \sigma^{\rm LO}(Q,Y)$, by
\begin{align} \label{eq:sigmaLO_2}
  \frac{\df\sigma^\LO}{\df Q^2 \df Y \df\Tau} &
 = \hat\sigma^\LO(Q,Y) \, f_a(x_a)\, f_b(x_b) \, \delta(\Tau)
\,,\qquad
  \hat\sigma^\LO(Q,Y) = \frac{\Msquared^\LO(Q, Y)}{2 x_a x_b \Ecm^4}
\,.\end{align}
Next, consider an additional real emission to the Born process.
Eq.~\eqref{eq:sigma2} yields
\begin{align} \label{eq:sigmaNLO_1}
 \frac{\df\sigma}{\df Q^2 \df Y \df\Tau} &
 = \int\! \frac{\df^d k}{(2\pi)^d}\, (2\pi) \delta_+(k^2)\,
   \frac{f_a(\zeta_a)\, f_b(\zeta_b)}{2 \zeta_a \zeta_b \Ecm^4}\,
   \Msquared(Q, Y; \{k\})\, \delta\bigl[\Tau - \hat\Tau(\{k\})\bigr]
\,,\end{align}
where we remind the reader that the incoming momenta $p_{a,b}$ are given by \eq{zeta_ab},
\begin{align} \label{eq:p_ab_2}
 p_a^\mu &= \zeta_a(k) \Ecm \frac{n^\mu}{2}
         = \biggl(k^- +  e^{+Y} \sqrt{Q^2 + k_T^2} \biggr) \frac{n^\mu}{2}
\,,\nn\\
 p_b^\mu &= \zeta_b(k) \Ecm \frac{\bn^\mu}{2}
         = \biggl(k^+ +  e^{-Y} \sqrt{Q^2 + k_T^2} \biggr) \frac{\bn^\mu}{2}
\,.\end{align}
From these solutions, we see the interesting feature that at subleading power, regardless of the type of final-state emission, the momenta entering both PDFs are modified.

Since we do not measure the azimuthal angle of $k$, it can be integrated over,
\begin{align}
 \int\! \frac{\df^dk}{(2\pi)^d}\, (2\pi) \delta_+(k^2)
= \frac{\Omega_{2-2\eps}}{4 (2\pi)^{d-1}} \int_0^\infty \frac{\df k^+ \df k^-}{(k^+ k^-)^\eps}
= \frac{(4\pi)^{-2+\eps}}{\Gamma(1-\eps)} \int_0^\infty \frac{\df k^+ \df k^-}{(k^+ k^-)^\eps}
\,.\end{align}
Eq.~\eqref{eq:sigmaNLO_1} simplifies to
\begin{align} \label{eq:sigmaNLO_2}
 \frac{\df\sigma}{\df Q^2 \df Y \df\Tau} &
 = \int_0^\infty \frac{\df k^+ \df k^-}{(k^+ k^-)^\eps}
   \frac{f_a(\zeta_a)\, f_b(\zeta_b)}{(4\pi)^2 2 \zeta_a \zeta_b \Ecm^4}\,
   \frac{(4\pi)^{\eps}}{\Gamma(1-\eps)}\, \Msquared(Q, Y; \{k\})\,
   \delta\bigl[\Tau - \hat\Tau(\{k\})\bigr]
\,.\end{align}
So far, this expression is still exact.
In the next step, we wish to expand the NLO cross section in $\lambda \sim \Tau/Q$.
When $\Tau$ is a \SCETI observable, we can use the EFT knowledge from \SCETI
to expand the momentum $k$ in both collinear and soft limits, as discussed in \sec{mom_route}.

At NLP, we need to expand \eq{sigmaNLO_2} consistently through $\cO(\lambda^2)$.
The $\cO(\lambda^2)$ power corrections arise from the following sources:
\begin{itemize}
 \item The incoming momenta $\zeta_{a,b}$: While collinear and soft limits yield
 quite different power expansions, both give a well-defined expansion in $\lambda$.
 We thus simply define the expansion
 \begin{align} \label{eq:zeta_ab_expansion}
  \zeta_{a,b} = x_{a,b} \biggl[ \frac{1}{z_{a,b}} + \Delta_{a,b}^{(2)} + \cO(\lambda^4) \biggr]
 \,,\end{align}
 where $z_{a,b} \sim \lambda^0$ and $\Delta_{a,b}^{(2)} \sim \lambda^2$.
 We have pulled out the Born momentum fractions $x_{a,b}$ and written $1/z_{a,b}$ as a fraction
 for later convenience.
 Explicit expressions can be obtained from \eqs{zeta_ab_soft}{zeta_ab_coll},
 and will be given below.
 \item PDFs: Since the momenta $\zeta_{a,b}$ enter the PDFs, these also have to be power expanded,
 \begin{align}
  f_{a,b}(\zeta_{a,b})
  &= f_{a,b}\biggl(\frac{x_{a,b}}{z_{a,b}}\biggr) + x_{a,b} \Delta_{a,b}^{(2)} \, f'_{a,b}\biggl(\frac{x_{a,b}}{z_{a,b}}\biggr) + \cO(\lambda^4)
  \nn\\&
  \equiv f_{a,b} + x_{a,b}\Delta_{a,b}^{(2)} \, f'_{a,b} + \cO(\lambda^4)
 \,.\end{align}
 \item Flux factor: Similar to the PDFs, we have to expand the flux factor
 \begin{align}
  \frac{1}{\zeta_a \zeta_b} = \frac{z_a z_b}{x_a x_b} \biggl[ 1 - z_a \Delta^{(2)} - z_b \Delta^{(2)}_b \biggr]
 \,.\end{align}
 \item Matrix element: The expansion of the matrix element depends both on the process and the considered limit.
 Here, we define the LP and NLP expansions by
 \begin{align} \label{eq:expansion_M2}
  \Msquared(Q, Y; \{k\}) = \Msquared^{(0)}(Q,Y; \{k\})) + \Msquared^{(2)}(Q,Y; \{k\}) + \cdots
 \,.\end{align}
 In the soft limit $\Msquared^{(0)} \sim \lambda^{-4}$ and $\Msquared^{(2)} \sim \lambda^{-2}$,
 while in the collinear limit $\Msquared^{(0)} \sim \lambda^{-2}$ and $\Msquared^{(2)} \sim \lambda^{0}$. In both cases we have the scaling $\int \df k^+ \df k^- A^{(2j)} \sim \lambda^{2j}$, which is why the soft and collinear corrections enter at the same order.

 For example, for a $2 \to 2$ process, the matrix element can be written in terms
 of the Mandelstam variables
 \begin{align} \label{eq:mandelstamgeneral}
  s_{ab} &= 2 p_a \cdot p_b = \frac{Q^2}{z_az_b}\left[ 1 + z_a \Delta^{(2)}_a + z_b \Delta^{(2)}_b  + \cO(\lambda^4) \right]
  \,,\nn \\
  s_{ak} &= -2 p_a \cdot k = -k^+ Q e^{+Y} \left(\frac{1}{z_a} +\Delta^{(2)}_a  + \cO(\lambda^4) \right)
  \,,\nn \\
  s_{bk} &= -2 p_b \cdot k = -k^- Q e^{-Y}\left(\frac{1}{z_b} +\Delta^{(2)}_b  + \cO(\lambda^4) \right)
 \,.\end{align}
 Since these terms now have a definite power counting, one can simply insert \eq{mandelstamgeneral}
 into $\Msquared(Q, Y; \{k\})$ and expand to the required order in $\lambda$.
 \item Measurement: Depending on the observable, the measurement function $\hat\Tau$ may also
 receive power corrections. Since $0$-jettiness is defined in terms of $n$, $\bar n$, $Q$ and $Y$, none of which receive power corrections in our approach, we do not have such corrections, and will therefore not write them explicitly in the following formulae.  More generally, these could be obtained from expanding $\delta\bigl[\Tau - \hat\Tau(\{k\})\bigr]$.
\end{itemize}
Inserting all these expansions into \eq{sigmaNLO_2} and expanding consistently to $\cO(\lambda^2)$, we obtain the LP result
\begin{align} \label{eq:sigmaLP}
 \frac{\df\sigma^{(0)}}{\df Q^2 \df Y \df\Tau} &
 = \int_0^\infty \frac{\df k^+ \df k^-}{(k^+ k^-)^\eps}
   \frac{z_a z_b \, f_a f_b}{(4\pi)^2 2 x_a x_b \Ecm^4}
   \frac{(4\pi)^{\eps}}{\Gamma(1-\eps)} \Msquared^{(0)}(Q, Y; \{k\}) \,
   \delta\bigl[\Tau - \hat\Tau(\{k\})\bigr]
\end{align}
and the NLP master formula
\begin{align} \label{eq:sigmaNLP}
 \frac{\df\sigma^{(2)}}{\df Q^2 \df Y \df\Tau} &
 = \int_0^\infty \frac{\df k^+ \df k^-}{(k^+ k^-)^\eps}
   \frac{z_a z_b}{(4\pi)^2 2 x_a x_b \Ecm^4}
   \frac{(4\pi)^{\eps}}{\Gamma(1-\eps)} \delta\bigl[\Tau - \hat\Tau(\{k\})\bigr]
   \nn\\&\quad\times \Bigl\{
   \Msquared^{(0)}(Q, Y; \{k\}) \Bigl[
    f_a f_b \Bigl(-z_a \Delta_a^{(2)} - z_b \Delta_b^{(2)} \Bigr)
   + x_a \Delta_a^{(2)} f'_a f_b + x_b \Delta_b^{(2)} f_a f'_b \Bigr]
   \nn\\&\qquad
   + f_a f_b \, \Msquared^{(2)}(Q, Y; \{k\})
   \Bigr\}
\,,\end{align}
where $z_{a,b} \equiv z_{a,b}(k)$ and $\Delta_{a,b}^{(2)} \equiv \Delta_{a,b}^{(2)}(k)$ are defined by \eq{zeta_ab_expansion}.
Note that the LP limits of the matrix elements are universal,
and hence \eq{sigmaLP} holds independently of the process,
i.e.\ it only depends on the observable $\Tau$.
Although the focus of this paper is on the power corrections,
in \app{app_LP} we provide a brief derivation of the leading-power terms.
Likewise, the $A^{(0)}$ term together with the square bracketed factor on the second line of \eq{sigmaNLP} is universal, such that all the process dependence arises from the last $A^{(2)}$ term.
We will discuss this in more detail in \sec{universality}.

In the following, we will evaluate \eq{sigmaNLP} in both the soft and collinear limit
for $0$-jettiness, \eq{Tau0}, whose measurement function for one emission is given by
\begin{equation} \label{eq:delta_Tau}
 \delta\bigl[\Tau - \hat\Tau(k)\bigr]
 = \Theta(\rho k^+ - \rho^{-1} k^-) \delta(\Tau - \rho^{-1} k^-)
 + \Theta(\rho^{-1} k^- - \rho k^+) \delta(\Tau - \rho k^+)
\,.\end{equation}
The value of $\rho$ depends on the specific definition of $\Tau$, as given in \eq{Njet_def}.

\subsection{Collinear Master Formula for 0-Jettiness}
\label{sec:sigma_NLO_NLP_coll}

The expansion of the incoming momenta $\zeta_{a,b}$ for an $n$-collinear emission
$k \sim (\lambda^2, 1, \lambda)$ has been given in \eq{zeta_ab_coll},
\begin{align} \label{eq:zeta_ab_coll2}
 \zeta_a(k) &= x_a \biggl[ \biggl(1 + \frac{k^- e^{-Y}}{Q}\biggr) + \frac{k_T^2}{2 Q^2} \biggr] + \cO(\lambda^4)
\,,\nn\\
 \zeta_b(k) &= x_b \biggl[ 1 + \biggl(\frac{k^+ e^{+Y}}{Q} + \frac{k_T^2}{2 Q^2} \biggr) \biggr] + \cO(\lambda^4)
\,,\end{align}
so the explicit expressions for the expansion \eq{zeta_ab_expansion} are
\begin{alignat}{3} \label{eq:auxiliary_coll}
 & z_a &&= \biggl(1 + \frac{k^- e^{-Y}}{Q}\biggr)^{-1}
\,,\quad
 && \Delta^{(2)}_a = \frac{k_T^2}{2 Q^2}
\,,\nn\\
 & z_b &&= 1
\,,\quad
 && \Delta^{(2)}_b = \biggl(\frac{k^+ e^{+Y}}{Q} + \frac{k_T^2}{2 Q^2} \biggr)
\,.\end{alignat}

Since an $n$-collinear emission satisfies $k^- \gg k^+$,
the 0-jettiness measurement \eq{delta_Tau} simplifies to
\begin{equation} \label{eq:delta_Tau_coll}
 \delta\bigl[\Tau - \hat\Tau(k)\bigr] = \delta(\Tau - \rho k^+)
\,.\end{equation}
Note that the integration in \eq{sigmaNLP} also includes the region $k^- \to 0$,
where the assumption $k^- \gg k^+$ is invalid.
Indeed, this region corresponds to the soft expansion.
It is guaranteed by the zero-bin subtraction procedure that
this overlap regime between the soft and collinear limits
is not double counted \cite{Manohar:2006nz}.
An important benefit of $0$-jettiness and our setup here is that the
zero-bin contribution that removes the overlap is scaleless and vanishes in
pure dimensional regularization, such that we do not need to consider it further.

Eq.~\ref{eq:delta_Tau_coll} fixes the $k^+$ integral in \eq{sigmaNLP}.
It is also useful to write the remaining $k^-$ integration in terms of $z_a$ using \eq{auxiliary_coll}, giving
\begin{align}
 k^- = Q e^Y \frac{1 - z_a}{z_a}
\,,\qquad
 \int_0^\infty \df k^- = \int_{x_a}^1 \frac{\df z_a}{z_a^2} \, Q e^Y
\,.\end{align}
Here the lower bound on the integration follows from the physical support of the PDF
$f_a(x_a/z_a)$.
Plugging back into \eq{sigmaNLP}, we obtain the $n$-collinear master formula
\begin{align} \label{eq:sigma_NLO_NLP_coll_FULL}
 \frac{\df\sigma_n^{(2)}}{\df Q^2 \df Y \df\Tau} &
 = \int_{x_a}^1 \frac{\df z_a}{z_a} \,
   \frac{1}{2 x_a x_b \Ecm^4}
   \frac{Q e^Y}{\rho} \biggl(\frac{Q \Tau e^Y}{\rho}\biggr)^{-\eps}
   \frac{z_a^\eps}{(1-z_a)^\eps}
   \frac{(4\pi)^{\eps}}{(4\pi)^2 \Gamma(1-\eps)}
   \nn\\&\quad\times \biggl\{
   \frac{\Tau e^Y}{Q \rho} \Msquared^{(0)}(Q, Y, \{k\}) \biggl[
    f_a\, f_b\, \frac{(1-z_a)^2 - 2}{2 z_a}
   + \frac{1-z_a}{2z_a}\, x_a  f'_a\, f_b + \frac{1+z_a}{2z_a}\, f_a\, x_b  f'_b \biggr]
   \nn\\&\qquad
   + f_a\, f_b \, \Msquared^{(2)}(Q, Y, \{k\})
   \biggr\}
\,,\end{align}
where $k$ is given by
\begin{align} \label{eq:k_collinear}
 k^\mu = Q e^Y \frac{1-z_a}{z_a} \frac{n^\mu}{2} + \frac{\Tau}{\rho} \frac{\bn^\mu}{2}
       + \sqrt{Q \Tau \frac{e^Y}{\rho} \frac{1-z_a}{z_a}} n_\perp^\mu
\,.\end{align}
It only remains to plug in the expansions of the matrix element $A^{(0)}$ and $A^{(2)}$ and to expand in $\eps$.
Note that even in the $n$-collinear case considered here, both the $n$ and $\bn$-collinear incoming momenta
receive power corrections, see \eq{p_ab_2}, leading to derivatives of both PDFs in \eq{sigma_NLO_NLP_coll_FULL}.

The analogous results for the $\bn$-collinear limit are obtained in the same manner, giving
\begin{align} \label{eq:sigma_NLO_NLP_nbcoll_FULL}
\frac{\df\sigma_{\bn}^{(2)}}{\df Q^2 \df Y \df\Tau} &
= \int_{x_b}^1 \frac{\df z_b}{z_b} \, \frac{1}{2 x_a x_b \Ecm^4}
   \frac{Q \rho}{e^Y} \biggl(\frac{Q \Tau \rho}{e^Y}\biggr)^{-\eps}
   \frac{z_b^\eps}{(1-z_b)^\eps}
   \frac{(4\pi)^{\eps}}{(4\pi)^2 \Gamma(1-\eps)}
   \nn\\&\quad\times \biggl\{
   \frac{\Tau \rho}{Q e^Y} \Msquared^{(0)}(Q, Y, \{k\}) \biggl[
   f_a f_b \frac{(1-z_b)^2 - 2}{2 z_b}
   + \frac{1+z_b}{2z_b}\, x_a f'_a\, f_b
   + \frac{1-z_b}{2z_b}\, f_a\, x_b f'_b \biggr]
   \nn\\&\qquad
   + f_a\, f_b \, \Msquared^{(2)}(Q, Y, \{k\})
   \biggr\}
\,,\end{align}
where $k$ is given by
\begin{align} \label{eq:k_bncollinear}
 k^\mu = \Tau \rho \, \frac{n^\mu}{2} + \frac{Q}{e^Y} \frac{1-z_b}{z_b}  \frac{\bn^\mu}{2}
       + \sqrt{Q \Tau \frac{\rho}{e^Y} \frac{1-z_b}{z_b}} n_\perp^\mu
\,.\end{align}
%

\subsection{Soft Master Formula for 0-Jettiness}
\label{sec:softmaster}

The expansion of the incoming momenta $\zeta_{a,b}$ for a soft emission
$k \sim (\lambda^2, \lambda^2, \lambda^2)$ has been given in \eq{zeta_ab_soft},
\begin{align}
 \zeta_a(k) &= x_a \biggl(1 +  \frac{k^- e^{-Y}}{Q} \biggr) + \cO(\lambda^4)
\,,\nn\\
 \zeta_b(k) &= x_b \biggl(1 +  \frac{k^+ e^{+Y}}{Q} \biggr) + \cO(\lambda^4)
\,,\end{align}
so the explicit expressions for the expansion \eq{zeta_ab_expansion} are
\begin{alignat}{3} \label{eq:auxiliary_soft}
 & z_a &&= 1
\,,\qquad
 && \Delta^{(2)}_a = \frac{k^- e^{-Y}}{Q}
\,,\nn\\
 & z_b &&= 1
\,,\qquad
 && \Delta^{(2)}_b = \frac{k^+ e^{+Y}}{Q}
\,.\end{alignat}
Plugging back into \eq{sigmaNLP}, we get
\begin{align} \label{eq:softmaster}
\frac{\df\sigma_s^{(2)}}{\df Q^2 \df Y \df\Tau}
&= \int_0^\infty \frac{\df k^+ \df k^-}{(k^+ k^-)^\eps}
   \frac{1}{(4\pi)^2 2 x_a x_b \Ecm^4}
   \frac{(4\pi)^{\eps}}{\Gamma(1-\eps)} \delta\bigl[\Tau - \hat\Tau(\{k\})\bigr]
\biggl\{ \frac{1}{Q} \Msquared^{(0)}(Q, Y, \{k\})
\nn\\&\qquad\times
\biggl[
   f_a\, f_b\, (-k^- e^{-Y} - k^+ e^{+Y})
   + k^- e^{-Y}\, x_a f'_a\, f_b + k^+ e^{+Y}\, f_a\, x_b f'_b \biggr]
\nn\\&\quad
   + f_a\, f_b \, \Msquared^{(2)}(Q, Y, \{k\})
\biggr\}
\,.\end{align}
Here, the measurement is given by \eq{delta_Tau},
\begin{equation}
 \delta\bigl[\Tau - \hat\Tau(k)\bigr]
 = \Theta(\rho k^+ - \rho^{-1} k^-) \delta(\Tau - \rho^{-1} k^-)
 + \Theta(\rho^{-1} k^- - \rho k^+) \delta(\Tau - \rho k^+)
\,.\end{equation}
We can further simplify \eq{softmaster} by utilizing the fact that
$\Msquared^{(0)}$ and $\Msquared^{(2)}$ have a well defined dependence on $k^+$ or $k^-$
because of power counting and mass dimension,
\begin{align} \label{eq:altMdef}
 \Msquared^{(0)}(Q, Y, \{k\}) = \frac{\altM^{(0)}(Q,Y)}{k^+ k^-}
\,,\quad
 \Msquared^{(2)}(Q, Y, \{k\}) = \frac{\altM_+^{(2)}(Q,Y)}{k^+} + \frac{\altM_-^{(2)}(Q,Y)}{k^-}
\,.\end{align}
Here the $\altM$'s are process-dependent expressions that depend on the Born measurements $Q$ and $Y$,
but are independent of both $k^+$ and $k^-$.
This implies that the $k^\pm$ integrals in \eq{softmaster} have the generic structure
\begin{align} \label{eq:I_rho}
 \int_0^\infty \frac{\df k^+ \df k^-}{(k^+ k^-)^\eps} \frac{\delta\bigl[\Tau - \hat\Tau(k)\bigr]}{(k^+)^\alpha (k^-)^\beta}
 = \rho^{\alpha-\beta} \Tau^{1-\alpha-\beta-2\eps} \biggl(\frac{1}{\eps+\alpha-1} + \frac{1}{\eps+\beta-1} \biggr)
\,.\end{align}
We then find the soft NLP master formula
\begin{align} \label{eq:sigma_NLO_NLP_soft}
\frac{\df\sigma_s^{(2)}}{\df Q^2 \df Y \df\Tau}
&= \frac{1}{(4\pi)^2 2 x_a x_b \Ecm^4}
   \frac{(4\pi)^{\eps}}{\Gamma(1-\eps)}
   \frac{1}{\eps} \frac{\Tau^{-2\eps}}{Q} \frac{1-2\eps}{1-\eps}\,
   \biggl\{ \altM^{(0)}(Q,Y)
\nn\\ & \qquad \times
\biggl[ f_a(x_a)\, f_b(x_b) \biggl(- \frac{\rho}{e^Y} -\frac{e^{Y}}{\rho} \biggr)
   + \frac{\rho}{e^Y}\, x_a f'_a(x_a)\, f_b(x_b) + \frac{e^{Y}}{\rho}\, f_a(x_a)\, x_b f'_b(x_b) \biggr]
\nn\\ & \quad
   +  f_a(x_a)\, f_b(x_b) \, \biggl[ \rho Q \, \altM_+^{(2)}(Q,Y) + \frac{Q}{\rho} \, \altM_-^{(2)}(Q,Y) \biggr]
\biggr\}
\,.\end{align}

\subsection{Universality of Power Corrections for 0-Jettiness}
\label{sec:universality}

Having derived our master formulas, in this section we comment on the universality of the power corrections.
In both the collinear and soft limits the power corrections arising from the derivatives of the PDFs
and from the expansion of the flux factor are proportional to the LP matrix element $\Msquared^{(0)}(Q,Y)$,
see \eqs{sigma_NLO_NLP_coll_FULL}{sigma_NLO_NLP_soft}.
Since the factorization properties of $\Msquared^{(0)}(Q,Y)$
are universal, most of the NLP corrections are universal as well,
in the sense that they essentially reduce to the LO cross section times a universal factor,
as we will make explicit below.
The only process-dependent piece arises from the NLP expansion $\Msquared^{(2)}(Q,Y)$ of the matrix element.
We stress that this limit is defined in our particular choice of Born measurements $Q^2$ and $Y$.
Using different observables, e.g.\ $q^\pm$, the NLP corrections to $\zeta_{a,b}$ in \eq{zeta_ab} would change,
inducing also a change of the NLP matrix element.

\subsubsection{Universality of Collinear Limit}
\label{sec:univ_coll}

We begin by considering the $n$-collinear limit of a real emission amplitude in detail.
We consider the Born process
\begin{equation}
 \kappa_a(q_a) + \kappa_b(q_b) \to L(q_a + q_b)
\,,\end{equation}
where $\kappa_i$ denotes all quantum numbers, including flavor,
of the incoming partons, and $L$ is the leptonic final state of momentum $q = q_a + q_b$.
The incoming momenta for the hard collision are given by
\begin{align}
 q_a^\mu = x_a \Ecm \frac{n^\mu}{2} = Q e^{+Y} \frac{n^\mu}{2}
\,,\qquad
 q_b^\mu = x_b \Ecm \frac{\bn^\mu}{2} = Q e^{-Y} \frac{\bn^\mu}{2}
\,.\end{align}
Now consider that parton $a$ arises from an $n$-collinear splitting of a parton with flavor $a'$,
\begin{equation}
 \kappa'_a(q'_a) + \kappa_b(q_b) \to L(q'_a + q_b - k) + \kappa_1(k)
\,.\end{equation}
To describe this at leading power, we only need the $\cO(\lambda^0)$ relations
for the momenta of the incoming partons, which can be read off from \eqs{zeta_ab_coll2}{auxiliary_coll},
\begin{align}
 {q'}_a^\mu &= \frac{q_a^\mu}{z_a} + \cO(\lambda^2) = \frac{Q e^Y}{z_a} \frac{n^\mu}{2} + \cO(\lambda^2)
\,.\end{align}
The $n$-collinear emission is given by \eq{k_collinear},
\begin{align}
 k^\mu = Q e^Y \frac{1-z_a}{z_a} \frac{n^\mu}{2} + \frac{\Tau}{\rho} \frac{\bn^\mu}{2}
       + \sqrt{Q \Tau \frac{e^Y}{\rho} \frac{1-z_a}{z_a}} n_\perp^\mu
\,.\end{align}
It follows that the leptonic momentum $q' = q'_a + q'_b - k = q + \cO(\lambda^2)$
is equal to the Born momentum $q = q_a + q_b$, and hence the collinear splitting does not
affect the leptonic phase space at LP.

The LP limit only exists if the splitting $\kappa'_a \to \kappa_a + \kappa'_1$ is allowed,
in which case it is given by the $\cO(\lambda^{-2})$ piece of the squared amplitude,
\begin{align}
 \Msquared_{a' b \to L k}(Q, Y, \{k\})
 = \frac{8\pi \as \muMS^{2\eps}}{Q e^Y k^+} P_{a a'}(z_a, \eps) \Msquared_{a b \to L}^\LO(Q, Y) + \cO(\lambda^0)
\,,\end{align}
where the $1/k^+$ gives rise to the $\lambda^{-2}$ behavior of the amplitude.
Here the $P_{aa'}$ are the $\eps$-dimensional splitting functions which are summarized in \app{app_LP}.

Recall that in our case, the measurement fixes $k^+ = \Tau/\rho$.
In the notation of \eq{expansion_M2}, we hence have for the LP matrix element
\begin{align} \label{eq:M2_collinear}
 \Msquared_{a' b \to L k}^{(0)}(Q, Y, \{k\})
 = 8\pi \as \muMS^{2\eps}\frac{\rho}{Q \Tau e^Y} P_{a a'}(z_a, \eps) \Msquared_{a b \to L}^\LO(Q, Y)
\,.\end{align}
These results enable us to explicitly give the universal part of the NLP result in the collinear limit.
Inserting into the collinear master formula \eq{sigma_NLO_NLP_coll_FULL}
and converting to the $\overline{\text{MS}}$ scheme, we find
\begin{align} \label{eq:sigma_NLO_NLP_coll_UNIV}
 &\frac{\df\sigma_n^{(2)}}{\df Q^2 \df Y \df\Tau}
\\&=
 \hat\sigma^\LO(Q,Y) \frac{\as}{4 \pi} \frac{e^Y}{Q \rho} \times
    \biggl(\frac{Q \Tau}{\mu^2} \frac{e^Y}{\rho}\biggr)^{-\eps} \frac{e^{\eps \gamma_E}}{\Gamma(1-\eps)}
   \int_{x_a}^1 \frac{\df z_a}{z_a} \,
   \frac{z_a^\eps}{(1-z_a)^\eps} P_{aa'}(z_a,\eps)
   \nn\\&\quad\times
   \biggl[ f_{a'}\biggl(\frac{x_a}{z_a}\biggr) f_b(x_b) \frac{(1-z_a)^2 - 2}{z_a}
   + \frac{1-z_a}{z_a} x_a f'_{a'}\biggl(\frac{x_a}{z_a}\biggr) f_b(x_b)
   + \frac{1+z_a}{z_a} f_{a'}\biggl(\frac{x_a}{z_a}\biggr) x_b f'_b(x_b) \biggr]
\nn\\\nn& \quad
 + \int_{x_a}^1 \frac{\df z_a}{z_a} \,
   \frac{f_a(x_a/z_a)\, f_b(x_b)}{2 x_a x_b \Ecm^4}
   \frac{Q e^Y}{\rho} \biggl(\frac{Q \Tau e^Y}{\rho}\biggr)^{-\eps}
   \frac{z_a^\eps}{(1-z_a)^\eps}
   \frac{(4\pi)^{\eps}}{(4\pi)^2 \Gamma(1-\eps)}\,
   \Msquared^{(2)}(Q, Y, \{k\})
\,.\end{align}
Here, we factored out the LO partonic cross section \eq{sigmaLO_2},
which is only possible because the collinear splitting leaves the leptonic
momentum invariant at LP.
We have made explicit the universal piece and nonuniversal components.
As already discussed, the full nonuniversal structure arises from the NLP matrix element
$\Msquared^{(2)}(Q, Y, \{k\})$ in the last line.

It would also be interesting to understand if there is a universal structure to $\Msquared^{(2)}(Q,Y)$.
This has recently been studied in  \refcite{Nandan:2016ohb} for pure $n$ gluon scattering amplitudes at the level of the Cachazo-He-Yuan scattering equations \cite{Cachazo:2013hca,Cachazo:2013iea}, where it was proven that in the subleading power collinear limits the tree-level amplitude factorizes into a convolution of the $n-1$ gluon integrand and a universal collinear kernel. It would be interesting to understand this at the level of the amplitude itself, as well as for fermions. Unlike at leading power, we do not expect that there are universal subleading power splitting functions that are simply functions of $z$, but there may exist splitting functions that involve differential or integral operators, as occurs in the soft limit at subleading power \cite{Low:1958sn,Burnett:1967km}. Understanding this will be particularly important for generalizing the calculation of the power corrections to more complicated processes.

\subsubsection{Universality of Soft Limit}
\label{sec:univ_soft}

As for the collinear case, the LP soft limit of the matrix element is universal.
Following similar steps as in \sec{univ_coll}, one can express the LP soft limit by
\begin{align} \label{eq:M2_soft_LP}
 \Msquared_{ab \to L k}^{(0)}(Q, Y; \{k\})
 = \frac{16 \pi \as \muMS^{2\eps} \mathbf{C}}{k^+ k^-} \times \Msquared_{a b \to L}^\LO(Q, Y)
\,,\end{align}
which only exists for $ab=gg, q\bar q$ and where $\mathbf{C} = C_A, C_F$ is the appropriate Casimir constant.
We thus obtain
\begin{align} \label{eq:sigma_NLO_NLP_soft_UNIV}
 \frac{\df\sigma_s^{(2)}}{\df Q^2 \df Y \df\Tau} &
 = \frac{\hat\sigma^\LO}{Q}
   \frac{\as \mathbf{C}}{\pi}
   \biggl[ \frac{1}{\eps} - \ln\frac{\Tau^2}{\mu^2} - 1 \biggr]
   \biggl[ f_a(x_a) f_b(x_b) \biggl(- \frac{\rho}{e^Y} -\frac{e^{Y}}{\rho} \biggr)
   \nn\\&\hspace{5cm}
   + \frac{\rho}{e^Y}\, x_a  f'_a(x_a)\, f_b(x_b) + \frac{e^{Y}}{\rho}\, f_a(x_a)\, x_b f'_b(x_b) \biggr]
\nn\\& \quad
   + \frac{f_a(x_a)\, f_b(x_b)}{(4\pi)^2 2 x_a x_b \Ecm^4}
   \frac{(4\pi)^{\eps}}{\Gamma(1-\eps)}
   \frac{1}{\eps} \Tau^{-2\eps} \frac{1-2\eps}{1-\eps}
   \biggl[ \rho |\altM_+^{(2)}(Q,Y)|^2 + \frac{1}{\rho} |\altM_-^{(2)}(Q,Y)|^2 \biggr]
\,.\end{align}
As for the collinear case, this emphasizes that the terms arising
from the expansion of the PDFs and flux factor are universal,
in the sense that they only depend on the universal LP soft limit of the amplitude.
The only nonuniversal contributions are $|\altM_{\pm}^{(2)}|^2$.
However, these terms can in fact be derived from universal formulae \cite{Low:1958sn,Burnett:1967km,DelDuca:1990gz} involving differential operators.
This has been recently studied in the threshold limit,
where one only requires soft contributions \cite{DelDuca:2017twk}.
However, when one is away from the threshold limit as considered here, one in general requires collinear contributions,
which as discussed above, are not (yet) known to be universal.

\section{Power Corrections at NLO for Color Singlet Production}
\label{sec:results_singlet}

In this section we give explicit results for the full NLP correction for $0$-jettiness at NLO
for Higgs and Drell-Yan production in all partonic channels.
Since we only consider cases that are $s$-channel processes at Born level, the LO matrix element only depends on $Q$ and one can factor out the LO partonic cross section $\hat\sigma^\LO(Q)$.
We write the NLP cross section as
\begin{align} \label{eq:sigma_NLP}
\frac{\df\sigma^{(2,n)}}{\df Q^2 \df Y \df \Tau}
&= \hat\sigma^\LO(Q) \, \Bigl(\frac{\as}{4\pi}\Bigr)^n
    \int_{x_a}^1 \frac{\df z_a}{z_a} \int_{x_b}^1 \frac{\df z_b}{z_b}
\biggl[
f_i\biggl(\frac{x_a}{z_a}\biggr) f_j\biggl(\frac{x_b}{z_b}\biggr) C_{f_i f_j}^{(2,n)}(z_a, z_b, \Tau)
\\\nn & \quad
+ \frac{x_a}{z_a} f'_i\biggl(\frac{x_a}{z_a}\biggr) f_j\biggl(\frac{x_b}{z_b}\biggr) C_{f_i' f_j}^{(2,n)}(z_a, z_b, \Tau)
+ f_i\biggl(\frac{x_a}{z_a}\biggr) \frac{x_b}{z_b} f'_j\biggl(\frac{x_b}{z_b}\biggr) C_{f_i f_j'}^{(2,n)}(z_a, z_b, \Tau)
\biggr]
\,,\end{align}
where as always
\begin{align}
 x_a = \frac{Q e^Y}{\Ecm} \,,\quad x_b = \frac{Q e^{-Y}}{\Ecm}
\,.\end{align}
We will always express the real emission amplitudes in terms of the Mandelstam variables
\begin{align}
 s_{ab} = 2 p_a \cdot p_b
\,,\quad
 s_{ak} = -2 p_a \cdot k
\,,\quad
 s_{bk} = -2 p_b \cdot k
\,.\end{align}
This allows us to straightforwardly obtain the LP and NLP expansion using \eq{mandelstamgeneral}.
We will give an explicit example of the derivation of the soft and collinear master formulas
for the $gg \to H g$ channel, and only summarize the results in the other channels.

\subsection{Gluon-Fusion Higgs Production}
\label{sec:ggH_results}

We begin by considering Higgs production in gluon fusion in the $m_t \to \infty$ limit.
At NLP, there are three different partonic channels, $gg\to Hg$, $q\bar q \to H g$ and $q g \to H q$,
which we consider separately. The calculation for $gg \to Hg$ is shown in full detail
as an illustration of our master formulae.
The LL power corrections were computed in \cite{Boughezal:2016zws,Moult:2017jsg}. Ref.~\cite{Moult:2017jsg} also computed the $q\bar q \to Hg$ NLL power corrections. 
The NLL power corrections for all partonic channels for gluon fusion Higgs were computed in \cite{Boughezal:2018mvf}.
We will compare with these results in \sec{compare_lit}.

Throughout this section we consider on-shell Higgs production,
for which the partonic cross section is given by
\begin{align}
 \hat\sigma^\LO(Q,Y)
 = \frac{\Msquared^\LO(Q,Y)}{2 x_a x_b \Ecm^4}
 = 2\pi \delta(Q^2 - m_H^2) \frac{|\cM_{gg\to H}^\LO(Q)|^2}{2 Q^2 \Ecm^2}
\,.\end{align}
The LO matrix element in $d=4-2\epsilon$ dimensions is given by
\cite{Dawson:1990zj,Djouadi:1991tka}
\begin{align}\label{eq:M2_gg_H}
 |\cM^\LO_{gg\to H}(Q)|^2 &= \frac{\as^2 Q^4}{576 \pi^2 v^2} \biggl(\frac{4\pi\muMS^2}{m_t^2}\biggr)^{2\eps} \frac{\Gamma^2(1+\eps)}{1-\eps}
\,.\end{align}

\subsubsection[\texorpdfstring{$gg \to Hg$}{gg -> Hg}]{\boldmath $gg \to Hg$}
\label{sec:collinear_example}

The spin- and color-averaged squared amplitude for $g(p_a) + g(p_b) \to H(q) + g(k)$ is given by \cite{Dawson:1990zj}
\begin{align} \label{eq:M2_ggHg}
 \Msquared_{gg \to H g}(Q, Y, \{k\}) &
 = \Msquared^\LO_{gg\to H}(Q) \times \frac{8 \pi \as C_A \muMS^{2\eps}}{ Q^4 (1-\eps)}
\nn\\ & \quad\times
   \biggl[ (1-2\eps) \frac{Q^8 + s_{ab}^4 + s_{ak}^4 + s_{bk}^4}{s_{ab} s_{ak} s_{bk}}
   + \frac{\eps}{2} \frac{(Q^4 + s_{ab}^2 + s_{ak}^2 + s_{bk}^2)^2}{s_{ab} s_{ak} s_{bk}} \biggr]
\,.\end{align}

\paragraph{\boldmath $n$-Collinear Limit}
Expanding \eq{M2_ggHg} using \eqs{mandelstamgeneral}{auxiliary_coll},
the LP and NLP limits of the matrix element are obtained as
\begin{align}
 \Msquared^{(0)}_{gg \to Hg}(Q, Y; \{k\}) &
 = 16 \pi \alpha_s C_A \muMS^{2 \epsilon} \Msquared^\LO_{gg\to H}(Q) \frac{ (1-z_a+z_a^2)^2 }{(1-z_a) z_a}\frac{\rho e^{-Y}}{Q \Tau}
\,,\\\nn
 \Msquared^{(2)}_{gg \to Hg}(Q, Y; \{k\}) &
 = 16 \pi \alpha_s  C_A \muMS^{2 \epsilon} \Msquared^\LO_{gg\to H}(Q)  \frac{1}{Q^2 z_a^2}
   \biggl[ 1 + 5 z_a^2 - z_a^3 + 2 z_a^4 - z_a^5 - 2 z_a^2 \frac{1}{1-\eps} \biggr]
\,.\end{align}
Since our scaling variable is $\lambda \sim \sqrt{\Tau/Q}$,
we clearly see that $\Msquared^{(0)} \sim \lambda^{-2}$ and $\Msquared^{(2)} \sim \lambda^0$,
as required at LP and NLP.

Inserting these expansions into \eq{sigma_NLO_NLP_coll_FULL} and converting to the $\overline{\text{MS}}$ scheme yields 
\begin{align}
 \frac{\df\sigma_n^{(2)}}{\df Q^2 \df Y \df\Tau} &
 = \hat\sigma^\LO_{gg\to H}(Q) \times  \frac{\as C_A}{\pi}
   \int_{x_a}^1 \frac{\df z_a}{z_a}
   \frac{1}{Q} \frac{e^Y}{\rho} \biggl( \frac{\Tau Q}{\mu^2} \frac{e^Y}{\rho} \biggr)^{-\eps}
   \frac{z_a^{\eps}}{(1-z_a)^{\eps}}
   \frac{e^{\eps \gamma_E}}{\Gamma(1-\eps)}
   \nn\\&\quad\times \biggl\{
   \frac{ (1-z_a+z_a^2)^2 }{(1-z_a) z_a}
   \biggl[ f_a \, f_b \frac{(1 - z_a)^2 - 2}{2 z_a}
   + x_a f'_a \, f_b \, \frac{1-z_a}{2 z_a}
   + f_a \, x_b f'_b \, \frac{1+z_a}{2 z_a} \biggr]
   \nn\\* &\hspace{1.cm}
   + f_a \, f_b \frac{1}{z_a^2}
   \biggl[ 1 + 5 z_a^2 - z_a^3 + 2 z_a^4 - z_a^5
  - 2 z_a^2 \frac{1}{1-\eps} \biggr]
  \biggr\}
\,.\end{align}
To expand this in $\eps$, we collect all powers of $(1-z_a)$ and then use the distributional identity
\begin{align}
 (1-z_a)^{-1-\epsilon} = -\frac{\delta(1-z_a)}{\epsilon} + \cL_0(1-z_a) + \cO(\epsilon)
\,,\end{align}
where $\cL_0(1-z)=1/(1-z)_+$ is the usual plus distribution.
We also combine the two separate $f_a f_b$ pieces, as at this level there is no use to further
distinguish the universal and process dependent pieces. This yields
\begin{align}
 \frac{\df\sigma_n^{(2)}}{\df Q^2 \df Y \df\Tau} &
 = \hat\sigma^\LO_{gg\to H}(Q) \times\frac{\as}{4 \pi}\,
   4 C_A\, \frac{e^Y}{Q\rho}
   \int_{x_a}^1 \frac{\df z_a}{z_a}
   \nn\\&\quad\times \biggl\{
     f_g\biggl(\frac{x_a}{z_a}\biggr) \, f_g(x_b)
     \biggl[ \biggl(\frac{1}{\eps} - \ln\frac{Q \Tau e^Y}{\mu^2 \rho} \biggr)\, \delta(1-z_a)
     \nn\\&\hspace{3.8cm}
      + \frac{1 - 2 z_a + 8 z_a^2 - 14 z_a^3 + 12 z_a^4 - 10 z_a^5 + 3 z_a^6}{2z_a^2} \cL_0(1-z_a)
      \biggr]
   \nn\\&\qquad
   + \frac{x_a}{z_a} f'_g\biggl(\frac{x_a}{z_a}\biggr) \, f_g(x_b) \, \frac{(1-z_a+z_a^2)^2}{2 z_a}
   \nn\\&\qquad
   + f_g\biggl(\frac{x_a}{z_a}\biggr) \, x_b f'_g(x_b) \, \biggl[
   \biggl(-\frac{1}{\eps}+ \ln\frac{Q \Tau e^Y}{\mu^2 \rho} \biggr)\, \delta(1-z_a)
   \nn\\&\hspace{4.3cm}
   + \frac{(1 + z_a)(1 - z_a + z_a^2)^2}{2 z_a^2}\, \cL_0(1-z_a) \biggr]
  \biggr\}
\,.\end{align}
Comparing to \eq{sigma_NLP}, we can read off the $n$-collinear kernels,
\begin{align} \label{eq:kernel_Cgg_n}
C_{f_g f_g,n}^{(2,1)}(z_a, z_b, \Tau)
&= 4C_A\, \frac{e^Y}{Q\rho} \biggl[
\biggl(\frac{1}{\eps} - \ln\frac{Q \Tau e^Y}{\mu^2 \rho} \biggr)\, \delta(1-z_a)
\nn\\& \quad
+ \frac{1 - 2 z_a + 8 z_a^2 - 14 z_a^3 + 12 z_a^4 - 10 z_a^5 + 3 z_a^6}{2z_a^2}\, \cL_0(1-z_a)\,
  \biggr]  \delta(1-z_b)
\,,\nn\\
C_{f'_g f_g,n}^{(2,1)}(z_a, z_b, \Tau)
&= 4C_A\, \frac{e^Y}{Q\rho}\, \frac{(1-z_a + z_a^2)^2}{2 z_a}\, \delta(1-z_b)
\,,\nn\\
C_{f_g f'_g,n}^{(2,1)}(z_a, z_b, \Tau)
&= 4C_A\, \frac{e^Y}{Q\rho}\, \biggl[
\biggl( -\frac{1}{\eps} + \ln\frac{Q \Tau e^Y}{\mu^2 \rho} \biggr)\, \delta(1-z_a)
\nn\\& \quad
+ \frac{(1+z_a)(1-z_a+z_a^2)^2}{2z_a^2}\, \cL_0(1-z_a) \biggr] \delta(1-z_b)
\,.\end{align}

\paragraph{Soft Limit}
To expand the matrix element in the soft limit, we use \eqs{mandelstamgeneral}{auxiliary_soft} to obtain
\begin{align}
 \Msquared_{gg\to H g}(Q, Y, \{k\}) = \Msquared^\LO_{gg\to H}(Q) \times
  16 \pi \alpha_s C_A \frac{ \muMS^{2 \epsilon}}{k^+ k^-} + \cO(\lambda^0)
\,.\end{align}
Note that the first term scales as $(k^+ k^-)^{-1} \sim \lambda^{-4}$,
while there is no $\cO(\lambda^{-2})$ component.
The NLP term in the expansion of the amplitude thus vanishes, and in the notation of \eq{altMdef} we have
\begin{align} \label{eq:altMggHg}
 \altM_{gg \to H g}^{(0)}(Q) = \Msquared^\LO_{gg\to H}(Q) \times 16 \pi \alpha_s C_A \muMS^{2 \epsilon}
\,,\qquad
 \altM_{gg \to H g}^{(2)}(Q) = 0
\,.\end{align}
Inserting into \eq{sigma_NLO_NLP_soft} and converting to the $\overline{\text{MS}}$ scheme yields
\begin{align}
 \frac{\df\sigma_s^{(2)}}{\df Q^2 \df Y \df\Tau}
&= \hat\sigma_{gg\to H}^\LO(Q) \frac{\alpha_s}{4\pi} \times 4C_A\,\frac{1}{Q}
   \biggl( \frac{1}{\eps} - \ln\frac{\Tau^2}{\mu^2} - 1 \biggr)
\\\nn&\quad\times
   \biggl[ f_g(x_a) f_g(x_b) \biggl(- \frac{\rho}{e^Y} -\frac{e^{Y}}{\rho} \biggr)
   + \frac{\rho}{e^Y}\, x_a f'_g(x_a)\, f_g(x_b)
   + \frac{e^{Y}}{\rho}\, f_g(x_a)\, x_b f'_g(x_b) \biggr]
\,.\end{align}
Since there is no NLP matrix element, one can also obtain this from the
universal expression for the soft limit in \eq{sigma_NLO_NLP_soft_UNIV}.
Comparing to \eq{sigma_NLP}, we can read off the soft kernel,
\begin{align} \label{eq:kernel_Cgg_s}
C_{f_g f_g,s}^{(2,1)}(z_a, z_b, \Tau)
&= 4C_A\,\frac{1}{Q}\biggl(-\frac{e^Y}{\rho} - \frac{\rho}{e^Y}\biggr)\,
   \biggl( \frac{1}{\eps} - \ln\frac{\Tau^2}{\mu^2} - 1 \biggr)\,
   \delta(1-z_a)\, \delta(1-z_b)\,
\,,\nn\\
C_{f'_g f_g,s}^{(2,1)}(z_a, z_b, \Tau)
&= 4C_A\,\frac{1}{Q}\frac{\rho}{e^Y}\,
   \biggl( \frac{1}{\eps} - \ln\frac{\Tau^2}{\mu^2} - 1 \biggr)\,
   \delta(1-z_a)\, \delta(1-z_b)\,
\,,\nn\\
C_{f_g f'_g,s}^{(2,1)}(z_a, z_b, \Tau)
&= 4C_A\,\frac{1}{Q}\frac{e^Y}{\rho}\,
   \biggl( \frac{1}{\eps} - \ln\frac{\Tau^2}{\mu^2} - 1 \biggr)\,
   \delta(1-z_a)\, \delta(1-z_b)
\,.\end{align}

\paragraph{Final Result}
Adding the $n$-collinear kernel \eq{kernel_Cgg_n}, the $\bn$-collinear kernel which follows from symmetry,
and the soft kernel \eq{kernel_Cgg_s}, all poles in $\eps$ cancel as expected, and we obtain
\begin{align} \label{eq:kernel_Cgg}
C_{f_g f_g}^{(2,1)}(z_a, z_b, \Tau)
&= 4 C_A\, \frac{e^Y}{Q\rho} \biggl[ \biggl(\ln\frac{\Tau \rho}{Q e^Y} + 1 \biggr) \delta(1-z_a)
\nn \\ & \quad
   + \frac{1 - 2 z_a + 8 z_a^2 - 14 z_a^3 + 12 z_a^4 - 10 z_a^5 + 3 z_a^6}{2z_a^2}\, \cL_0(1-z_a)
   \biggr] \delta(1-z_b)
\nn \\ & \quad
   + \biggl( a \leftrightarrow b \,,\, \frac{\rho}{e^Y} \to \frac{e^Y}{\rho} \biggr)
\,,\nn\\
C_{f'_g f_g}^{(2,1)}(z_a, z_b, \Tau)
&= 4 C_A\, \frac{\rho}{Qe^Y}\, \delta(1-z_a) \biggl[
   \biggl( - \ln\frac{\Tau e^Y}{Q \rho} - 1  \biggr) \delta(1-z_b)
   \nn\\&\hspace{4cm}
   + \frac{(1+z_b)(1-z_b+z_b^2)^2}{2z_b^2}\, \cL_0(1-z_b) \biggr]
\nn\\& \quad
+ 4C_A \frac{e^Y}{Q\rho} \frac{(1-z_a + z_a^2)^2}{2 z_a} \delta(1-z_b)
\,,\nn\\
C_{f_g f_g'}^{(2,1)}(z_a, z_b, \Tau)
&= 4 C_A\, \frac{e^Y}{Q \rho} \biggl[
   \biggl( - \ln\frac{\Tau \rho}{Q e^Y} - 1  \biggr) \delta(1-z_a)
   \nn\\&\hspace{2.cm}
   + \frac{(1+z_a)(1-z_a+z_a^2)^2}{2z_a^2}\, \cL_0(1-z_a) \biggr] \delta(1-z_b)
\nn\\& \quad
+ 4C_A \frac{\rho}{Q e^Y} \delta(1-z_a) \frac{(1-z_b + z_b^2)^2}{2 z_b}
\,.\end{align}
Substituting these results into \eq{sigma_NLP} yields the NLP cross section for $gg\to Hg$ at NLO.

\subsubsection[\texorpdfstring{$gq \to Hq$}{gq -> Hq}]{\boldmath $gq \to Hq$}

The $gq\to Hq$ channel has power corrections at both LL and NLL.
The spin- and color-averaged squared amplitude for $g(p_a) + q(p_b) \to H(q) + q(k)$ is given by \cite{Dawson:1990zj}
\begin{align}
 \Msquared_{gq \to Hq}(Q, Y, \{k\})
 = - \Msquared^\LO_{gg\to H}(Q) \times 8 \pi \as C_F \muMS^{2\eps} \frac{1}{Q^4 s_{bk}}
  \Bigl[ s_{ab}^2 + s_{ak}^2 - \eps (s_{ab}+s_{ak})^2 \Bigr]
\,.\end{align}

\paragraph{Soft Limit}
The LP soft limit vanishes, since a leading-power soft interaction (which is eikonal) cannot change a $n$-collinear quark into a $n$-collinear gluon and soft quark. However this does occur at NLP in the soft expansion and yields
\begin{align}
 \Msquared^{(2)}_{gq \to Hq}(Q, Y, \{k\}) &
 = \Msquared^\LO_{gg\to H}(Q) \times 8 \pi \as C_F \muMS^{2\eps} \frac{1-\eps}{Q k^- e^{-Y}}
\,,\end{align}
and the soft kernel is given by
\begin{align} \label{eq:kernel_Cqg_s}
C_{f_g f_q , s}^{(1,2)}(z_a, z_b, \Tau)
= 2 C_F\, \frac{e^Y}{Q\rho}\,
 \biggl( \frac{1}{\eps} - \ln\frac{\Tau^2}{\mu^2} - 2 \biggr)
 \delta(1-z_a)\, \delta(1-z_b)
\,.\end{align}

\paragraph{\boldmath $\bn$-Collinear Limit}
The $\bn$-collinear limit has both a LP and NLP contribution, given by
\begin{align}
 \Msquared^{(0)}_{gq \to Hq}(Q, Y, \{k\}) &
 = \Msquared^\LO_{gg\to H}(Q)  \times 8 \pi \as C_F \frac{e^Y}{\rho}
   \frac{1 + (1-z_b)^2 - \eps z_b^2}{Q \Tau z_b}
\,, \nn \\
 \Msquared^{(2)}_{gq \to Hq}(Q, Y, \{k\}) &
 = \Msquared^\LO_{gg\to H}(Q) \times 4 \pi \as C_F \muMS^{2\eps}
   \frac{4 - z_b^3 + z_b^4 - \eps z_b^2 (4 - z_b + z_b^2)}{Q^2 z_b^2}
\,.\end{align}
The $\bn$-collinear kernel is obtained as
\begin{align} \label{eq:kernel_Cgq_bn}
C_{f_g f_q , \bn}^{(1,2)}(z_a, z_b, \Tau)
&= C_F\, \frac{\rho}{Qe^Y}\,\delta(1-z_a)\, \frac{2 - 2 z_b + 5 z_b^2 - 5 z_b^3 + 2 z_b^4}{z_b^2}
\,,\nn\\
C_{f'_g f_q , \bn}^{(1,2)}(z_a, z_b, \Tau)
&= C_F\, \frac{\rho}{Qe^Y}\, \delta(1-z_a)\, \frac{(1+z_b)[1 + (1-z_b)^2]}{z_b^2}
\,,\nn\\
C_{f_g f_q' , \bn}^{(1,2)}(z_a, z_b, \Tau)
&= C_F\, \frac{\rho}{Qe^Y}\, \delta(1-z_a)\, \frac{(1-z_b)[1 + (1-z_b)^2]}{z_b}
\,.\end{align}

\paragraph{\boldmath $n$-Collinear Limit}
The $n$-collinear limit vanishes at LP. The NLP expansion of the matrix element gives
\begin{align}
 \Msquared^{(2)}_{gq \to Hq}(Q, Y, \{k\}) &
 = \Msquared^\LO_{gg\to H}(Q)  \times \frac{8 \pi \as C_F \muMS^{2\eps} (1-\eps)}{Q^2 (1-z_a) z_a}
\,.\end{align}
The only nonvanishing kernel is
\begin{align} \label{eq:kernel_Cgq_n}
C_{f_g f_q , n}^{(1,2)}(z_a, z_b, \Tau)
&= 2C_F\, \frac{e^Y}{Q\rho} \biggl[
\biggl( \frac{-1}{\eps} + \ln\frac{Q \Tau e^Y}{\mu^2 \rho} + 1 \biggr)\,\delta(1-z_a)
+ \frac{\cL_0(1-z_a)}{z_a} \biggr] \delta(1-z_b)
\,.\end{align}

\paragraph{Final Result}
Combining the $n$-collinear, $\bn$-collinear, and soft kernels, the $1/\eps$ pole vanishes,
and we obtain the final results,
\begin{align} \label{eq:kernel_Cgq}
C_{f_g f_q}^{(1,2)}(z_a, z_b, \Tau)
&= 2C_F\, \frac{e^Y}{Q\rho} \biggl[
   \biggl( -\ln\frac{\Tau \rho}{Q e^Y} - 1\biggr)\,\delta(1-z_a)
   + \frac{\cL_0(1-z_a)}{z_a} \biggr] \delta(1-z_b)
\nn\\& \quad
+ C_F\, \frac{\rho}{Qe^Y}\, \delta(1-z_a)\, \frac{2 - 2 z_b + 5 z_b^2 - 5 z_b^3 + 2 z_b^4}{z_b^2}
\,,\nn\\
C_{f'_g f_q}^{(1,2)}(z_a, z_b, \Tau)
&= C_F\, \frac{\rho}{Qe^Y}\, \delta(1-z_a)\, \frac{(1+z_b)[1 + (1-z_b)^2]}{z_b^2}
\,,\nn\\
C_{f_g f_q'}^{(1,2)}(z_a, z_b, \Tau)
&= C_F\, \frac{\rho}{Qe^Y}\, \delta(1-z_a)\, \frac{(1-z_b)[1 + (1-z_b)^2]}{z_b}
\,.\end{align}
Substituting these results into \eq{sigma_NLP} yields the NLP cross section for $gq\to Hq$ at NLO.

\subsubsection[\texorpdfstring{$qg \to Hq$}{qg -> Hq}]{\boldmath $qg \to Hq$}

The final results needed in \eq{sigma_NLP} for $qg \to H q$ follow from \eq{kernel_Cgq}
by flipping $a \leftrightarrow b$, $e^Y/\rho \leftrightarrow \rho/e^Y$ and $f_g \leftrightarrow f_q$,
\begin{align} \label{eq:kernel_Cqg}
C_{f_q f_g}^{(1,2)}(z_a, z_b, \Tau)
&= 2C_F\, \frac{\rho}{Qe^Y}\, \delta(1-z_a) \biggl[
\biggl( - \ln\frac{\Tau e^Y}{Q \rho} - 1 \biggr)\, \delta(1-z_b) + \frac{\cL_0(1-z_b)}{z_b} \biggr]
\nn \\ & \quad
+ C_F\, \frac{e^Y}{Q\rho}\, \frac{2 - 2 z_a + 5 z_a^2 - 5 z_a^3 + 2 z_a^4}{z_a^2}\, \delta(1-z_b)
\,,\nn\\
C_{f'_q f_g}^{(1,2)}(z_a, z_b, \Tau)
&= C_F\, \frac{e^Y}{Q\rho}\, \frac{(1-z_a)[1 + (1-z_a)^2]}{z_a}\, \delta(1-z_b)
\,,\nn\\
C_{f_q f_g'}^{(1,2)}(z_a, z_b, \Tau)
&= C_F\, \frac{e^Y}{Q\rho}\, \frac{(1+z_a)[1 + (1-z_a)^2]}{z_a^2}\, \delta(1-z_b)
\,.\end{align}

\subsubsection[\texorpdfstring{$q\bar q\to Hg$}{qqbar -> Hg}]{\boldmath $q\bar q\to Hg$}

The $q\bar q\to Hg$ channel first contributes at NLL.
It was first given in \cite{Moult:2017jsg} and then in \cite{Boughezal:2018mvf}, which agreed,
but we reproduce it here for completeness.
The squared matrix element, including the average on the initial state spin and colors, is  given by \cite{Dawson:1990zj}
\begin{align}
 \Msquared_{q\bar q \to Hg}(Q, Y, \{k\}) &
 = \Msquared^\LO_{gg\to H}(Q) \times \frac{64\pi}{3} \as C_F \muMS^{2\eps} \frac{1-\eps}{Q^4 s_{ab}}
   \bigl[ s_{ak}^2 + s_{bk}^2 - \eps (s_{ak}+s_{bk})^2 \bigr]
\,,\end{align}
With our choice of Born measurements, the soft limit vanishes both at LP and NLP, leaving only the collinear NLP correction.
The LP collinear limit also vanishes, leaving only the NLP $n$-collinear limit
\begin{equation}
 \Msquared^{(2)}_{q\bar q \to Hg}(Q, Y, \{k\}) = \Msquared^\LO_{gg\to H}(Q) \times
   \frac{64\pi}{3} \as C_F \muMS^{2\eps} (1-\eps)^2 \frac{(1-z_a)^2}{Q^2 z_a}
\,,\end{equation}
and the $\bn$-collinear result is obtained by replacing $z_a \leftrightarrow z_b$.
Combining both, we obtain the kernel for \eq{sigma_NLP}
\begin{align} \label{eq:kernel_Cqq}
C_{f_q f_{\bar q}}^{(2,1)}(z_a, z_b, \Tau)
= \frac{16 C_F}{3}\,\frac{1}{Q} \biggl[ \frac{e^Y}{\rho}\, \frac{(1-z_a)^2}{z_a}\, \delta(1-z_b)
   + \frac{\rho}{e^Y}\, \delta(1-z_a)\, \frac{(1-z_b)^2}{z_b} \biggr]
\,.\end{align}

\subsection{Drell-Yan Production}

We now consider the Drell-Yan process $p p \to Z/\gamma^* \to l^+ l^-$ , and for brevity denote it as $p p \to V$.
At NLO we have the partonic channels $q\bar q\to Vg$ and $qg \to Vq$.
The LL power corrections for these channels were calculated to NNLO in \cite{Moult:2016fqy,Boughezal:2016zws}.

For Drell-Yan, it is important to be able to include off-shell effects.
The LO partonic cross section as a function of the leptonic invariant mass $Q$ is given by
\begin{align}
 \hat\sigma^\LO(Q) = \frac{4\pi \alpha_{em}^2}{3 N_c Q^2 \Ecm^2}
 \biggl[ Q_q^2 + \frac{(v_q^2 + a_q^2)(v_l^2 + a_l^2) - 2 Q_q v_q v_l (1 - m_Z^2/Q^2)}{(1-m_Z^2/Q^2)^2
         + m_Z^2 \Gamma_Z^2 / Q^4} \biggr]
\,.\end{align}
Here, $v_{l,q}$ and $a_{l,q}$ are the standard vector and axial couplings of the leptons and quarks to the $Z$ boson, and we have integrated over the $l^+ l^-$ phase space.

\subsubsection[\texorpdfstring{$q\bar q\to Vg$}{qqbar -> Vg}]{\boldmath $q\bar q\to Vg$}

We first consider the partonic channel $q\bar q\to Vg$. The squared amplitude is given by \cite{Gonsalves:1989ar}
\begin{align} \label{eq:M2_qqbar_Zg}
 |\cM_{q\bar q\to Vg}|^2 &
 = |\cM_{q\bar q\to V}|^2 \times \frac{8 \pi \as C_F \muMS^{2\eps}}{Q^2}
  \left[ (1-\epsilon) \left(\frac{s_{ak}}{s_{bk}}+\frac{s_{bk}}{s_{ak}}\right)+\frac{2 s_{ab} Q^2}{s_{ak} s_{bk}} -2\epsilon \right]
\,.\end{align}

\paragraph{Soft Limit}
With our setup, the soft limit of the matrix element has no NLP correction,
\begin{align}
 \Msquared_{q\bar q\to Vg}(Q,Y,\{k\})
 = \Msquared^\LO_{q \bar q \to V}(Q) \times \frac{16 \pi \as C_F \muMS^{2\eps}}{k^+ k^-} + \cO(\lambda^0)
\,,\end{align}
and the soft kernels are given by
\begin{align}
 C_{f_q f_{\bar q}, s}^{(2,1)}(z_a, z_b, \Tau) &
 = 4 C_F \biggl(-\frac{e^Y}{Q \rho} - \frac{\rho}{Q e^Y} \biggr)
   \biggl( \frac{1}{\eps} - \ln\frac{\Tau^2}{\mu^2} - 1 \biggr)
   \delta(1-z_a)\, \delta(1-z_b)
\,,\nn\\
  C_{f'_q f_{\bar q}, s}^{(2,1)}(z_a, z_b, \Tau) &
 = 4 C_F\, \frac{\rho}{Q e^Y}\,
   \biggl( \frac{1}{\eps} - \ln\frac{\Tau^2}{\mu^2} - 1 \biggr)
   \delta(1-z_a)\, \delta(1-z_b)
\,,\nn\\
  C_{f_q f'_{\bar q}, s}^{(2,1)}(z_a, z_b, \Tau) &
 = 4 C_F\, \frac{e^Y}{Q \rho}\,
   \biggl( \frac{1}{\eps} - \ln\frac{\Tau^2}{\mu^2} - 1 \biggr)
   \delta(1-z_a)\, \delta(1-z_b)
\,.\end{align}

\paragraph{Collinear Limit}
The $n$-collinear expansion of the matrix element yields
(at NLP, we only need $\eps\to0$)
\begin{align}
 \Msquared^{(0)}_{q\bar q\to Vg}(Q,Y,\{k\})
 &= \Msquared^\LO_{q \bar q \to V}(Q) \times 8 \pi \as C_F \muMS^{2\eps} \frac{\rho}{e^Y} \frac{1 + z_a^2 - \eps(1-z_a)^2}{Q \Tau (1-z_a)}
\,,\\
 \Msquared^{(2)}_{q\bar q\to Vg}(Q,Y,\{k\})
 &= \Msquared^\LO_{q \bar q \to V}(Q) \times 4 \pi \as C_F \frac{1 - z_a + z_a^2 - z_a^3}{Q^2 z_a}
\,.\end{align}
The $n$-collinear kernel is
\begin{align}
 C_{f_q f_{\bar q}, n}^{(2,1)}(z_a, z_b, \Tau) &
  = 4 C_F\, \frac{e^Y}{Q \rho}\, \biggl[
    \biggl(\frac{1}{\eps} - \ln\frac{Q \Tau e^Y}{\mu^2 \rho} \biggr)
    \delta(1-z_a)
  \nn\\*&\hspace{2cm}
    + \frac{1}{2}(z_a - 2)(1+z_a^2)\, \cL_0(1-z_a) \biggr] \delta(1-z_b)
\,,\nn\\
 C_{f'_q f_{\bar q}, n}^{(2,1)}(z_a, z_b, \Tau) &
  = 4 C_F\, \frac{e^Y}{Q \rho}\, \frac{1+z_a^2}{4} \delta(1-z_b)
\,,\nn\\
 C_{f_q f'_{\bar q}, n}^{(2,1)}(z_a, z_b, \Tau) &
  =  4 C_F\, \frac{e^Y}{Q \rho}\, \biggl[
    \biggl( - \frac{1}{\eps} + \ln\frac{Q \Tau e^Y}{\mu^2 \rho} \biggr)
    \delta(1-z_a)
  \nn\\&\hspace{2cm}
   + \frac{(1+z_a)(1+z_a^2)}{4z_a}\, \cL_0(1-z_a) \biggr] \delta(1-z_b)
\,.\end{align}

\paragraph{Final Result}
Adding the $n$, $\bn$ and $s$ kernel, we get
\begin{align}
 C_{f_q f_{\bar q}}^{(2,1)}(z_a, z_b, \Tau) &
  = 4 C_F\, \frac{e^Y}{Q \rho}\, \biggl[ \biggl(\ln\frac{\Tau \rho}{Q e^Y} + 1 \biggr)\, \delta(1-z_a)
    + \frac{1}{2} (z_a - 2)(1+z_a^2)\, \cL_0(1-z_a) \biggr] \delta(1-z_b)
   \nn\\&\quad  + \biggl( \frac{e^Y}{\rho} \to \frac{\rho}{e^Y} \,, a \leftrightarrow b \biggr)
\,,\nn\\
 C_{f'_q f_{\bar q}}^{(2,1)}(z_a, z_b, \Tau) &
  = 4 C_F\, \frac{\rho}{Q e^Y}\, \delta(1-z_a)\biggl[ \biggl( -\ln\frac{\Tau e^Y}{Q \rho} - 1\biggr)\, \delta(1-z_b)
  + \frac{(1+z_b)(1+z_b^2)}{4z_b}\, \cL_0(1-z_b) \biggr]
  \nn\\& \quad
  + 4 C_F\, \frac{e^Y}{Q \rho}\, \frac{1+z_a^2}{4}\, \delta(1-z_b)
\,,\nn\\
 C_{f_q f'_{\bar q}}^{(2,1)}(z_a, z_b, \Tau) &
 = 4 C_F \frac{e^Y}{Q \rho} \biggl[ \biggl( -\ln\frac{\Tau \rho}{Q e^Y} - 1\biggr)\, \delta(1-z_a)
  + \frac{(1+z_a)(1+z_a^2)}{4z_a}\, \cL_0(1-z_a) \biggr] \delta(1-z_b)
  \nn\\& \quad
  + 4 C_F\, \frac{\rho}{Q e^Y}\, \delta(1-z_a)\, \frac{1+z_b^2}{4}
\,.\end{align}
Substituting these results into \eq{sigma_NLP} yields the NLP cross section for $q\bar q\to Vg$ at NLO.

\subsubsection[\texorpdfstring{$qg\to Vq$}{qg -> Vq}]{\boldmath $qg\to Vq$}

Next we consider the partonic channel $qg\to Vq$. The squared amplitude is given by \cite{Gonsalves:1989ar}
\begin{align}
 \Msquared_{qg \to Vq}(Q,Y,\{k\})
 = - \Msquared^\LO_{q \bar q \to V}(Q) \times \frac{8 \pi \as T_F \muMS^{2\eps}}{Q^2 (1-\eps)}
  \left[ (1-\epsilon) \left(\frac{s_{ab}}{s_{bk}}+\frac{s_{bk}}{s_{ab}}\right)+\frac{2 s_{ak} Q^2}{s_{ab} s_{bk}} -2\epsilon \right]
\,.\end{align}

\paragraph{Soft Limit}
The LP soft limit vanishes, and the NLP soft expansion is given by
\begin{align}
 \Msquared^{(2)}_{qg \to Vq}(Q,Y,\{k\})
 =  \Msquared^\LO_{q \bar q \to V}(Q) \times 8 \pi \as T_F \muMS^{2\eps} \frac{e^Y}{Q k^-}
\,.\end{align}
The soft kernel is given by
\begin{align}
 C_{f_q f_g,s}^{(2,1)} =  2 T_F \frac{e^Y}{Q \rho}
  \biggl( \frac{1}{\eps} - \ln\frac{\Tau^2}{\mu^2} - 1 \biggr) \delta(1-z_a)\, \delta(1-z_b)
\,.\end{align}

\paragraph{\boldmath $n$-Collinear Limit}
The $n$-collinear limit does not contribute at LP, since the LP interaction
can not change the $\bn$-collinear gluon into a $\bn$-collinear antiquark.
The NLP matrix element is given by
\begin{align}
 \Msquared^{(2)}_{qg \to Vq}(Q,Y,\{k\}) &
 =  \Msquared^\LO_{q \bar q \to V}(Q)  \times 8 \pi \as T_F
 \frac{1 + (1-z_a)^2 -\eps z_a^2}{(1-\eps) (1-z_a)Q^2}
\,,\end{align}
and the collinear kernel is
\begin{align}
 C_{f_q f_g,n}^{(2,1)}(z_a, z_b, \Tau) &
 = 2 T_F\, \frac{e^Y}{Q \rho}\, \biggl[ \biggl(\frac{-1}{\eps} + \ln\frac{Q \Tau e^Y}{\mu^2 \rho} \biggr)\,
   \delta(1-z_a)
   \nn\\&\hspace{2cm}
   + [1 + (1-z_a)^2]\, \cL_0(1-z_a) \biggr] \delta(1-z_b)
\,.\end{align}

\paragraph{\boldmath $\bn$-Collinear Limit}
The $\bn$-collinear limit is IR finite, so we work in $d=4$,
\begin{align}
  \Msquared^{(0)}_{qg \to Vq}(Q,Y,\{k\}) &
 =  \Msquared^\LO_{q \bar q \to V}(Q)  \times 8\pi\as T_F \frac{e^Y}{\rho} \frac{1 - 2 z_b + 2 z_b^2}{Q \Tau}
\,,\\
  \Msquared^{(2)}_{qg \to Vq}(Q,Y,\{k\}) &
 =  \Msquared^\LO_{q \bar q \to V}(Q)  \times 4\pi \as T_F \frac{1+z_b+4z_b^2-8z_b^3+4z_b^4}{Q^2 z_b}
\,.\end{align}
The $\bn$-collinear kernel is given by
\begin{align}
 C_{f_q f_g,\bn}^{(2,1)}(z_a, z_b, \Tau) &= T_F\, \frac{\rho}{Q e^Y}\, \delta(1-z_a) \, (1-z_b)(1 + 8 z_b - 6 z_b^2)
\,,\nn\\
 C_{f'_q f_g,\bn}^{(2,1)}(z_a, z_b, \Tau) &= T_F\, \frac{\rho}{Q e^Y}\, \delta(1-z_a) \, \frac{(1+z_b)(1 - 2 z_b + 2 z_b^2)}{z_b}
\,,\nn\\
 C_{f_q f_g',\bn}^{(2,1)}(z_a, z_b, \Tau) &= T_F\, \frac{\rho}{Q e^Y}\, \delta(1-z_a) \, (1 - z_b)(1-2 z_b + 2 z_b^2)
\,.\end{align}

\paragraph{Final Result}
Adding the $s,n,\bn$ kernels, the pole in $\eps$ cancels and we get
\begin{align} \label{eq:kernel_Cqg_DY}
 C_{f_q f_g}^{(2,1)}(z_a, z_b, \Tau) &=
  2 T_F\, \frac{e^Y}{Q \rho}\, \biggl[
  \biggl( -\ln\frac{\Tau \rho}{Q e^Y} - 1 \biggr)\, \delta(1-z_a) + [1 + (1-z_a)^2]\, \cL_0(1-z_a) \biggr] \delta(1-z_b)
  \nn\\& \quad
  + T_F\, \frac{\rho}{Q e^Y}\, \delta(1-z_a) \, (1-z_b)(1 + 8 z_b - 6 z_b^2)
\,,\nn\\
 C_{f'_q f_g}^{(2,1)}(z_a, z_b, \Tau)
 &= T_F\, \frac{\rho}{Q e^Y}\, \delta(1-z_a) \, \frac{(1+z_b)(1 - 2 z_b + 2 z_b^2)}{z_b}
\,,\nn\\
 C_{f_q f_g'}^{(2,1)}(z_a, z_b, \Tau)
 &= T_F\, \frac{\rho}{Q e^Y}\, \delta(1-z_a) \, (1 - z_b)(1-2 z_b + 2 z_b^2)
\,.\end{align}
Substituting these results into \eq{sigma_NLP} yields the NLP cross section for $qg\to Vq$ at NLO.

\subsubsection[\texorpdfstring{$g q\to Vq$}{gq -> Vq}]{\boldmath $g q\to Vq$}

For completeness, we also give the explicit results for the $gq\to Vq$ channel,
which can easily be obtained from \eq{kernel_Cqg_DY} by flipping $a \leftrightarrow b$,
$e^Y/\rho \leftrightarrow \rho/e^Y$ and $f_q \leftrightarrow f_g$,
\begin{align} \label{eq:kernel_Cgq_DY}
 C_{f_g f_q}^{(2,1)}(z_a, z_b, \Tau) &=
  2 T_F\, \frac{\rho}{Q e^Y}\, \delta(1-z_a) \biggl[
  \biggl( -\ln\frac{\Tau e^Y}{Q \rho} - 1 \biggr)\, \delta(1-z_b) + [1 + (1-z_b)^2]\, \cL_0(1-z_b) \biggr]
  \nn\\ & \quad
  + T_F\, \frac{e^Y}{Q \rho}\, (1-z_a)(1 + 8 z_a - 6 z_a^2)\, \delta(1-z_b)
\,,\nn\\
 C_{f'_g f_q}^{(2,1)}(z_a, z_b, \Tau)
 &= T_F\, \frac{e^Y}{Q \rho}\, (1 - z_a)(1-2 z_a + 2 z_a^2)\, \delta(1-z_b)
\,,\nn\\
 C_{f_g f'_q}^{(2,1)}(z_a, z_b, \Tau)
 &= T_F\, \frac{e^Y}{Q \rho}\, \frac{(1+z_a)(1 - 2 z_a + 2 z_a^2)}{z_a}\, \delta(1-z_b)
\,.\end{align}

\section{Comparison with Integrated Results in the Literature}
\label{sec:compare_lit}

In this section, we compare our NLO results to previous results in the literature.
The LL results presented by a subset of the present authors in \refscite{Moult:2016fqy,Moult:2017jsg} fully agree with the results obtained in this paper.

The results in \refscite{Boughezal:2016zws,Boughezal:2018mvf} are given only integrated over the color-singlet rapidity $Y$, and hence take quite a different form at the integrand level.
To compare to them, we integrate our results over $Y$,
which allows us to use integration by parts to bring our results into the same integrated
form as those in \refscite{Boughezal:2016zws,Boughezal:2018mvf}.
For leptonic $\Tau$, whose definition involves $Y$, we find that \refcite{Boughezal:2018mvf} uses a different definition, and hence we cannot make a meaningful comparison.
For hadronic $\Tau$, whose definition is independent of $Y$, we find explicit agreement
for the LL results after integrating over $Y$.

At NLL, the results obtained here for the power corrections differential in $Y$, for both the leptonic and hadronic definitions and all partonic channels, are new. After integrating over $Y$ we find almost complete
agreement with the hadronic results of \refcite{Boughezal:2018mvf}, up to a relatively simple
term.%
\footnote{This missing term has been confirmed by the authors of \refcite{Boughezal:2018mvf}
and was corrected in their version 2.}

Since there are a number of differences in our treatment compared to \refscite{Boughezal:2016zws,Boughezal:2018mvf}, we provide a detailed comparison in this section.
In \sec{phasespace_NLO} we discuss our different treatments of the NLO phase space and of the Born measurements, and show that the rapidity dependence cannot be easily reconstructed from the results in \refscite{Boughezal:2016zws,Boughezal:2018mvf}.
In \sec{compare_results_lit} we provide an explicit comparison of the results for the $gg \to Hg$ channel integrated over rapidity at LL and NLL, both analytically and numerically.

\subsection{Treatment of the NLO Phase Space}
\label{sec:phasespace_NLO}

The derivation in \refcite{Boughezal:2018mvf} differs from ours here (and that in \refscite{Moult:2016fqy,Moult:2017jsg}) in that it is not differential in the rapidity $Y$.
To explore the differences arising from this, we give a brief derivation of the NLO phase space following the same steps as \refcite{Boughezal:2018mvf}.
Note that in the following we always work with an on-shell process,
in contrast to our more general setup in \sec{master_formula}.
We also only consider the case $k^+ < k^-$, since the case $k^+ > k^-$ follows by symmetry.

We start with the expression for the NLO phase space as given in \refcite{Boughezal:2018mvf},
\begin{align} \label{eq:PS_NLO_FP}
\frac{\df\text{PS}_{\text{NLO}}}{\df\Tau}
&= \frac{\Tau^{-\epsilon} (4\pi \muMS^2)^{-\epsilon}  }{8\pi \Gamma(1-\epsilon)}
\int\! \df \xi_a \df\xi_b \, \frac{ f_g(\xi_a) f_g(\xi_b)}{2\xi_a \xi_b\Ecm^2} \biggl(  \frac{Q_a \xi_a}{x_a} \biggr)^{1-\epsilon}
\nn \\ &\quad\times
\int \df z_a (1-z_a)^{-\epsilon}\delta \Bigl( \xi_a \xi_b z_a \Ecm^2  -m_H^2 -\frac{Q_a \xi_a}{x_a}\Tau \Bigr)
\,,\end{align}
where $s = \Ecm^2$, $Q_a$ is defined in \eq{Tau0_2} and $x_a$ arises from the $\Tau$ measurement.

We can derive a similar expression in our notation, including in addition the rapidity measurement
as done in our main derivation. Denoting the incoming momenta at NLO by $q'_{a,b}$, we have from \eq{sigma1}
\begin{align}
 \frac{\df\text{PS}^\NLO}{\df Y \df \Tau} &
 = \int_0^1 \df \xi_a \df \xi_b \frac{f_a(\xi_a) f_b(\xi_b)}{2 \xi_a \xi_b \Ecm^2}
   \muMS^{2\eps} \int\frac{\df^d k}{(2\pi)^d} (2\pi) \delta_+(k^2)
   \int\frac{\df^d q}{(2\pi)^d} (2\pi) \delta_+(q^2 - Q^2)
   \nn\\&\quad\times
   \, (2\pi)^d \delta(q'_a + q'_b - q - k)
   \delta\biggl(Y - \frac{1}{2} \ln\frac{q^-}{q^+}\biggr)
   \delta[\Tau - \hat\Tau(k)]
\nn\\&
 = \frac{1}{8\pi} \frac{(4\pi \muMS^2)^\eps}{\Gamma(1-\eps)}
   \int_0^1 \df \xi_a \df \xi_b \frac{f_a(\xi_a) f_b(\xi_b)}{2 \xi_a \xi_b \Ecm^2}
   \int_0^\infty \frac{\df k^+ \df k^-}{(k^+ k^-)^\eps} \delta[\Tau - \hat\Tau(k)]
   \nn\\&\quad\times
   \delta(\xi_a \xi_b \Ecm^2 - \xi_a \Ecm k^+ - \xi_b \Ecm k^- - Q^2) \,
   \delta\biggl(Y - \frac{1}{2} \ln\frac{\xi_a \Ecm - k^-}{\xi_b \Ecm - k^+}\biggr)
\,.\end{align}
As in \eq{PS_NLO_FP}, we assume that $k^+ < k^-$ to set $\hat\Tau(k) = \rho k^+$, which gives
\begin{align}
 \frac{\df\text{PS}^\NLO}{\df Y \df \Tau} &
 = \frac{1}{8 \pi} \frac{(4\pi \muMS^2)^{\eps}}{\Gamma(1-\eps)}
   \int_0^1 \df \xi_a \df \xi_b \frac{f_a(\xi_a) f_b(\xi_b)}{2 \xi_a \xi_b \Ecm^2}
   \int_0^\infty \frac{\df k^-}{\rho} \biggl( \frac{\rho}{\Tau k^-} \biggr)^\eps
   \\\nn& \quad \times
   \delta(\xi_a \xi_b \Ecm^2 - Q^2 - \xi_a \Ecm \Tau/\rho - \xi_b \Ecm k^-) \,
   \delta\biggl(Y - \frac{1}{2} \ln\frac{\xi_a \Ecm - k^-}{\xi_b \Ecm - \Tau/\rho}\biggr)
\,.\end{align}
Following \refcite{Boughezal:2018mvf}, we now change variables via $k^- = \xi_a \Ecm (1-z_a)$,
\begin{align} \label{eq:PS_NLO_1}
 \frac{\df\text{PS}^\NLO}{\df Y \df \Tau} &
 = \frac{\Tau^{-\eps}}{8\pi} \frac{(4\pi \muMS^2)^{\eps}}{\Gamma(1-\eps)}
   \int_0^1 \df \xi_a \df \xi_b \frac{f_a(\xi_a) f_b(\xi_b)}{2 \xi_a \xi_b \Ecm^2}
   \biggl( \frac{\xi_a \Ecm}{\rho} \biggr)^{1-\eps}
   \int\df z_a \, (1-z_a)^{-\eps}
   \nn\\&\quad\times
   \delta(z_a \xi_a \xi_b \Ecm^2 - \xi_a \Ecm \Tau/\rho - Q^2) \,
   \delta\biggl(Y - \frac{1}{2} \ln\frac{z_a \xi_a}{\xi_b - \frac{\Tau}{\rho \Ecm}}\biggr)
\,.\end{align}
Up to the rapidity measurement from the final $\delta$ function, we find complete agreement with \eq{PS_NLO_FP}
if we identify
\begin{align} \label{eq:Qa}
\rho \equiv \rho(Y) = \frac{x_a \Ecm}{Q_a(x_a)}
\,.\end{align}
At this step, our treatment differs from the one in \refcite{Boughezal:2018mvf}.
Since we explicitly implement measurement $\delta$ functions for both $Q$ and $Y$, we can uniquely solve
for $\xi_a$ and $\xi_b$ in terms of $Q$ and $Y$ or equivalently $x_a$ and $x_b$,
\begin{align} \label{eq:xi_ab}
 \xi_a &= \frac{ e^{+Y}}{2 z_a^2 \Ecm} \biggl[  \frac{\Tau e^{Y}}{\rho} (1-z_a) + \sqrt{ \biggl(\frac{\Tau e^Y}{\rho}\biggr)^2 (1-z_a)^2 + 4 Q^2 z_a^2} \biggr]
\,,\nn\\
 \xi_b &= \frac{ e^{-Y}}{2 z_a \Ecm} \biggl[  \frac{\Tau e^{Y}}{\rho} (1+z_a) + \sqrt{ \biggl(\frac{\Tau e^Y}{\rho}\biggr)^2  (1-z_a)^2 + 4 Q^2 z_a^2} \biggr]
\,.\end{align}
This holds for both $\rho = 1$ and $\rho = e^Y$.
This is equivalent to \eq{zeta_ab} (where we used the notation $\zeta_{a,b}$ instead of $\xi_{a,b}$ here).
The reason this expression looks different is just because in \eq{zeta_ab} we performed this step before fixing $k^+$ in terms of
$\Tau$ and before changing variables from $k^-$ to $z_a$ via $k^- = \xi_a \Ecm (1-z_a)$.

Following a similar strategy as in \sec{master_formula}, one can now replace $\xi_{a,b}$ in \eq{PS_NLO_1} by the solution \eq{xi_ab},
take the Jacobian from solving the $\delta$ functions into account, and then simply expand in $\Tau$.
The main difference to the derivation in \sec{master_formula} is that here, one directly expands the phase space in $\Tau$, while in \sec{master_formula} we expanded in terms of the generic power-counting parameter $\lambda$.

In \refcite{Boughezal:2018mvf}, there is only the $Q^2$ measurement but no rapidity measurement, i.e.\ $Y$ is implicitly integrated over.
Hence, there is only one constraint for the two variables $\xi_a,\xi_b$, whose solution is not unique.
They choose to perform the variable transformation from $\xi_{a,b}$ to new variables $\tilde x_{a,b}$ defined by
\begin{align}\label{eq:xi_a}
\xi_a = \frac{\tilde x_a^2 \tilde x_b \Ecm^2}{z_a \tilde x_a \tilde x_b \Ecm^2 - Q_a(\tilde x_a) \Tau}
\,, \qquad
 \xi_b &= \tilde x_b
\,.\end{align}
We write $\tilde x_{a,b}$ here to distinguish these from the Born variables $x_{a,b} = Q e^{\pm Y}/\Ecm$ that appear in the Born-projected momenta in \eq{Born}.
While they satisfy $\tilde x_a \tilde x_b = Q^2/\Ecm^2$ due to the $Q^2$ measurement constraint, $(1/2)\ln(\tilde x_a / \tilde x_b)$ is not equal to the rapidity $Y$, which would require the solution
in \eq{xi_ab}.

In \refcite{Boughezal:2018mvf}, the $\tilde x_{a,b}$ defined by \eq{xi_a}
also enter in the definition of the 0-jettiness measure in \eq{Tau0_0} in place of $x_{a,b}$.
As a result, the nonhadronic $\Tau$ definition in \refcite{Boughezal:2018mvf} is not the same as the usual leptonic $\Tau$ with $\rho = e^Y$ that we use.
Their hadronic definition is the same as ours, as it has no $x_{a,b}$ dependence.
Therefore in the following we restrict our comparison to the hadronic definition.

We also note that one cannot easily recover
the rapidity dependence from the integrands of the final results in \refcite{Boughezal:2018mvf}.
To see this explicitly, consider inserting the rapidity measurement by comparing \eqs{PS_NLO_FP}{PS_NLO_1}, which gives
\begin{align}\label{eq:Y_insert}
 \delta\biggl(Y - \frac{1}{2} \ln\frac{z_a \xi_a}{\xi_b - \frac{\Tau}{\rho \Ecm}}\biggr)
 &= \delta\biggl(Y - \frac{1}{2} \ln\frac{z_a \xi_a^{(0)}}{\xi_b^{(0)}}\biggr)
 \\\nn&\quad
   + \frac{\Tau}{2} \delta'\biggl(Y - \frac{1}{2} \ln\frac{z_a \xi_a^{(0)}}{\xi_b^{(0)}}\biggr)
     \biggl( \frac{{\xi'}_b^{(0)}}{\xi_b^{(0)}} - \frac{{\xi'}_a^{(0)}}{\xi_a^{(0)}}
            - \frac{1}{ \xi_b^{(0)}\rho \Ecm} \biggr)
 + \cO(\Tau^2)
\,.\end{align}
On the right-hand side we have carried out the power expansion about $\Tau\to 0$ and the superscript $^{(0)}$ denotes the results for these variables at LP, while $\xi_a^{\prime(0)} = d\xi_a/d\Tau \big|_{\Tau\to 0}$, etc. This accounts for the fact that in general the $\xi_{a,b}$ can depend on $\Tau$ themselves.
Equation \eqref{eq:Y_insert} shows that one cannot use the LP expression
$\delta[Y - (1/2)\ln(z_a \xi_a^{(0)}/\xi_b^{(0)})] = \delta[Y - (1/2)\ln(\tilde x_a/\tilde x_b)]$
to recover the $Y$ dependence from the $\tilde x_{a,b}$ dependence of the results in \refcite{Boughezal:2018mvf}, as this does not account for the additional power corrections
induced by the $Y$ measurement in the second line of \eq{Y_insert}.
This implies that the results in \refcite{Boughezal:2018mvf} and also those in \refcite{Boughezal:2016zws}
cannot be used when being differential in rapidity or integrated over bins of rapidity, but only integrated over all $Y$. This was also confirmed to us by the authors.

\subsection[Explicit Comparison to Results in the Literature for \texorpdfstring{$gg\to Hg$}{gg -> Hg}]{Explicit Comparison to Results in the Literature for \boldmath $gg\to Hg$}
\label{sec:compare_results_lit}

Our final results take a quite different form than those in \refscite{Boughezal:2016zws,Boughezal:2018mvf}.
For us, both $\xi_a$ and $\xi_b$ receive power corrections resulting in derivatives for both PDFs.
In contrast, the variable transformation in \eq{xi_a} for the case of $k^+ < k^-$
does not yield power corrections for $\xi_b$ and hence no derivatives of $f_b$,
while the expansion of $\xi_a$ yields derivatives of $f_a$ (and vice versa for $k^+ > k^-$).
Due to this different form, one cannot directly compare the integrands of the two results,
but one needs to use integration by parts to bring the results into the same form,
as we will now show explicitly. In particular, we will show that the results of \refscite{Moult:2016fqy,Moult:2017jsg}, obtained also here, do agree with the results of \refscite{Boughezal:2016zws, Boughezal:2018mvf} at LL when integrating over all $Y$.

Integrating our result over $Y$, and transforming the integration variables to $x_{a,b} = Q e^{\pm Y}/\Ecm$, we obtain from \eq{sigma_NLP}
\begin{align} \label{eq:sigma_NLP_NLO_1}
 \frac{\df\sigma^{(2,1)}}{\df \Tau} &
 = \frac{\as}{4\pi} \int_0^1 \df x_a \df x_b \,
   2\pi \delta(x_a x_b \Ecm^2 - m_H^2) \frac{|\cM_{gg\to H}^\LO(m_H)|^2}{2 x_a x_b \Ecm^2}
   \int_{x_a}^1 \frac{\df z_a}{z_a} \int_{x_b}^1 \frac{\df z_b}{z_b} \,
   \nn\\&\quad\times
    \biggl[
    f_i\biggl(\frac{x_a}{z_a}\biggr) f_j\biggl(\frac{x_b}{z_b}\biggr) C_{f_i f_j}^{(2,1)}(z_a, z_b, \Tau)
   + \frac{x_a}{z_a} f'_i\biggl(\frac{x_a}{z_a}\biggr) f_j\biggl(\frac{x_b}{z_b}\biggr) C_{f_i' f_j}^{(2,1)}(z_a, z_b, \Tau)
   \nn\\&\hspace{1cm}
   + \frac{x_b}{z_b} f_i\biggl(\frac{x_a}{z_a}\biggr) f'_j\biggl(\frac{x_b}{z_b}\biggr) C_{f_i f_j'}^{(2,1)}(z_a, z_b, \Tau)
   \biggr]
\,.\end{align}

We will show the integration by parts explicitly for the $f_i f'_j$ piece.
Let us denote the piece we wish to integrate by parts by $D^{(2,1)}$,
which can be chosen freely.
To integrate over $Y_1 < Y < Y_2$, we switch the integration variables $x_a, x_b$ back to $Q^2$ and $Y$, use that
\begin{align}
 \frac{x_b}{z_b} f'_j\biggl(\frac{x_b}{z_b}\biggr)
 = \frac{Q e^{-Y}}{\Ecm z_b} f'_j\biggl(\frac{Q e^{-Y}}{\Ecm z_b}\biggr)
 = - \frac{\df}{\df Y} f_j\biggl(\frac{Q e^{-Y}}{\Ecm z_b}\biggr)
\,,\end{align}
and integrate by parts with respect to $Y$.
Combining the resulting pieces with those in \eq{sigma_NLP_NLO_1}, we find
\begin{align} \label{eq:sigma_NLP_NLO_2}
 \frac{\df\sigma^{(2,1)}}{\df \Tau} &
 = \frac{\as}{4\pi} \int_0^1 \df x_a \df x_b \,
   2\pi \delta(x_a x_b \Ecm^2 - m_H^2) \frac{|\cM_{gg\to H}^\LO(m_H)|^2}{2 x_a x_b \Ecm^2}
   \int_{x_a}^1 \frac{\df z_a}{z_a} \int_{x_b}^1 \frac{\df z_b}{z_b} \,
   \nn\\&\quad\times
   \biggl\{
    f_i\biggl(\frac{x_a}{z_a}\biggr) f_j\biggl(\frac{x_b}{z_b}\biggr)
    \biggl[ C_{f_i f_j}^{(2,1)}(z_a, z_b, \Tau) + \frac{\df}{\df Y} D^{(2,1)}(z_a, z_b, \Tau) \biggr]
   \nn\\&\hspace{1cm}
   + \frac{x_a}{z_a} f'_i\biggl(\frac{x_a}{z_a}\biggr) f_j\biggl(\frac{x_b}{z_b}\biggr)
    \biggl[ C_{f_i' f_j}^{(2,1)}(z_a, z_b, \Tau) + D^{(2,1)}(z_a, z_b, \Tau) \biggr]
   \nn\\&\hspace{1cm}
   + \frac{x_b}{z_b} f_i\biggl(\frac{x_a}{z_a}\biggr) f'_j\biggl(\frac{x_b}{z_b}\biggr)
     \Bigl[ C_{f_i f_j'}^{(2,1)}(z_a, z_b, \Tau) - D^{(2,1)}(z_a, z_b, \Tau) \Bigr]
   \biggr\}
\nn\\&
- \frac{\as}{4\pi} \int\df Q^2 \,
   2\pi \delta(Q^2 - m_H^2) \frac{|\cM_{gg\to H}^\LO(m_H)|^2}{2 Q^2\Ecm^2}
   \int_{\frac{Q e^Y}{\Ecm}}^1 \frac{\df z_a}{z_a} \int_{\frac{Q e^{-Y}}{\Ecm}}^1 \frac{\df z_b}{z_b}
   \nn\\&\qquad\times
   f_i\biggl(\frac{Q e^Y}{\Ecm z_a}\biggr)  f_j\biggl(\frac{Q e^{-Y}}{\Ecm z_b}\biggr)
   D^{(2,1)}(z_a, z_b, \Tau) \bigg|_{Y=Y_1}^{Y=Y_2}
\,.\end{align}
The dependence on $D^{(2,1)}$ exactly cancels in this expression.
We can choose $D^{(2,1)}$ freely to obtain different forms of the $Y$-integrated  result.
The last term in \eq{sigma_NLP_NLO_2} is the boundary contribution, which vanishes as $Y_{1,2} \to \pm \infty$, i.e.\ only if one is fully inclusive in $Y$.
They do in general contribute when placing acceptance cuts on $Y$.

We now work out explicitly the required integration by parts both at LL and NLL
to bring our results into the integrated form as given in \refscite{Boughezal:2016zws,Boughezal:2018mvf}.
For concreteness, we focus on the $gg\to Hg$ channel. For the reasons mentioned
earlier, we can only compare the results for the hadronic $\Tau$ definition.

\subsubsection{Comparison at LL}

At LL, our results in \eq{kernel_Cgg} simplify to
\begin{align} \label{eq:kernel_Cgg_LL}
 C_{f_g f_g}^{(2,1),\text{LL}}(z_a, z_b, \Tau) &
 = 4 C_A \biggl[
   \frac{e^Y}{Q \rho} \ln\frac{\Tau \rho}{Q e^Y} +
   \frac{\rho}{Q e^Y} \ln\frac{\Tau e^Y}{Q \rho} \biggr]
   \delta(1-z_a) \delta(1-z_b)
\,,\nn\\
 C_{f'_g f_g}^{(2,1),\text{LL}}(z_a, z_b, \Tau) &
 =  - 4 C_A \frac{\rho}{Q e^Y} \ln\frac{\Tau e^Y}{Q \rho} \delta(1-z_a) \delta(1-z_b)
\,,\nn\\
 C_{f_g f_g'}^{(2,1),\text{LL}}(z_a, z_b, \Tau) &
 = - 4 C_A \frac{e^Y}{Q \rho} \ln\frac{\Tau \rho}{Q e^Y} \delta(1-z_a) \delta(1-z_b)
\,.\end{align}
These agree with the earlier results obtained by a subset of the current authors in \refscite{Moult:2016fqy,Moult:2017jsg}.
Note that for strict LL accuracy, one can also write the logarithms as
$\ln(\Tau/Q) \pm \ln (\rho/e^Y)$ and only keep the $\ln(\Tau/Q)$ at LL, while
including the $\pm\ln(\rho/e^Y)$ pieces in the NLL contributions.
(This is the convention used in \refscite{Moult:2016fqy, Moult:2017jsg} and
in \sec{results}.)
Here, we keep them as part of the LL result, as they are relevant for the comparison with \refcite{Boughezal:2018mvf}.

Up to a trivial change in notation, the LL result given in \refcite{Boughezal:2018mvf}
for hadronic $\Tau$ is
\begin{align} \label{eq:sigma_ggHg_LL_had_lit}
 \frac{\df \sigma^\text{NLP\cite{Boughezal:2018mvf}}_\text{LL}}{\df\Tau^\hadcm} &
 = \frac{\as C_A}{\pi} \int_0^1 \df \tilde x_a \df \tilde x_b \, 2\pi \delta(\tilde x_a \tilde x_b \Ecm^2 - m_H^2)\,
   \frac{|\cM^\LO_{gg\to H}(m_H)|^2}{2 \tilde x_a \tilde x_b \Ecm^2}
   \\\nn&\quad\times
   \biggl[
    - \tilde x_a f'_g(\tilde x_a) f_g(\tilde x_b) \, \frac{\tilde x_a \Ecm}{m_H^2} \ln\frac{\Tau^\hadcm}{\tilde x_a \Ecm}
    - \tilde x_b f_g(\tilde x_a) f'_g(\tilde x_b) \, \frac{\tilde x_b \Ecm}{m_H^2} \ln\frac{\Tau^\hadcm}{\tilde x_b \Ecm}
   \biggr]
\,.\end{align}
As discussed before, the $\tilde x_{a,b}$ here are not equal to the Born variables $x_{a,b}$.

Inserting our LL result in \eq{kernel_Cgg_LL} with $\rho = 1$ into \eq{sigma_NLP_NLO_1}, we have
\begin{align} \label{eq:sigma_ggHg_had_LL_1}
 \frac{\df\sigma^{(2,1)}_\text{LL}}{\df \Tau^\hadcm} &
 = \frac{\as C_A}{\pi} \int_0^1 \df x_a \df x_b \, 2\pi \delta(x_a x_b \Ecm^2 - m_H^2)\,
   \frac{|\cM^\LO_{gg\to H}(m_H)|^2}{2 x_a x_b \Ecm^2}
   \nn\\&\quad\times
   \biggl[
    f_g(x_a) f_g(x_b) \biggl( \frac{e^Y}{m_H} \ln\frac{\Tau^\hadcm e^{-Y}}{m_H} + \frac{e^{-Y}}{m_H} \ln\frac{\Tau^\hadcm e^Y}{m_H} \biggr)
   \nn\\&\qquad
   - x_a f'_g(x_a) f_g(x_b) \frac{e^{-Y}}{m_H} \ln\frac{\Tau^\hadcm e^Y}{m_H}
   - x_b f_g(x_a) f'_g(x_b) \frac{e^{Y}}{m_H} \ln\frac{\Tau^\hadcm e^{-Y}}{m_H}
   \biggr]
\,,\end{align}
where $e^Y = \sqrt{x_a/x_b}$.
At the integrand level, the two results clearly have a different form, as was also remarked in \refscite{Moult:2017jsg,Boughezal:2018mvf}.

To show explicitly that \eqs{sigma_ggHg_LL_had_lit}{sigma_ggHg_had_LL_1} do agree,
we integrate by parts to move the $f_g f_g$ contribution in \eq{sigma_ggHg_had_LL_1}
into the $f_g f'_g$ and $f'_g f_g$ terms. Using \eq{sigma_NLP_NLO_2}, we can achieve this by choosing
\begin{align}
 D^{(2,1)}(z_a, z_b, \Tau^\hadcm)
&= 4 C_A \biggl( - \frac{e^Y}{m_H} \ln\frac{\Tau^\hadcm e^{-Y}}{m_H}
                 + \frac{e^{-Y}}{m_H} \ln\frac{\Tau^\hadcm e^Y}{m_H} \biggr)
   \delta(1-z_a)\, \delta(1-z_b)
\,.\end{align}
Integrating over $Y_1 < Y < Y_2$ and using that $e^Y = \sqrt{x_a / x_b}$ and $m_H = \sqrt{x_a x_b} \Ecm$, \eq{sigma_ggHg_had_LL_1} becomes
\begin{align} \label{eq:sigma_ggHg_had_LL}
 \frac{\df\sigma^{(2,1)}_\text{LL}(Y_1, Y_2)}{\df \Tau^\hadcm} &
 = \frac{\as C_A}{\pi} \int_0^1 \df x_a \df x_b
   \, 2\pi \delta(x_a x_b \Ecm^2 - m_H^2)\, \frac{|\cM_{gg\to H}^\LO(m_H)|^2}{2 x_a x_b \Ecm^2}
   \\* &\qquad\times
   \biggl[
   - x_a f'_g(x_a) f_g(x_b) \frac{x_a \Ecm}{m_H^2} \ln\frac{\Tau^\hadcm}{x_a \Ecm}
   - x_b f_g(x_a) f'_g(x_b) \frac{x_b \Ecm}{m_H^2} \ln\frac{\Tau^\hadcm}{x_b \Ecm}
   \biggr]
\nn\\& \quad
+ \frac{\as C_A}{\pi}  \frac{2\pi|\cM_{gg\to H}^\LO(m_H)|^2}{2 m_H^2 \Ecm^2}
   f_g\biggl(\frac{m_H e^Y}{\Ecm}\biggr) f_g\biggl(\frac{m_H e^{-Y}}{\Ecm}\biggr)
\nn\\& \qquad \times
   \biggl[ \frac{e^{Y}}{m_H} \ln\frac{\Tau^\hadcm e^{-Y}}{m_H} -
     \frac{e^{-Y}}{m_H} \ln\frac{\Tau^\hadcm e^{Y}}{m_H} \biggr]
   \bigg|_{Y_1}^{Y_2}
\nn\\\nn&\quad
 + \frac{\as C_A}{\pi}\! \int_0^1\!\! \df x_a \df x_b
   \, 2\pi \delta(x_a x_b \Ecm^2\! - m_H^2)\, \frac{|\cM_{gg\to H}^\LO(m_H)|^2}{2 x_a x_b \Ecm^2}
   f_g(x_a) f_g(x_b) \frac{e^Y\! +\! e^{-Y}}{m_H}
\,.\end{align}
The first two lines exactly reproduce \eq{sigma_ggHg_LL_had_lit}.
The following two lines are the boundary term from integration by parts,
which vanishes as $Y_{1,2} \to \pm \infty$.
The last line is a NLL effect and can be neglected for the LL comparison.
(It is induced by the integration by parts acting on the
$Y$ dependence kept inside the argument of the logarithms.)
Therefore, the two expressions in \eqs{sigma_ggHg_LL_had_lit}{sigma_ggHg_had_LL_1}
agree at LL and at integrated level if and only if one integrates over all rapidity.

\begin{figure}[t]
\centering
\includegraphics[width=0.6\textwidth]{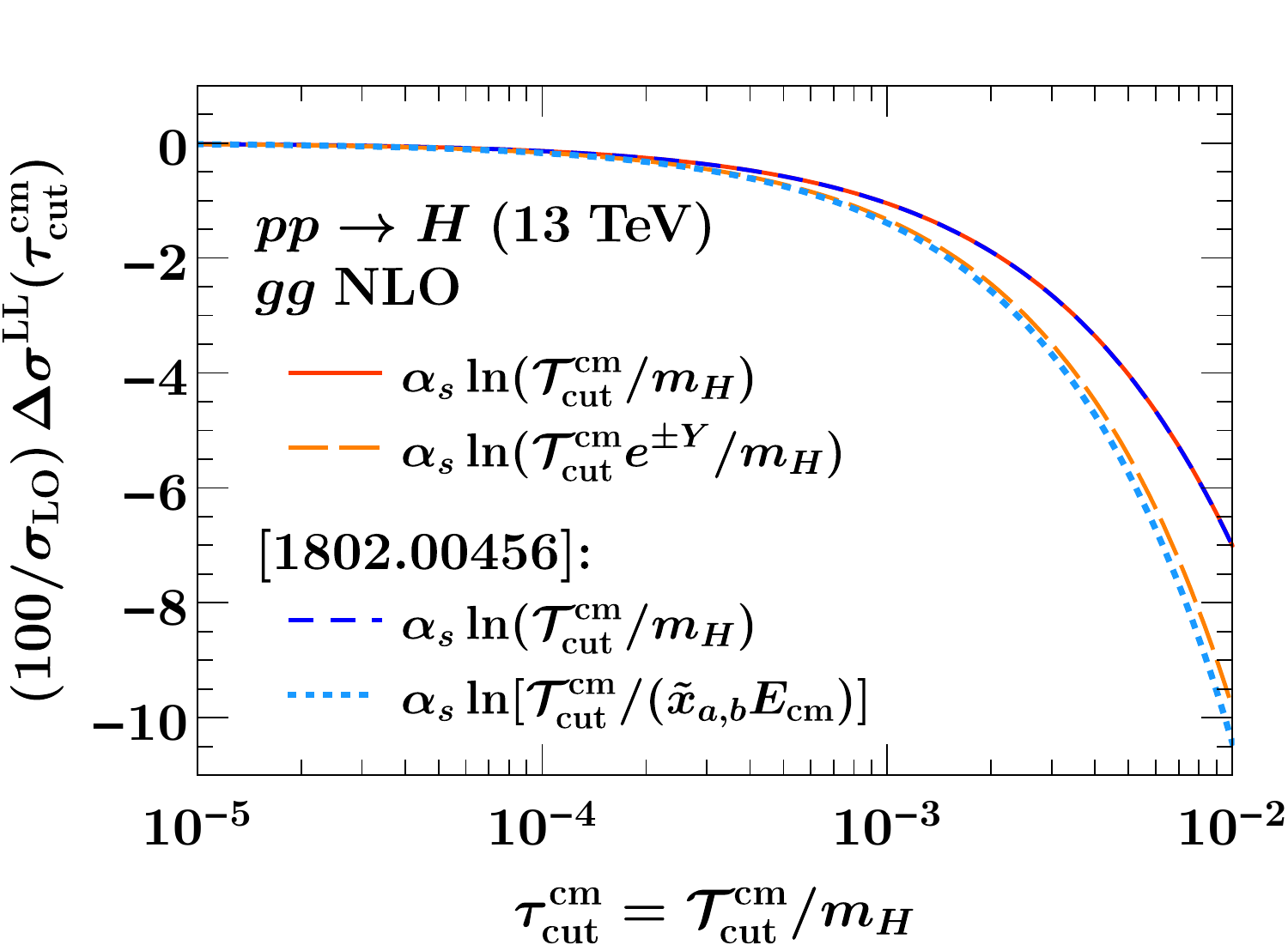}
\caption{Comparison of the $Y$-integrated LL power correction for hadronic $\Tau$ for $gg\to Hg$.
The solid red and blue dashed curves show the LL results keeping only $\ln(\Tau/m_H)$.
In the long-dashed orange and dotted light blue curves we keep all $\ln(\Tau e^{\pm Y}/m_H)$ or $\ln[\Tau/(\tilde x_{a,b} \Ecm)]$ terms.
In both cases, our result in \eq{sigma_ggHg_had_LL_1} and the result
of \refcite{Boughezal:2018mvf} in \eq{sigma_ggHg_LL_had_lit} agree. The small difference
in the second case arises due to the fact that $e^{\pm Y}/m_H$ is not exactly the same as
$\tilde x_{a,b} \Ecm$.}
\label{fig:comparison_cumulant_Taucm}
\end{figure}

To illustrate this numerically, the $Y$-integrated results are compared in \fig{comparison_cumulant_Taucm}.
First note that the hadronic LL results in \eq{sigma_ggHg_had_LL_1} do not exactly correspond to those previously given in \refscite{Moult:2016fqy,Moult:2017jsg}.
This is due to the formally NLL terms proportional to $\ln(\rho/e^Y)$, discussed below \eq{kernel_Cgg_LL}, which are dropped in the strict LL results in \refscite{Moult:2016fqy,Moult:2017jsg},
but are kept in \eq{sigma_ggHg_had_LL_1}. The analogous NLL terms proportional to $\ln(\tilde x_{a,b}\Ecm)$
are also kept in \refscite{Boughezal:2016zws,Boughezal:2018mvf} and \eq{sigma_ggHg_LL_had_lit}.
Dropping these NLL terms in \eqs{sigma_ggHg_LL_had_lit}{sigma_ggHg_had_LL_1}, our and their LL results defined in terms of the same $\ln(\Tau/m_H)$ agree exactly,
as shown by the solid red and blue dashed curves in \fig{comparison_cumulant_Taucm}.%
\footnote{%
In the first version of \refcite{Boughezal:2018mvf} an analogous numerical comparison showed a disagreement
between their integrated LL results and our corresponding result from \refcite{Moult:2017jsg}. This was only due to an incorrect comparison. We thank the authors of \refcite{Boughezal:2018mvf} for confirming this.}
The long-dashed orange and dotted blue curves in \fig{comparison_cumulant_Taucm} show the
results when using instead $\ln(\Tau e^{\pm Y}/m_H)$ or $\ln[\Tau/(\tilde x_{a,b} \Ecm)]$ to multiply
the LL coefficients. The observed difference to the solid red/dashed blue strict LL result has the size of a typical NLL contribution. There is also a very small difference between the long-dashed orange and dotted blue results due to the fact that $e^{\pm Y}/m_H$ is not exactly the same as $\tilde x_{a,b} \Ecm$. This difference is exactly accounted for by the last line in \eq{sigma_ggHg_had_LL}.

\subsubsection{Comparison at NLL}
\label{sec:NLLcompare}

We now extend our comparison of the $Y$-integrated results to NLL,
focusing again only on the $g g \to H g$ channel,
which contains all possible complications.
The full NLL result of \refcite{Boughezal:2018mvf} can be written as
\begin{align} \label{eq:sigma_ggHg_NLL_lit}
 \frac{\df \sigma^\text{NLP\cite{Boughezal:2018mvf}}}{\df\Tau^\hadcm} &
 = \frac{\as C_A}{\pi} \int_0^1 \df \tilde x_a \df \tilde x_b \, 2\pi \delta(\tilde x_a \tilde x_b \Ecm^2 - m_H^2)
   \frac{|\cM^\LO_{gg\to H}(m_H)|^2}{2 \tilde x_a \tilde x_b \Ecm^2}
   \int_{\tilde x_a}^1 \frac{\df z_a}{z_a} \frac{\tilde x_a \Ecm}{m_H^2}
   \nn\\&\quad\times
   \biggl\{
   f_g\biggl(\frac{\tilde x_a}{z_a}\biggr) f_g(\tilde x_b) \biggl[
    \biggl(\frac{(1-z_a+z_a^2)^2}{z_a^2} - 1 \biggr) \cL_0(1-z_a)
    + \frac{3 z_a^2 + 1 - z_a + z_a^3}{z_a^2} \biggr]
   \nn\\\nn&\hspace{1cm}
   + \frac{\tilde x_a}{z_a} f'_g\biggl(\frac{\tilde x_a}{z_a}\biggr) f_g(\tilde x_b) \biggl[
     - \delta(1-z_a) \ln\frac{\Tau^\hadcm}{\tilde x_a \Ecm}
     + \frac{(1-z_a + z_a^2)^2}{z_a^2} \cL_0(1-z_a)
   \biggr]
   \nn\\&\hspace{1cm}
 - f_g\biggl(\frac{\tilde x_a}{z_a}\biggr) \, \tilde x_b f'_g(\tilde x_b) \delta(1-z_a) \biggr\}
 \nn\\& \quad
   + (a \leftrightarrow b)
\,.\end{align}
To bring our result into this same form,
we need to integrate by parts twice, first with respect to $Y$ as shown in \eq{sigma_NLP_NLO_2},
and then with respect to $z_a$.
The details of this calculation are given in \app{ggHg_NLPs}.
The final result is shown in \eq{sigma_NLL_comparison_4} and is given by the result of
\refcite{Boughezal:2018mvf} in \eq{sigma_ggHg_NLL_lit} plus an extra contribution,
\begin{align} \label{eq:sigma_NLL_comparison}
\frac{\df\sigma^{(2,1)}}{\df \Tau^\hadcm}
 &= \frac{\df \sigma^\text{NLP\cite{Boughezal:2018mvf}}}{\df\Tau^\hadcm}
\\* \nn & \quad
 + 2\frac{\as C_A}{\pi} \int_0^1 \df x_a \df x_b \,
   2\pi \delta(x_a x_b \Ecm^2 - m_H^2) \frac{|\cM_{gg\to H}^\LO(m_H)|^2}{2 x_a x_b \Ecm^2}
   f_g(x_a) f_g(x_b)\, \frac{e^Y + e^{-Y}}{m_H}
\,.\end{align}
The two results should agree exactly upon integration, and we have not been able to find a source for this discrepancy. 
As discussed in more detail in
\sec{results}, the numerical comparison with MCFM provides a strong confirmation of our result.
The numerical extraction of the integrated NLL coefficient yields $-0.460 \pm 0.026$, which agrees
well with our analytic predicted value of $-0.466$ (see table~\ref{tab:NLLresults_H} below).
Dropping the term in the final line of \eq{sigma_NLL_comparison} would instead predict the value $ -1.669$.%
\footnote{
We recently received confirmation from the authors of \refcite{Boughezal:2018mvf} that after rechecking their calculation they identified a missing term, and now agree with our result for $\df\sigma^{(2,1)}/\df \Tau^\hadcm$.}

\section{Numerical Results}
\label{sec:results}

In this section we study our results numerically, including the size of the power corrections and the rapidity dependence. We also compare our analytic results for the $\cO(\alpha_s)$ NLP power corrections with the full nonsingular spectrum obtained numerically from the LO $V+$jet and $H+$jet calculations in MCFM8~\cite{Campbell:1999ah, Campbell:2010ff, Campbell:2015qma, Boughezal:2016wmq}.
In \refscite{Moult:2016fqy,Moult:2017jsg}, the NLP corrections were extracted numerically
by using a fit of the known form of their logarithmic structure to the nonsingular spectrum from MCFM8.
In \refscite{Moult:2016fqy,Moult:2017jsg}, these fits were carried out for the leptonic definition.
Here, we have in addition performed the fits also for the hadronic definition.
We find excellent agreement between the analytically predicted values
and the numerically extracted values for all coefficients, i.e., for the LL and NLL coefficients
in all partonic channels for both the leptonic and hadronic definition.
This provides a strong and independent cross check for the correctness of the analytic NLL results obtained here. By comparing the complete nonsingular spectrum with our NLP result, we can also assess the importance of power corrections beyond NLP.

The NLO power corrections for each partonic channel are extracted from the nonsingular spectrum by using the fit function
\begin{align} \label{eq:fitfun}
F_\mathrm{NLO}(\tau)
&= \frac{\df}{\df\ln\tau}\Bigl\{ \tau \bigl[(a_1 + b_1 \tau + c_1 \tau^2) \ln\tau
+ a_0 + b_0\tau + c_0\tau^2 \bigr] \Bigr\}
\,,\end{align}
with $\tau \equiv \Tau_0/m_Z$ for $Z$ production and $\tau \equiv \Tau_0/m_H$ for Higgs production. Details of the fitting procedure have been
described already in \refscite{Moult:2016fqy,Moult:2017jsg}, so we do not repeat them here.
A key point is that in order to obtain a precise and unbiased fit result for the to-be extracted $a_i$ coefficients, it is crucial to include the higher-power $b_i$ and $c_i$ terms in \eq{fitfun}, and to carefully choose the fit range and verify the stability of the fit, as was
done in~\refscite{Moult:2016fqy,Moult:2017jsg}. At the level of precision the $a_i$ are extracted,
this is essential since the full nonsingular cross section includes the complete set
of power corrections and if the $b_i$ and $c_i$ terms were
neglected, these higher-power corrections would be absorbed by the $a_i$ terms
in the fit, rendering their numerically extracted values meaningless.
To obtain a precise extraction of the NLL coefficient $a_0$, we fix the LL coefficient
$a_1$ in the fit to its analytic result.

The relevant coefficients for our NLP comparison at NLO are the LL coefficient $a_1$ and the NLL coefficient $a_0$. For leptonic $\Tau$ they
were extracted for Drell-Yan in \refcite{Moult:2016fqy} and for gluon-fusion Higgs in \refcite{Moult:2017jsg} and for the hadronic $\Tau$ we have obtained them here.
Depending on the partonic channel, the uncertainties on the fitted
coefficients range from $0.08\%$ to $2.3\%$ for leptonic $\Tau$ and from $0.6\%$ to $5.7\%$
for hadronic $\Tau$. The latter has larger uncertainties because its power corrections are
larger, requiring the fit to be restricted to smaller $\Tau$ values where the
uncertainties in the nonsingular data are larger.

\subsection{Drell-Yan Production}

We first consider Drell-Yan production, taking $pp\to Z/\gamma^*$
at $E_{\rm cm} = 13\TeV$. We use the MMHT2014 NNLO PDFs~\cite{Harland-Lang:2014zoa} with
fixed scales $\mu_r = \mu_f = m_Z$, and $\alpha_s(m_Z) = 0.118$. We fix $Q=m_Z$,
integrate over the vector-boson rapidity,
and work in the narrow-width approximation for the $Z$-boson. The NLP corrections
for the leptonic $\Tau$ definition were numerically extracted in \refcite{Moult:2016fqy}.
The results for both the leptonic and hadronic definitions for all partonic channels are collected and
compared to our analytic predictions in table~\ref{tab:NLLresults_DY}.
We find excellent agreement within the fit uncertainties in all cases.

\begin{table}[ht!]
\centering
\begin{tabular}{c|ll}
\hline\hline
NLO $\Tau_0^{\rm lep}$  $q \bar q \to Zg$ & $a_1$ & $a_0$
\\ \hline
fitted~\cite{Moult:2016fqy} & $+0.25366\pm 0.00131$  & $+0.13738 \pm 0.00057$
\\
analytic & $+0.25509$ & $+0.13708$
\\ \hline
NLO $\Tau_0^{\rm lep}$  $qg+gq \to Zq$ &  $a_1$ & $a_0$
\\ \hline
fitted~\cite{Moult:2016fqy} & $-0.27697 \pm 0.00113$ &  $-0.40062\pm 0.00052$
\\
analytic & $-0.27720$ & $-0.40105$
\\
\hline\hline
NLO $\Tau_0^\hadcm$ $q \bar q \to Zg$ & $a_1$ & $a_0$
\\ \hline
fitted & $+1.4188 \pm 0.0614$  & $-2.4808 \pm 0.0176$
\\
analytic & $+1.3935$ & $-2.4806$
\\ \hline\hline
NLO $\Tau_0^\hadcm$  $qg+gq \to Zq$ &  $a_1$ & $a_0$
\\ \hline
fitted & $-2.2981 \pm  0.0442$ &  $+4.0991 \pm 0.0132$
\\
analytic & $-2.3224$ & $+4.0965$
\\ \hline\hline
\end{tabular}
\caption{Comparison between our analytic predictions and the fitted results for the LL $a_1$ and NLL $a_0$ coefficients in Drell-Yan production. These fitted values for $a_1$ and $a_0$ with the leptonic definition and the analytic results for $a_1$ were already given in \refcite{Moult:2016fqy}.}
\label{tab:NLLresults_DY}
\end{table}

\begin{figure*}[t!]
\includegraphics[width=0.5\textwidth]{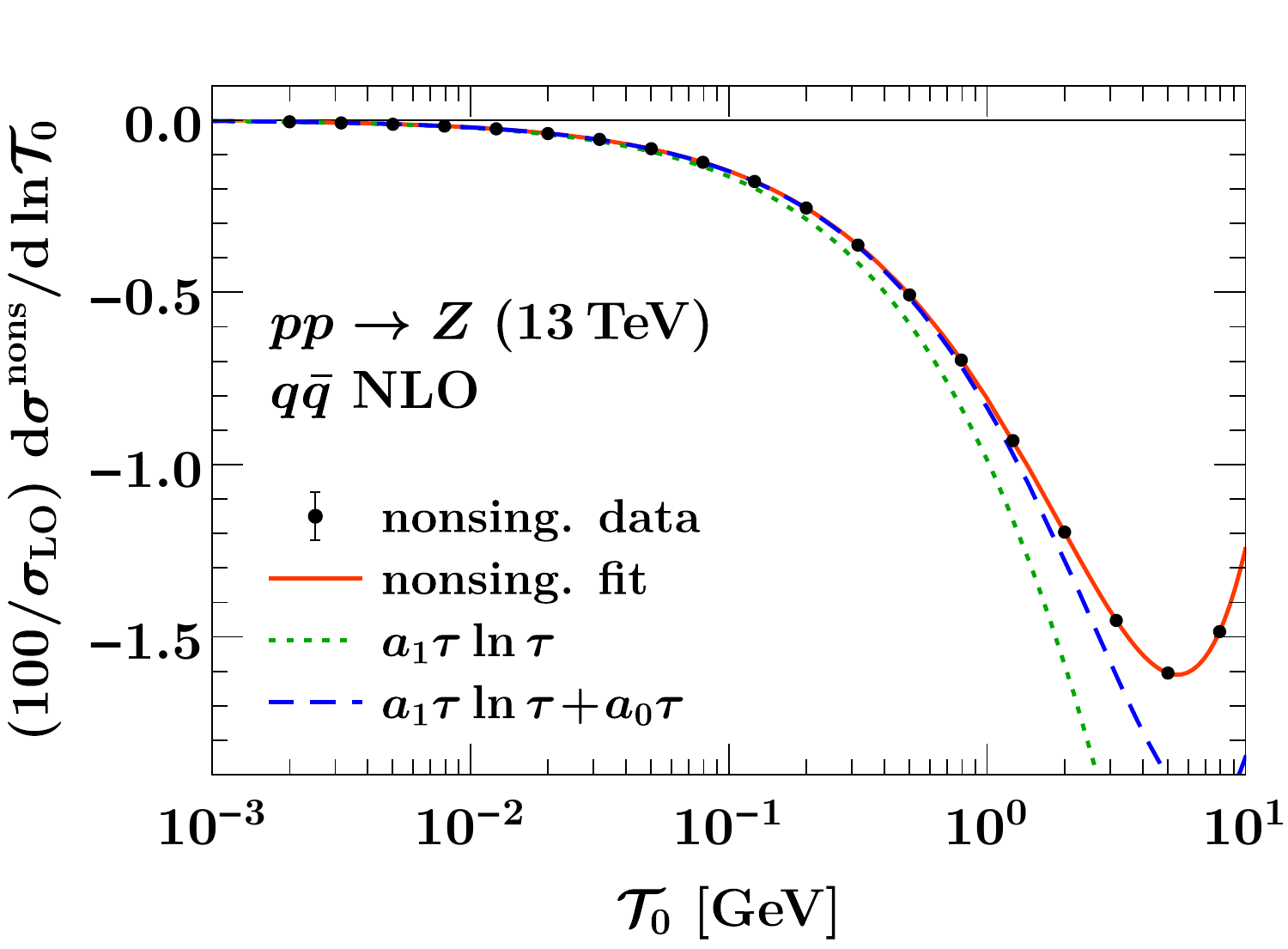}%
\hfill
\includegraphics[width=0.5\textwidth]{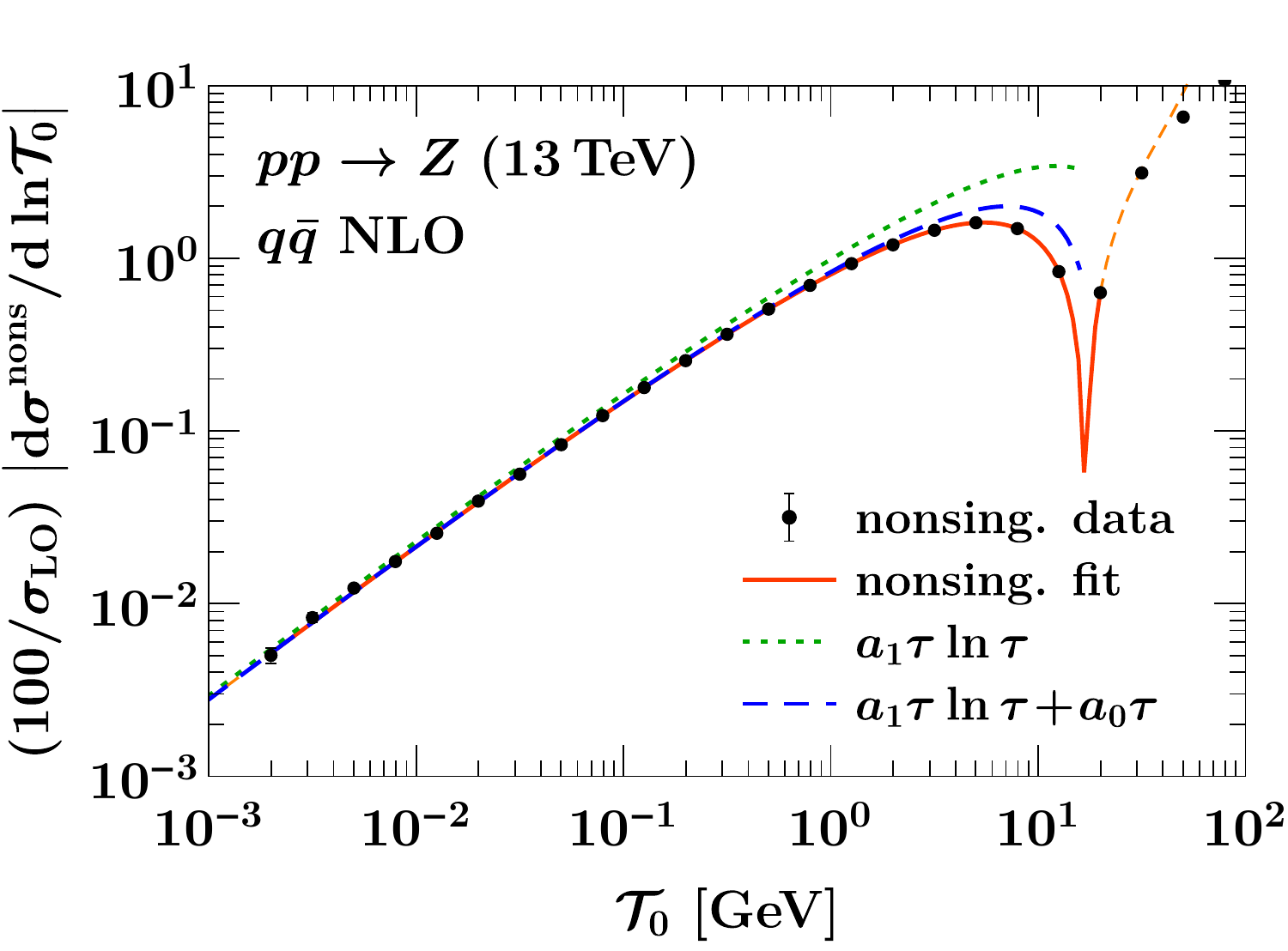}%
\\
\includegraphics[width=0.5\textwidth]{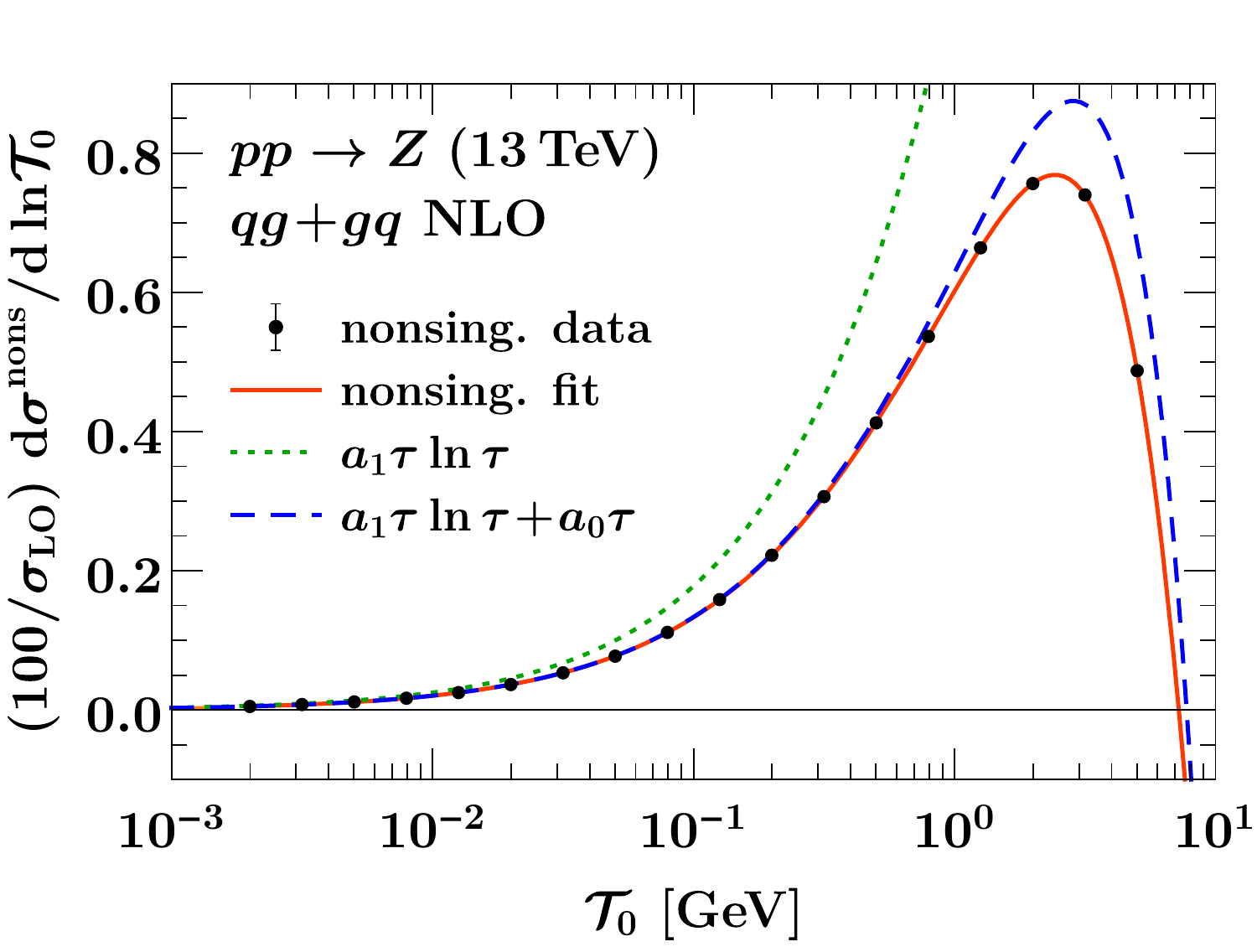}%
\hfill
\includegraphics[width=0.5\textwidth]{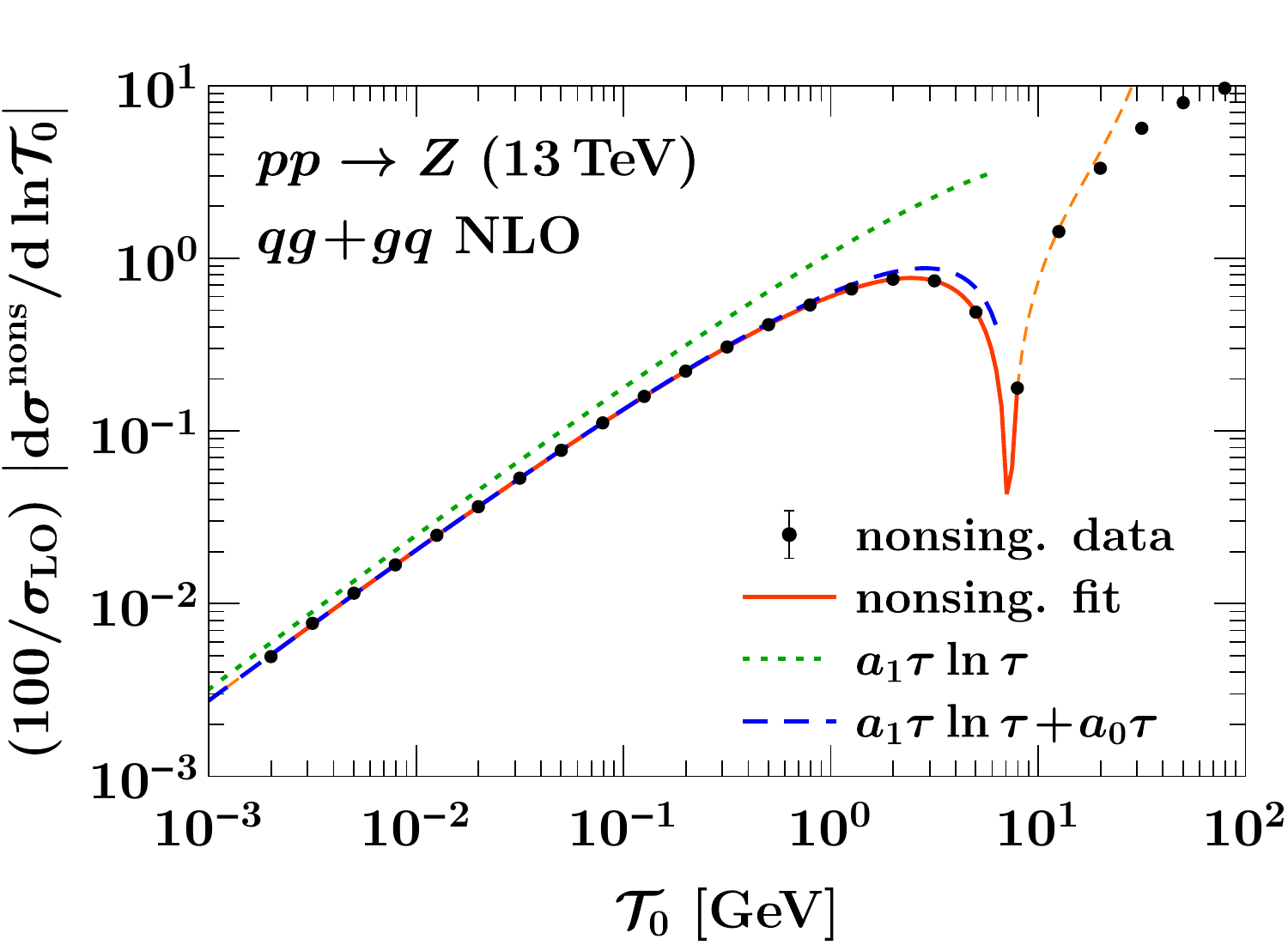}%
\caption{The $\cO(\alpha_s)$ nonsingular corrections for $Z$ production for the $q\bar q$ channel (top row) and the $qg+gq$ channel (bottom row). A fit to the nonsingular data is shown by the solid red curve. The LL and NLL results are shown by green dotted and blue dashed curves, respectively. In all cases, the NLL approximation provides an excellent approximation to the complete nonsingular cross section.}
\label{fig:fitNLO}
\end{figure*}

\begin{figure*}[t]
\includegraphics[width=0.5\textwidth]{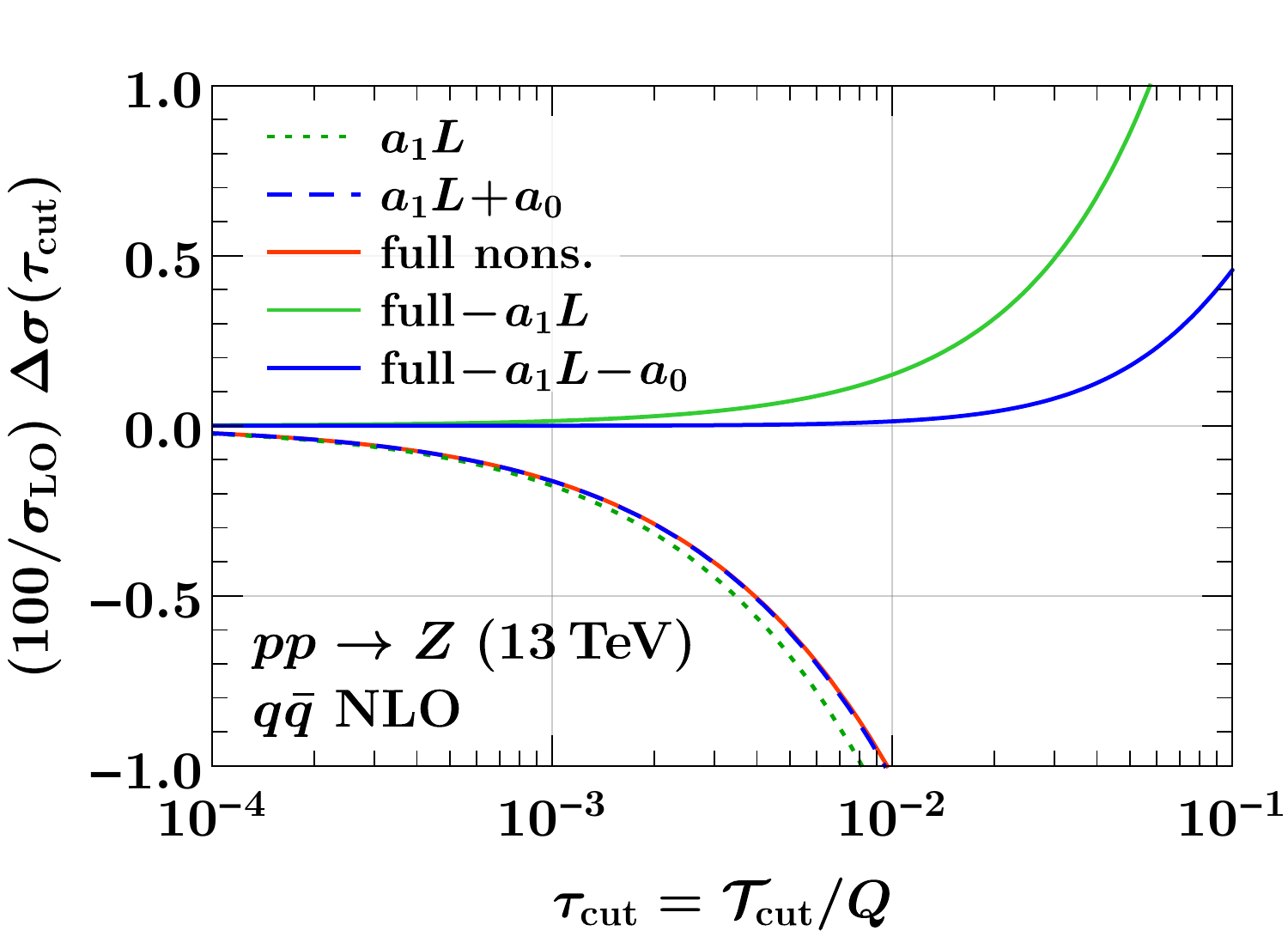}%
\hfill
\includegraphics[width=0.5\textwidth]{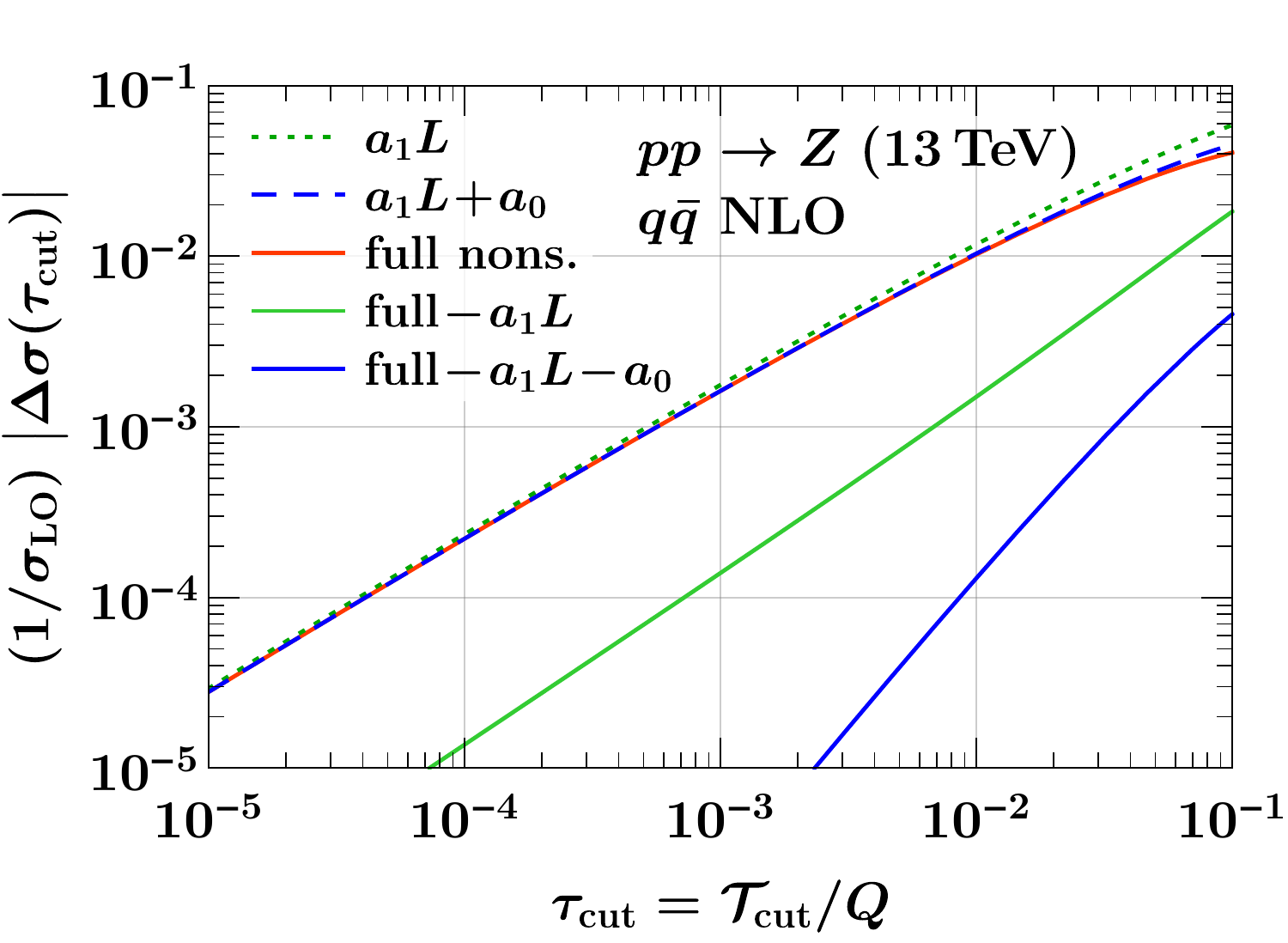}%
\\
\includegraphics[width=0.5\textwidth]{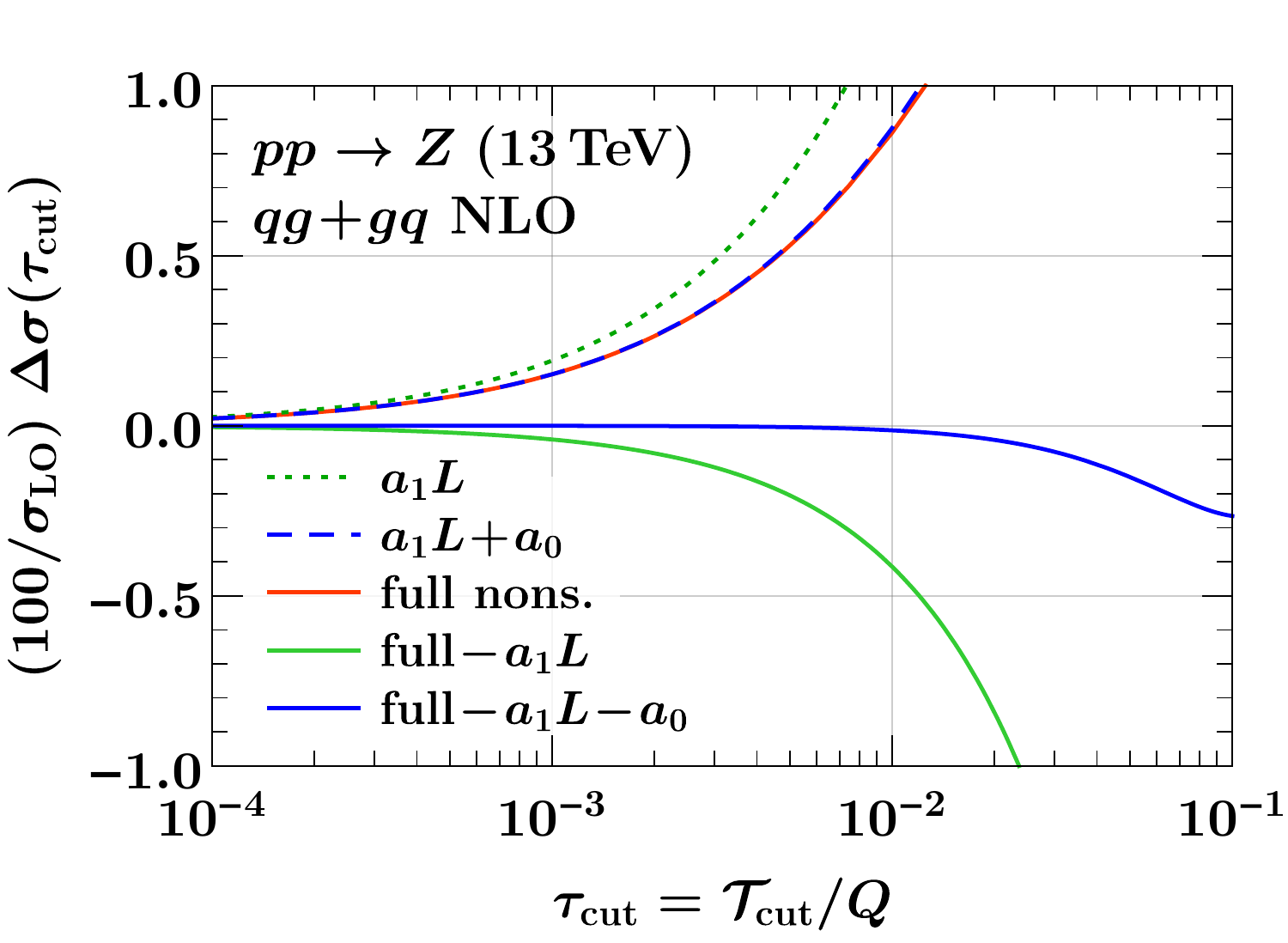}%
\hfill
\includegraphics[width=0.5\textwidth]{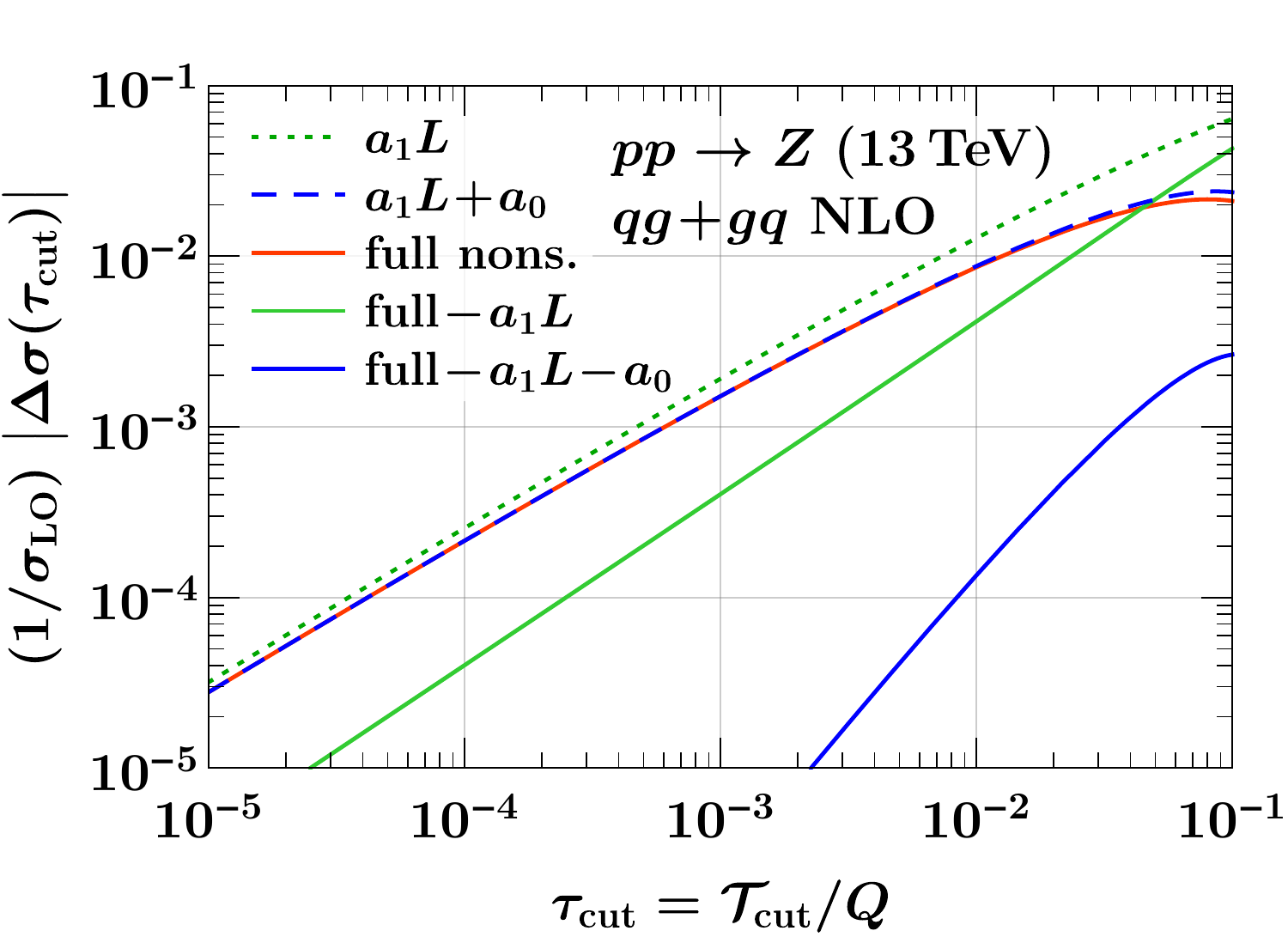}%
\caption{The power corrections for the cumulative $\Delta\sigma(\tau_\cut)$ at $\cO(\alpha_s)$ for $Z$ production in the $q\bar q$ channel (top row) and $qg+gq$ channel (bottom row). In both cases, after the inclusion of the NLL power corrections, $\Delta\sigma(\tau_\cut)$ is reduced by a factor of 100 or more
for $\tau_\cut < 10^{-2}$.}
\label{fig:cumulantNLO}
\end{figure*}

\begin{figure*}[t!]
\includegraphics[width=0.5\textwidth]{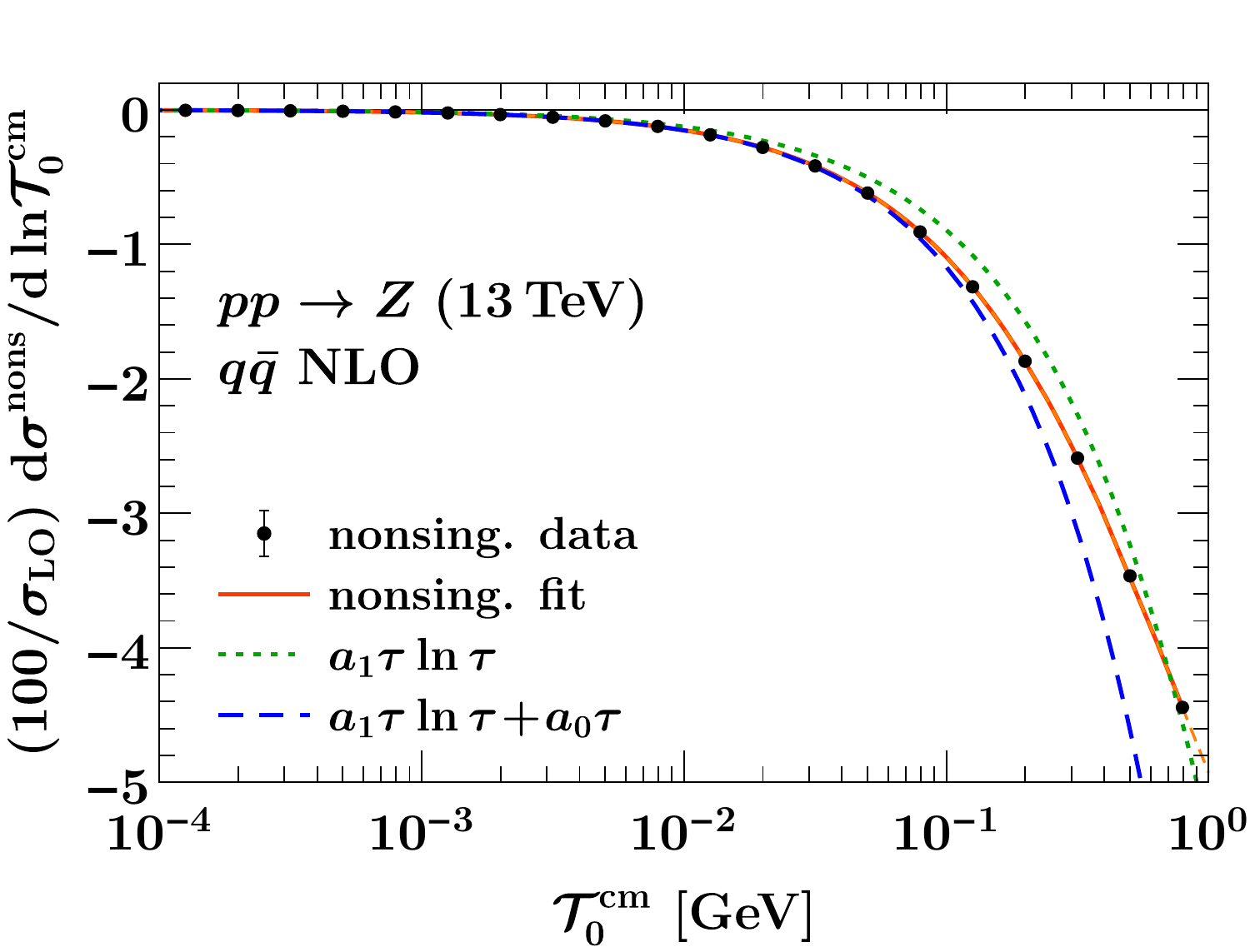}%
\hfill
\includegraphics[width=0.5\textwidth]{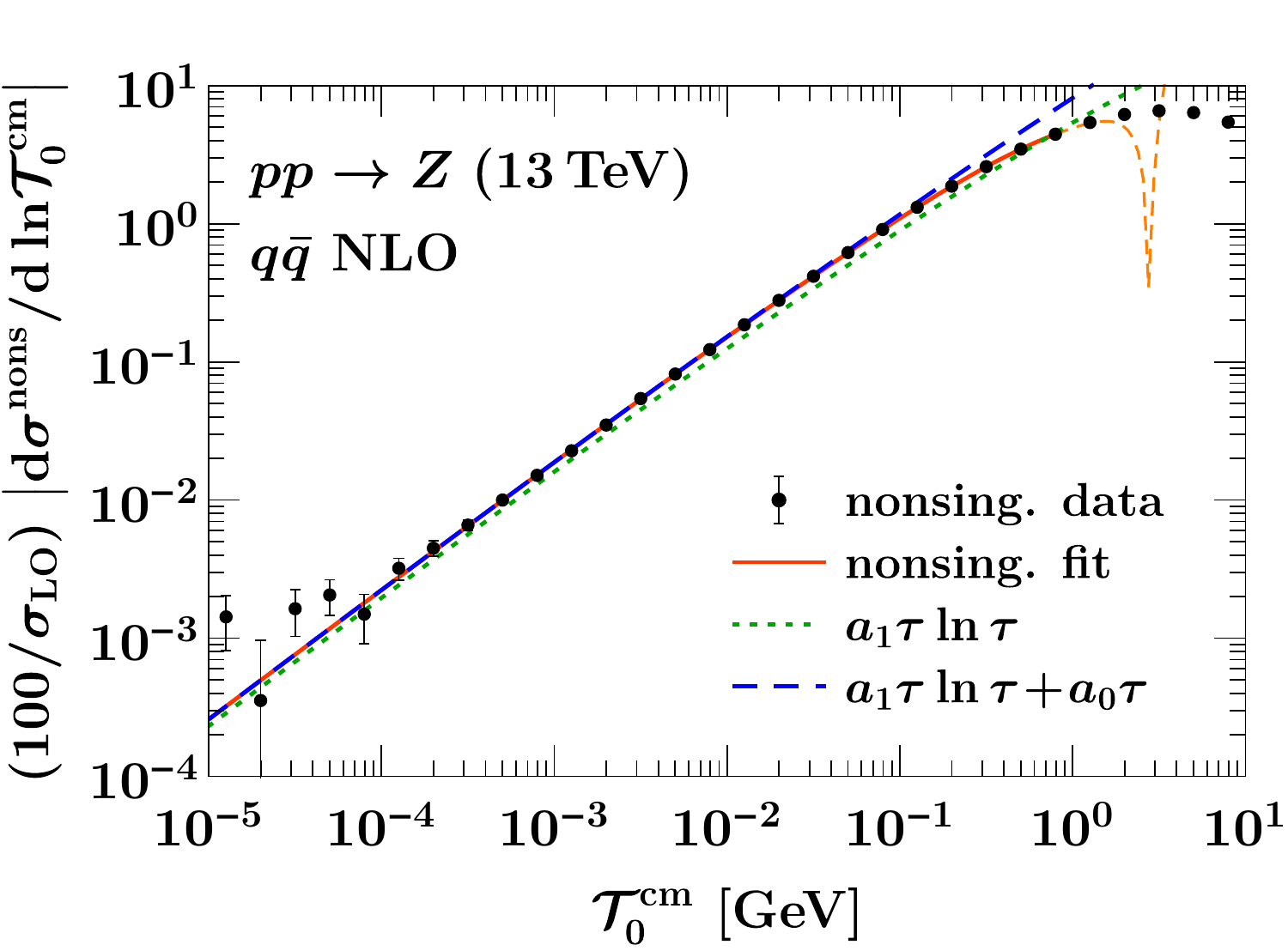}%
\\
\includegraphics[width=0.5\textwidth]{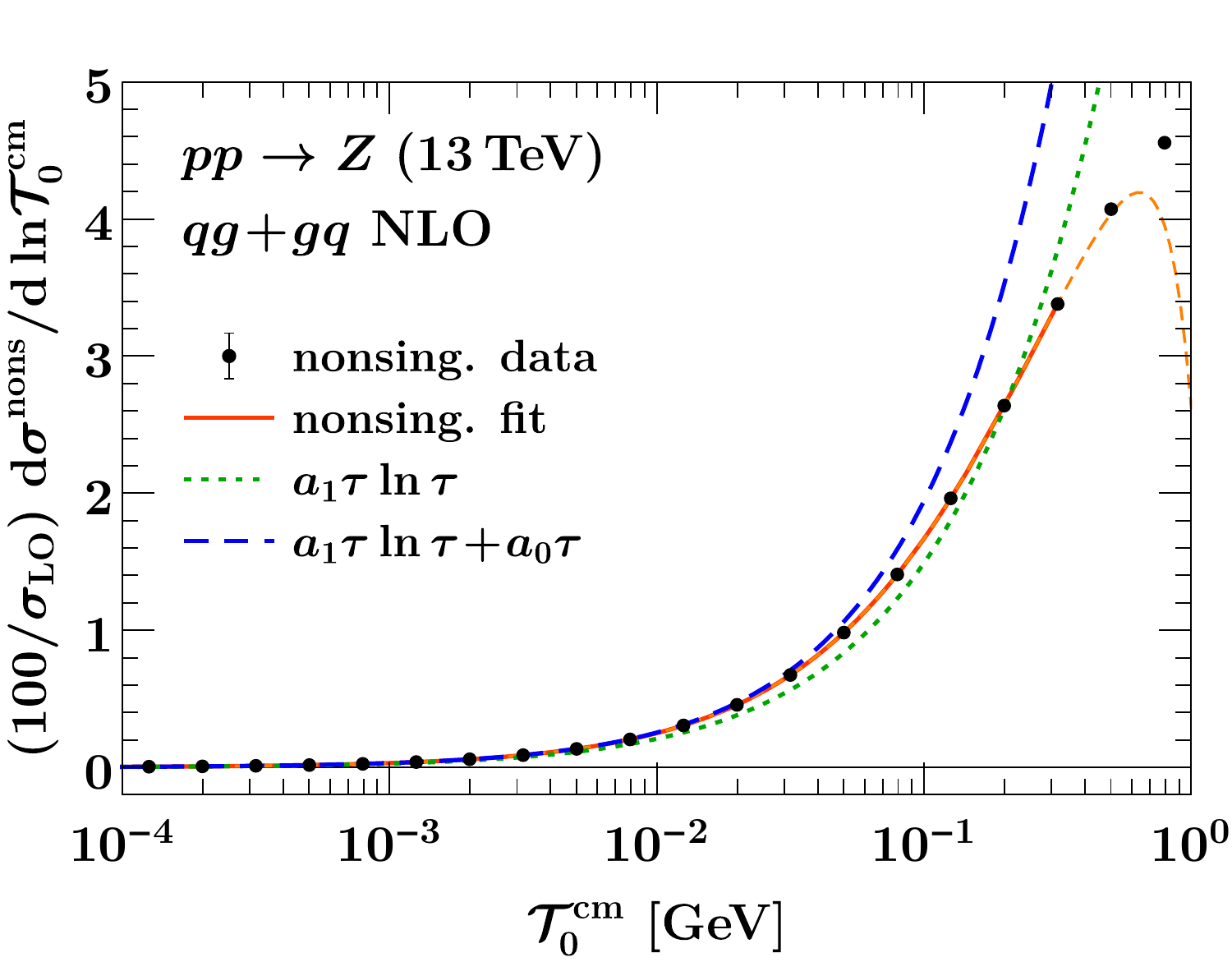}%
\hfill
\includegraphics[width=0.5\textwidth]{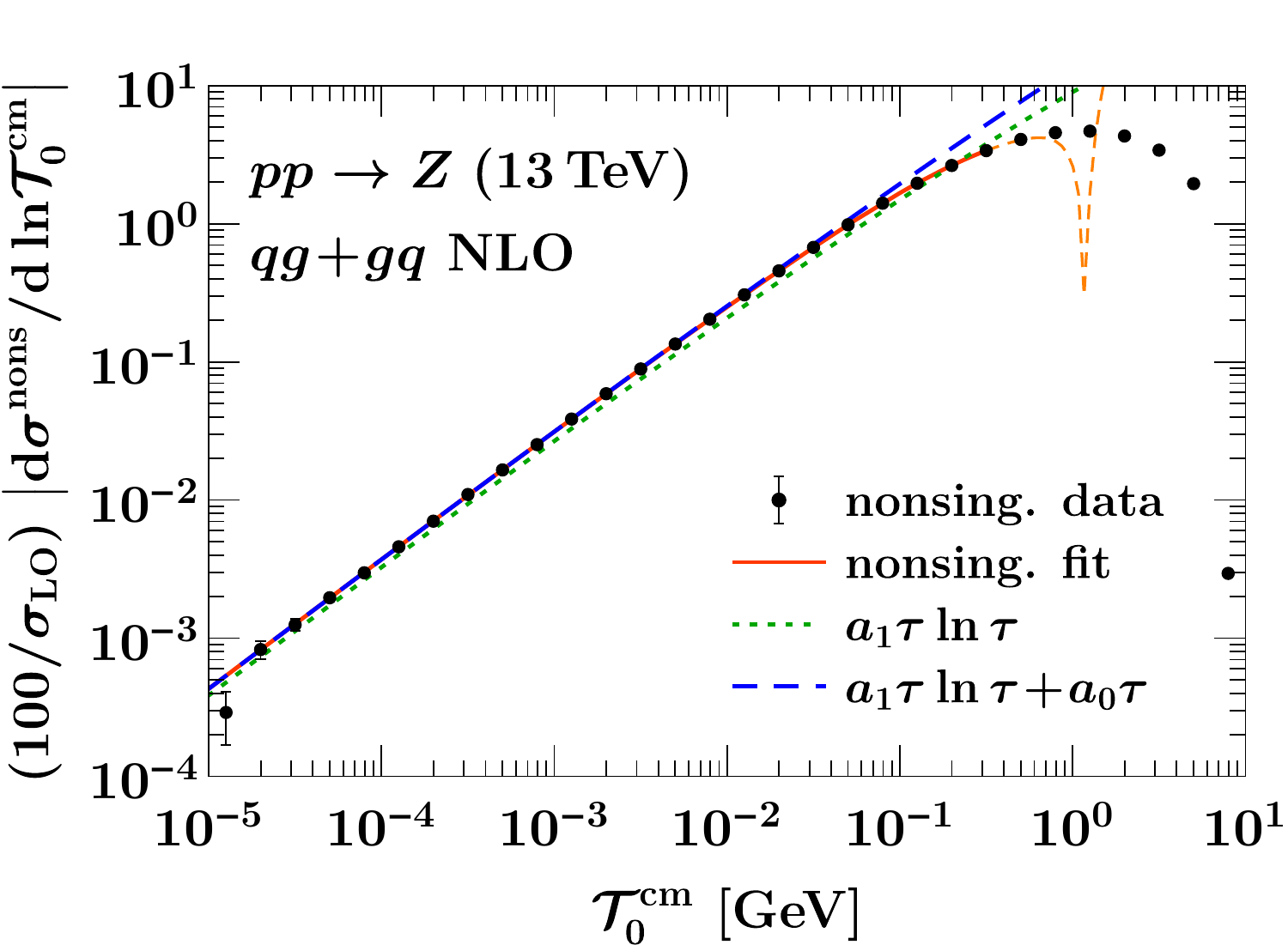}%
\caption{Same as \fig{fitNLO} for the hadronic $\Tau$ definition.}
\label{fig:fitNLOcm}
\end{figure*}

\begin{figure*}[t]
\includegraphics[width=0.5\textwidth]{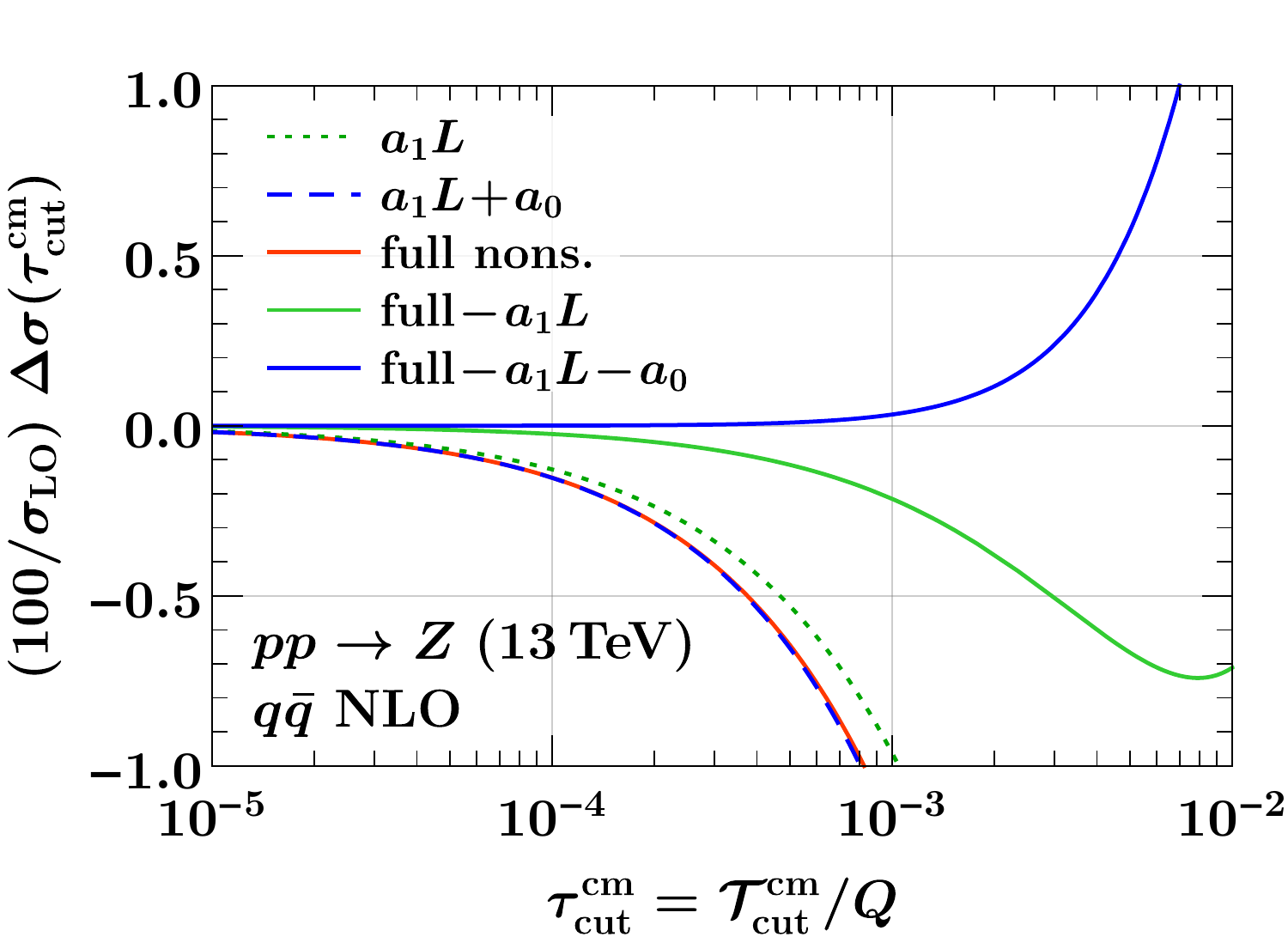}%
\hfill
\includegraphics[width=0.5\textwidth]{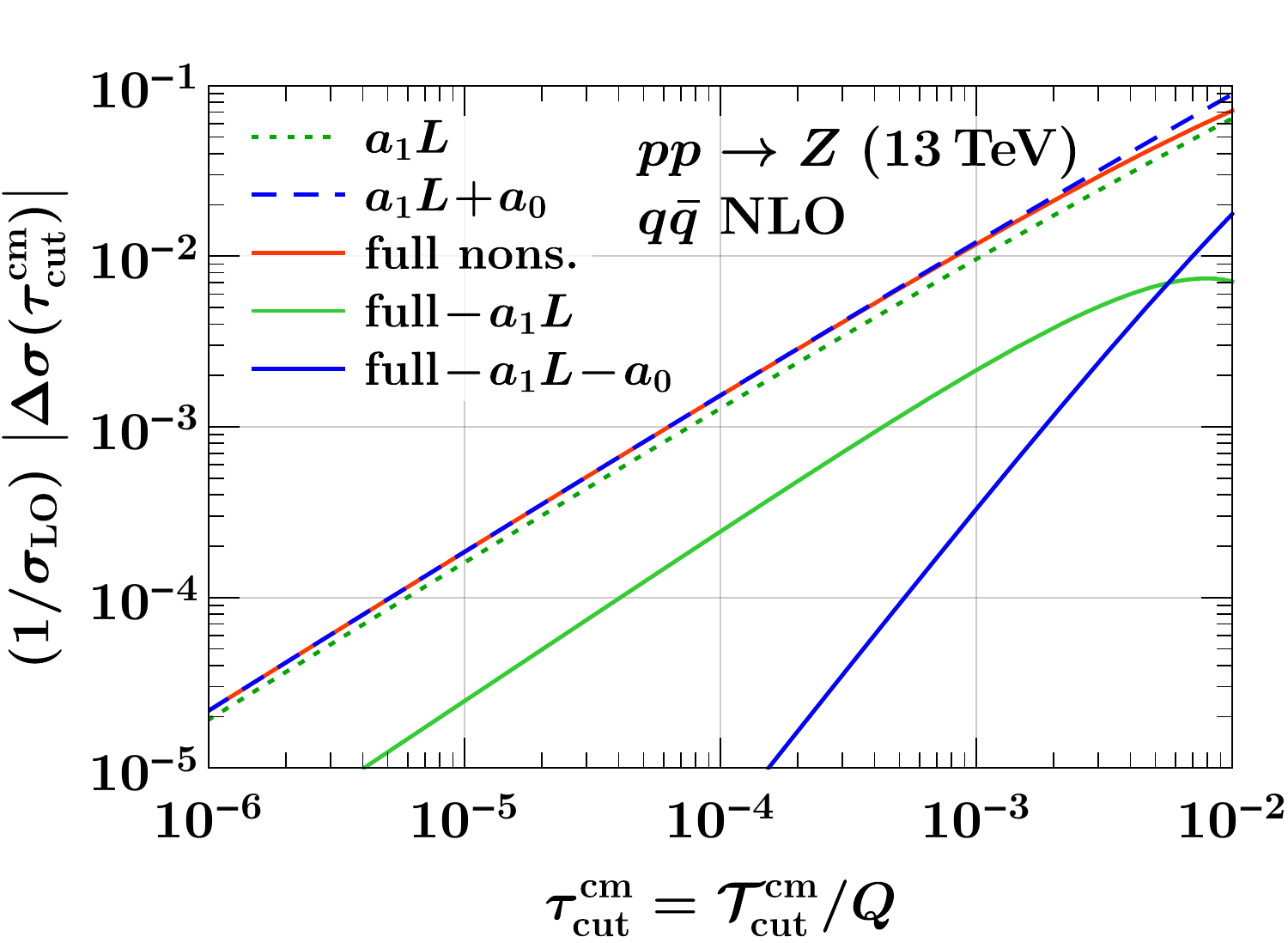}%
\\
\includegraphics[width=0.5\textwidth]{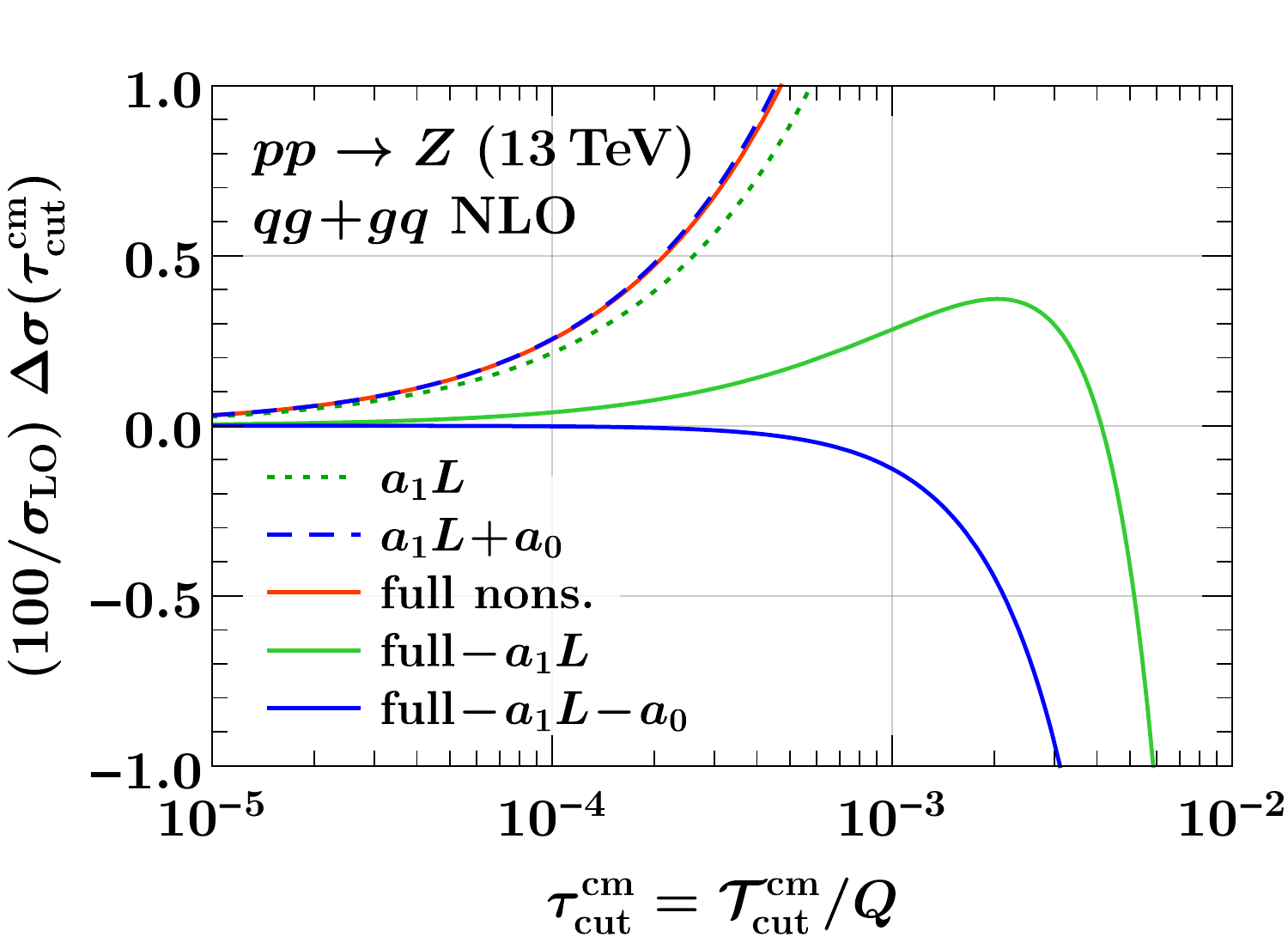}%
\hfill
\includegraphics[width=0.5\textwidth]{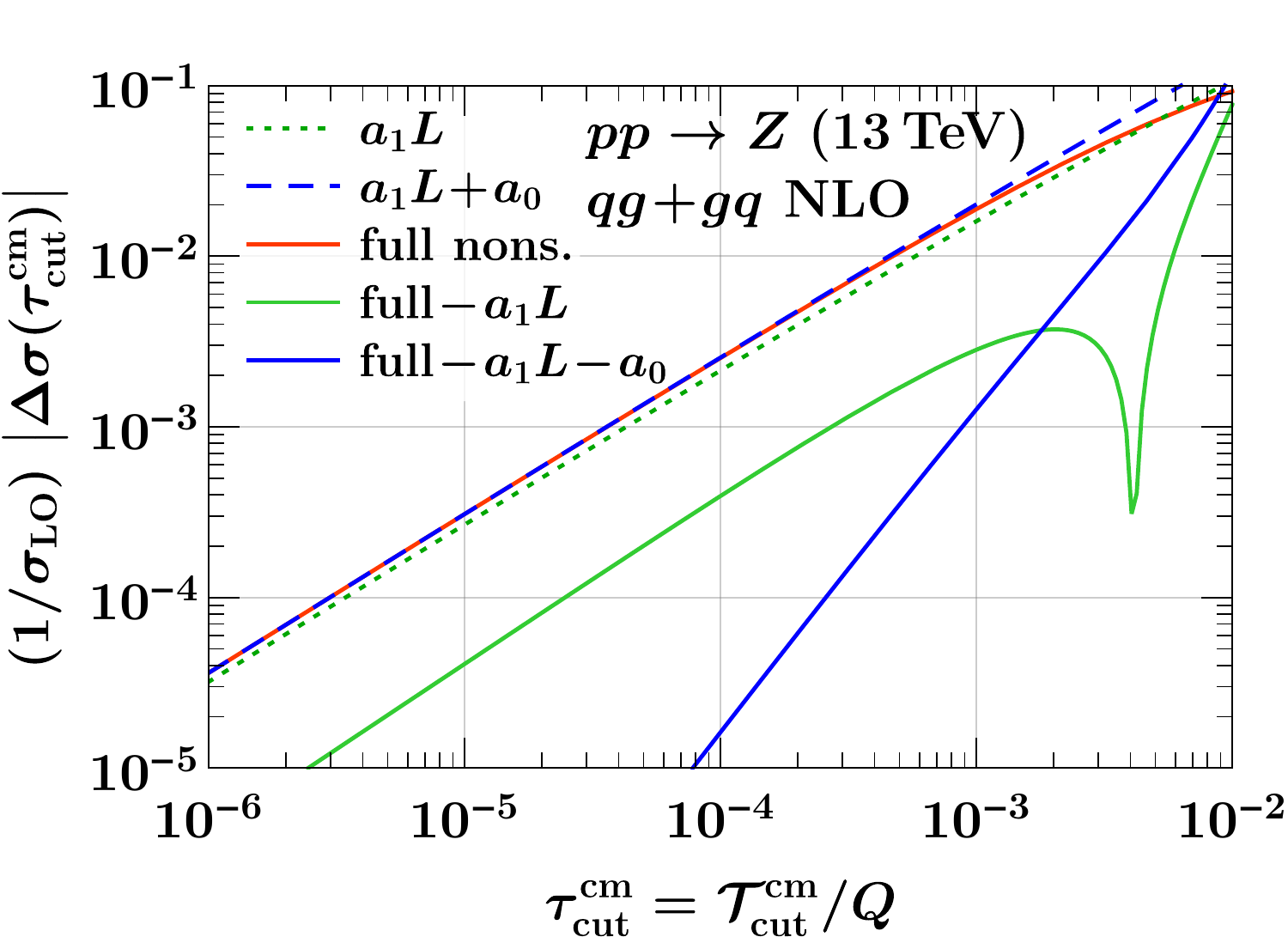}%
\caption{Same as \fig{cumulantNLO} for the hadronic $\Tau$ definition.}
\label{fig:cumulantNLOcm}
\end{figure*}

\begin{figure*}[t!]
\includegraphics[width=0.49\textwidth]{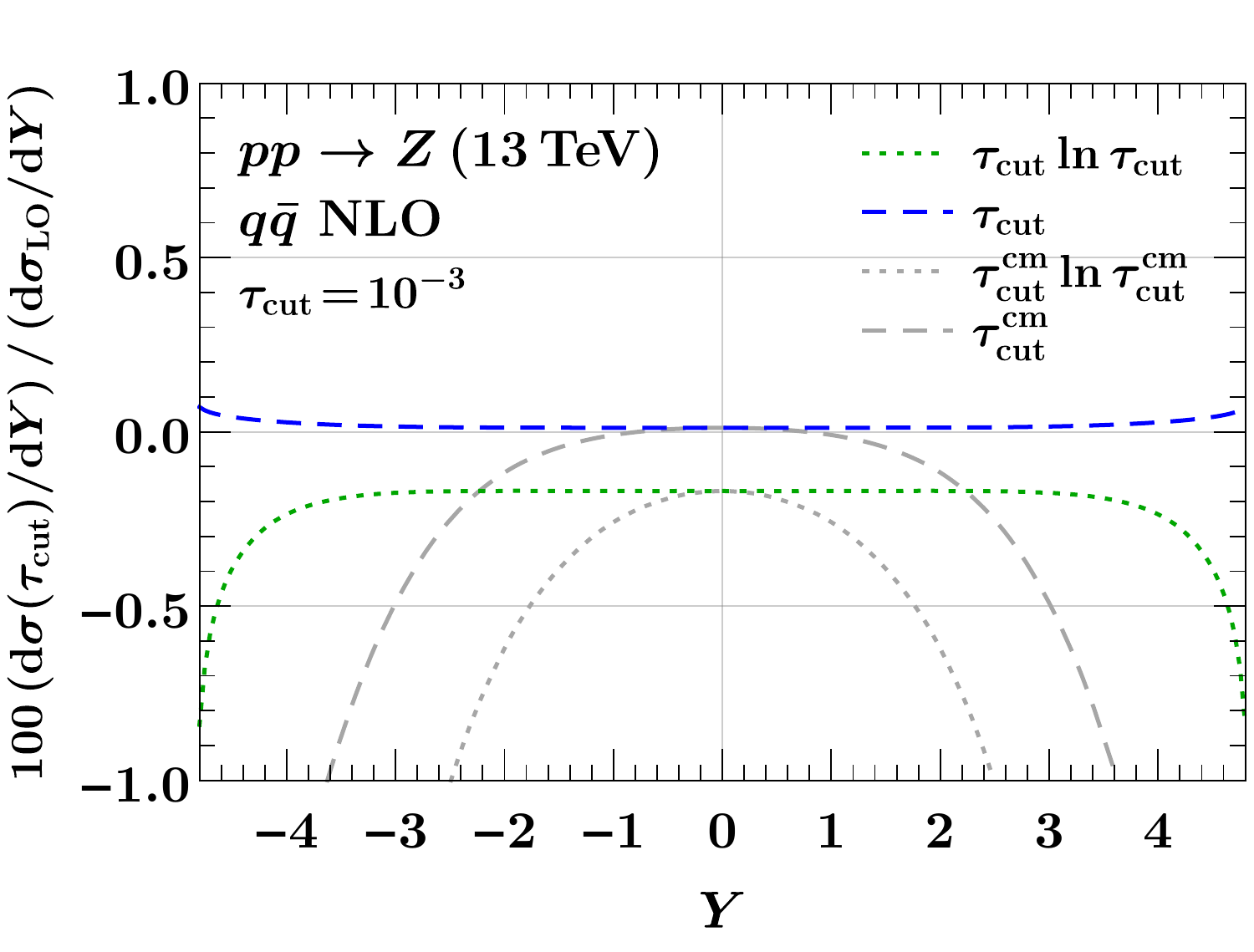}%
\hfill
\includegraphics[width=0.49\textwidth]{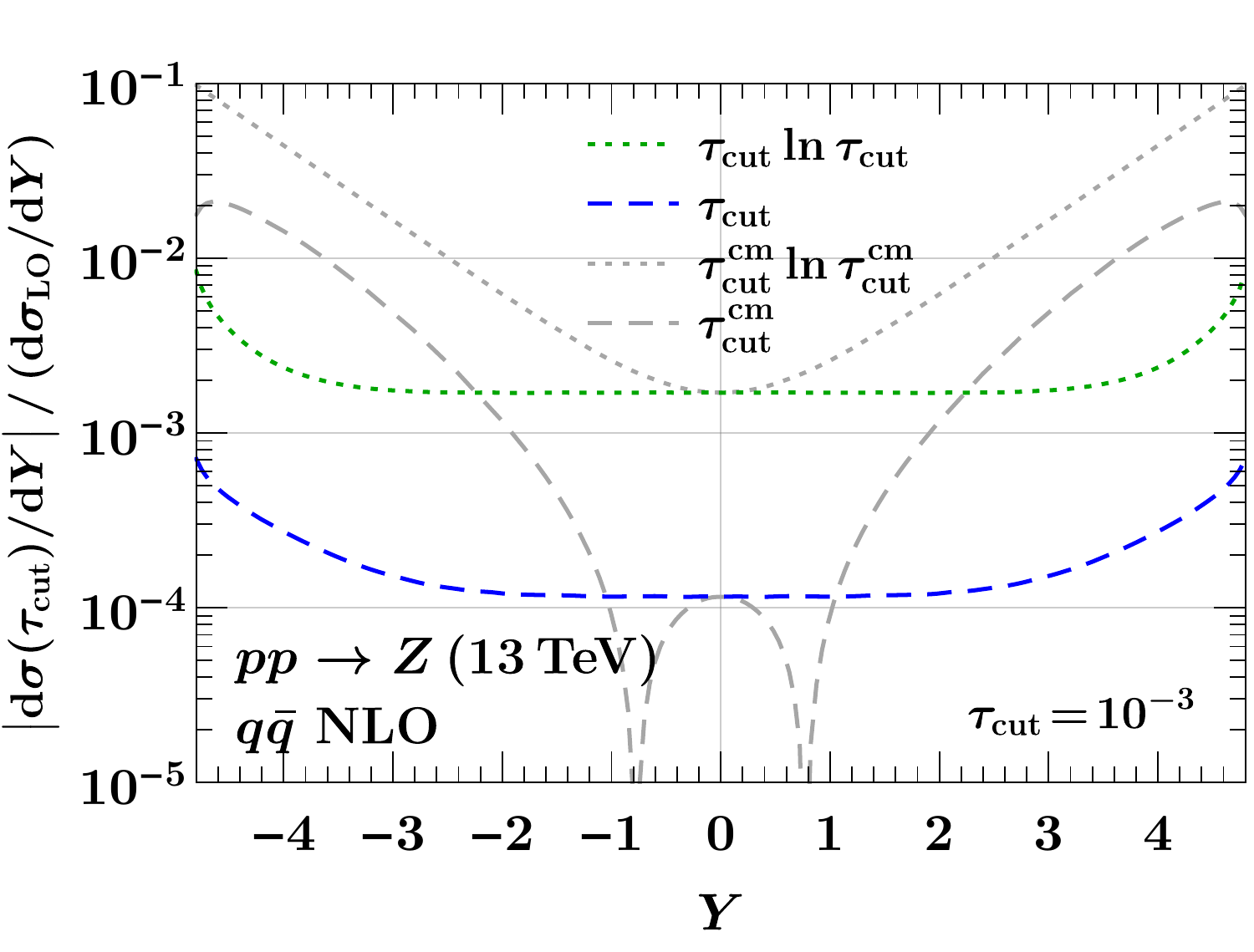}%
\\
\includegraphics[width=0.49\textwidth]{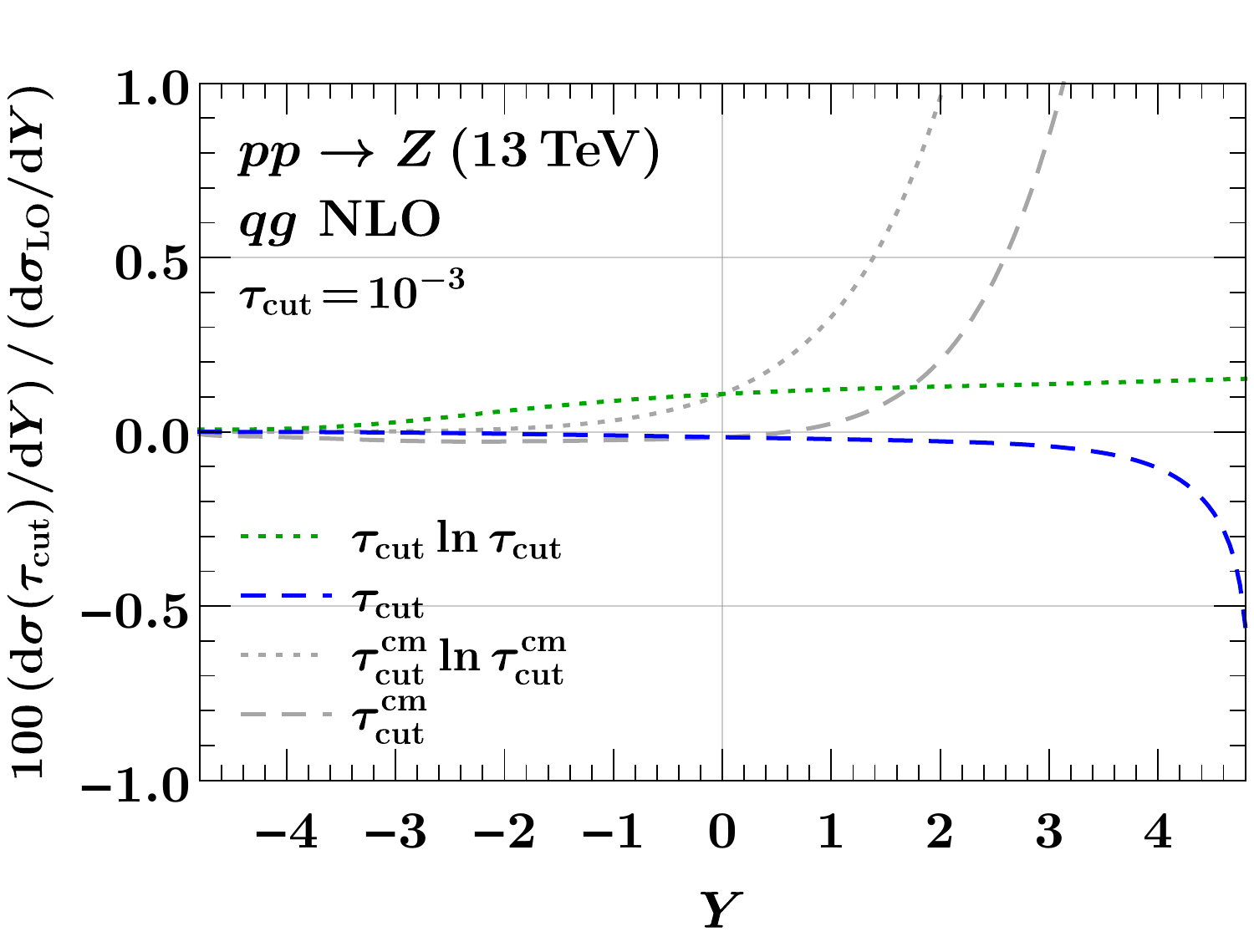}%
\hfill
\includegraphics[width=0.49\textwidth]{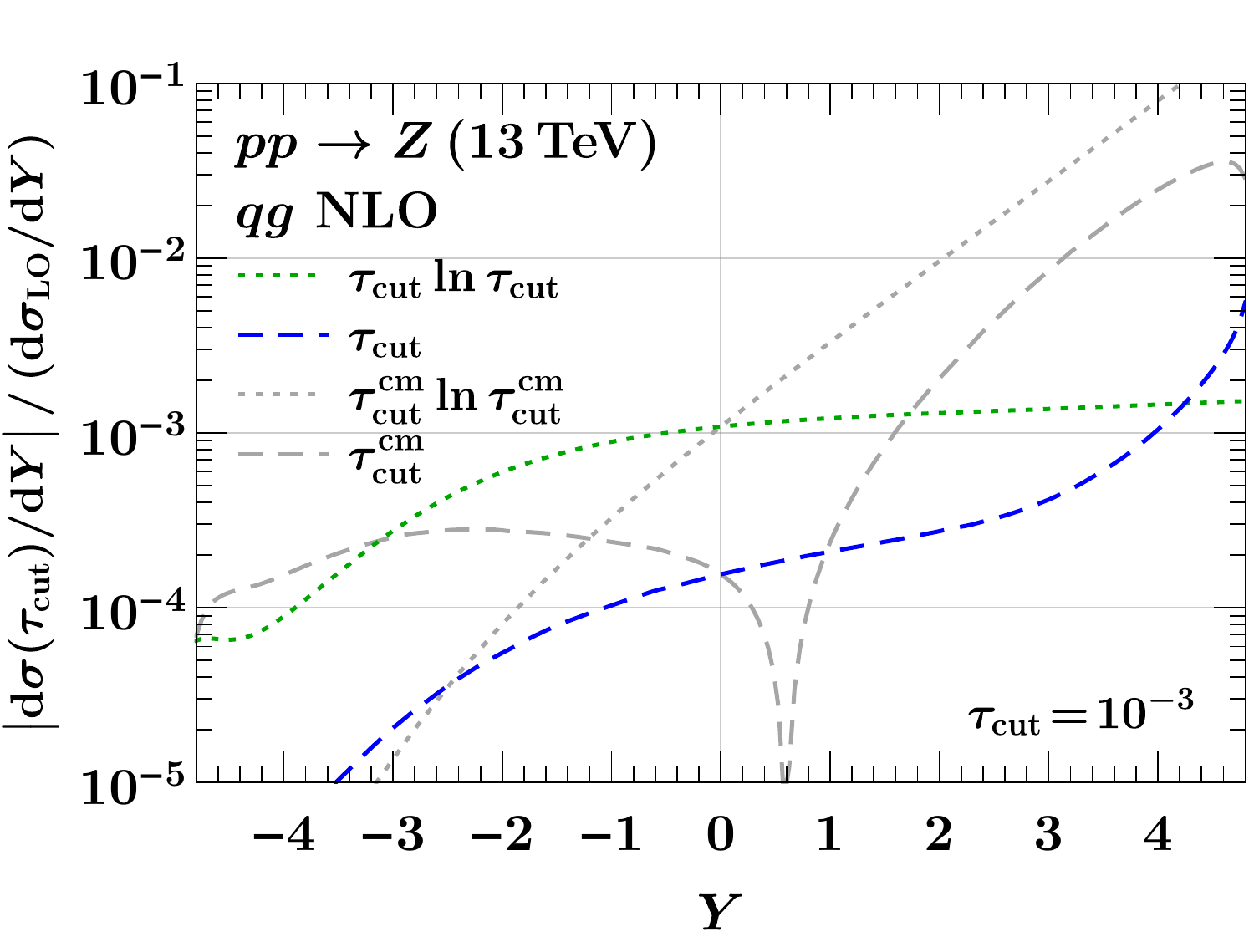}%
\caption{The NLO NLP corrections as a function of rapidity at fixed $\tau_\cut=10^{-3}$ for $Z$ production for the $q\bar q$ channel (top row) and the $qg$ channel (bottom row). The LL and NLL coefficients for leptonic $\Tau$ are shown by the green dotted and blue dashed curves and for hadronic $\Tau$ by the dotted and dashed gray curves.}
\label{fig:dsigdY_DY}
\end{figure*}

In \fig{fitNLO} we show the complete NLO nonsingular contributions as black dots, as well as a fit to their form with the solid red curve. Given the agreement in table~\ref{tab:NLLresults_DY} between our analytic $a_0$ and the earlier fit result for $a_0$, we have fixed $a_0$ to the analytic result, and redone the fit using \eq{fitfun} to obtain this red curve. The red curve from this fit is fully consistent with the earlier fit result from \refcite{Moult:2016fqy}. The dashed orange curve in \fig{fitNLO} is the extension of the fit function beyond its fit range. In dotted green and dashed blue we show our analytic predictions. We see that with the inclusion of the NLL power
corrections, we obtain an excellent description of the full nonsingular cross
section up to nearly $\Tau_0 \sim 1$ GeV. This is quite remarkable, and shows
that additional higher-order power correction terms are truly suppressed.

In \fig{cumulantNLO} we show a plot of the corresponding residual power corrections for
the cumulant, $\Delta\sigma(\tau_\cut)$, on both a linear scale (left)
and logarithm scale (right). The solid red curve shows the full power corrections,
the solid green curves show the remaining power
corrections after including $a_1$ in the subtractions,
and the solid blue curve those after including $a_1$ and $a_0$ in the subtractions.
We see that with the inclusion of the full NLL power corrections, we achieve
more than a factor of $100$ reduction in the residual power corrections as
compared with the leading-power result at NLO. Both partonic
channels have similarly sized power corrections and show a fast convergence of
the power expansion. The fact that the blue curve in the logarithmic plot exhibits
a steeper slope than the red and green curves is due to its $\ord{\tau_\cut^2}$ scaling
corresponding to a next-to-next-to-leading power correction. This provides a nice
visualization that our results correctly capture the complete NLP contribution.

The analogous results for the fitted nonsingular spectrum and the residual
power corrections $\Delta\sigma(\tau^\hadcm_\cut)$ for the hadronic $\Tau$ definition
are shown in \figs{fitNLOcm}{cumulantNLOcm}.
As expected, the power corrections
are substantially larger for $\Tau^\hadcm$ than for the leptonic definition. To obtain
similarly sized power corrections, one has to go to about an order of magnitude
smaller values of $\Tau^\hadcm$. Apart from the overall enhancement, the qualitative
behavior of the LL and NLL contributions and the different partonic channels is
the same. This is expected from our analytic results, which show that the coefficients
for both definitions have essentially the same structure and primarily differ in the overall
factors of $e^{\pm Y}$ leading to the rapidity enhancement for the hadronic definition
already observed in \refscite{Moult:2016fqy, Moult:2017jsg}.

In \fig{dsigdY_DY} we show the rapidity dependence of the NLP corrections at fixed
$\tau_\cut = 10^{-3}$ for both leptonic and hadronic $\Tau$ normalized to the LO
rapidity spectrum. We can clearly see the exponential enhancement for the hadronic definition at large $\abs{Y}$. For the $qg$ channel,
the asymmetric behavior in rapidity is expected from its analytic result.
The result for the $gq$ channel corresponds to taking $Y\to -Y$, such that their sum
is symmetric in rapidity. While the leptonic definition does not suffer from the
exponential enhancement of the hadronic definition, it still exhibits a substantial increase at
large positive $Y$ in the $qg$ channel, as well as a suppression at
large negative $Y$.  This is due to the substantially different $x$-dependence
of the quark-gluon luminosity (and its derivative) compared to the
$q\bar q$ luminosity in the LO result to which we normalize. Knowing the
NLL contribution to the power corrections differential in rapidity enables one
to explicitly account for this effect in the subtractions.

\FloatBarrier

\subsection{Gluon-Fusion Higgs Production}

Next, we consider gluon-fusion Higgs production. We take $pp\to H$
at $E_{\rm cm} = 13\TeV$ with an on-shell, stable Higgs boson with $m_H = 125\GeV$,
integrated over all $Y$. We use the MMHT2014 NNLO PDFs~\cite{Harland-Lang:2014zoa}, with
fixed scales $\mu_r = \mu_f = m_H$, and $\alpha_s(m_H) = 0.1126428$. The NLP power corrections for this configuration for the leptonic $\Tau$ definition were extracted numerically in \refcite{Moult:2017jsg}.
The results for both leptonic and hadronic definitions for all partonic channels
are collected and compared to our analytic predictions in table~\ref{tab:NLLresults_H}.
In all cases, excellent agreement is observed within the fit uncertainties.

\begin{table}[h]
\centering
\begin{tabular}{c|ll}
\hline\hline
NLO $\Tau_0^{\rm lep}$  $gg \to Hg$ & $a_1$ & $a_0$
\\ \hline
fitted~\cite{Moult:2017jsg} & $+0.60936 \pm 0.00600$  & $+0.18241 \pm 0.00425$
\\
analytic & $+0.60400$ & $+0.18627$
\\ \hline
NLO $\Tau_0^{\rm lep}$  $gq+qg \to Hq$ &  $a_1$ & $a_0$
\\ \hline
fitted~\cite{Moult:2017jsg} & $-0.03733 \pm 0.00066$  & $-0.42552 \pm 0.00032$
\\
analytic & $-0.03807$ & $-0.42576$
\\ \hline
NLO $\Tau_0^{\rm lep}$  $q\bar q \to Hg$ &  $a_1$ & $10^3\, a_0$
\\ \hline
fitted~\cite{Moult:2017jsg} & -- & $+4.90060 \pm 0.00013$
\\
analytic & -- & $+4.90047$
\\
\hline\hline
NLO $\Tau_0^\hadcm$ $gg \to Hg$ & $a_1$ & $a_0$
\\ \hline
fitted & $+1.5436 \pm 0.0585$  & $-0.45954 \pm 0.02606$
\\
analytic & $+1.5225$ & $-0.46646$
\\ \hline\hline
NLO $\Tau_0^\hadcm$  $gq+qg \to Hq$ &  $a_1$ & $a_0$
\\ \hline
fitted & $-0.06606 \pm  0.00161$ &  $-0.33932 \pm 0.00194$
\\
analytic & $-0.06498$ & $-0.34068$
\\ \hline\hline
NLO $\Tau_0^\hadcm$  $q\bar q \to Hg$ &  $a_1$ & $10^3\, a_0$
\\ \hline
fitted & -- &  $+6.13445 \pm 0.00015$
\\
analytic & -- & $+6.13448$
\\ \hline\hline
\end{tabular}
\caption{Comparison between our analytic predictions and the fitted results for the LL $a_1$ and NLL $a_0$ coefficients in Higgs production. These fitted values for $a_1$ and $a_0$ with the leptonic definition and the analytic results for $a_1$ were already given in \refcite{Moult:2017jsg}.}
\label{tab:NLLresults_H}
\end{table}

\begin{figure*}[t!]
\includegraphics[width=0.5\textwidth]{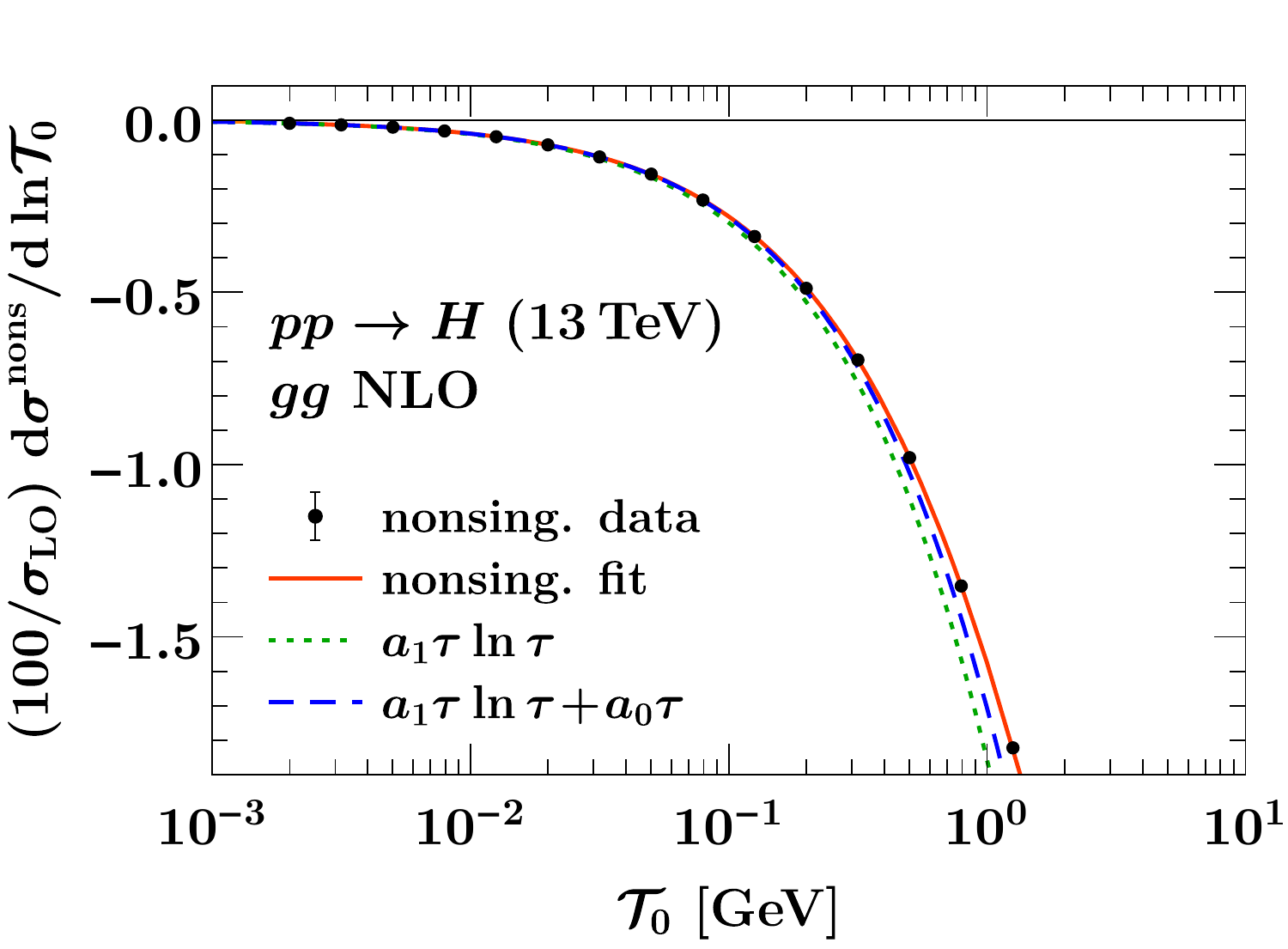}%
\hfill
\includegraphics[width=0.5\textwidth]{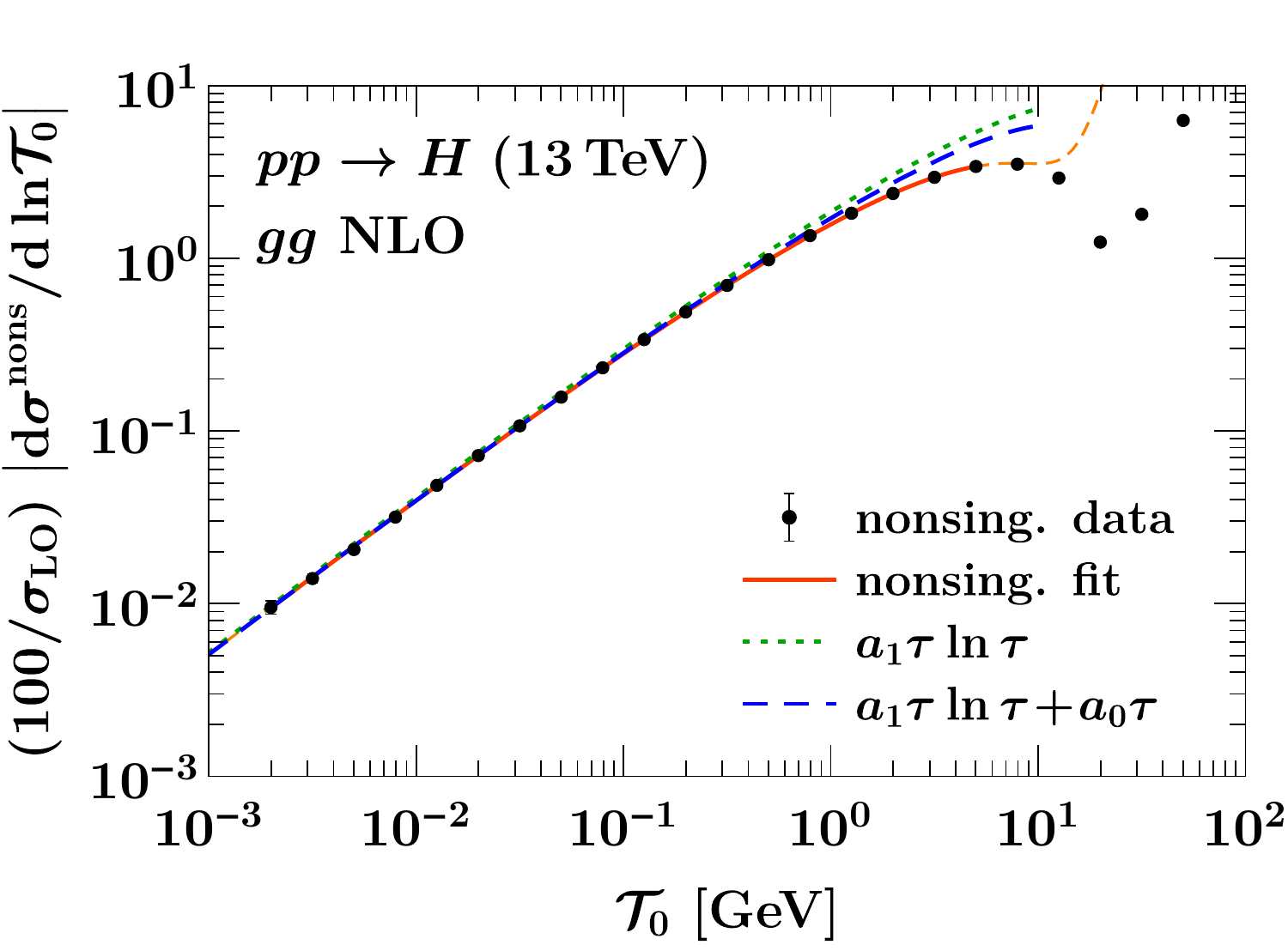}%
\\
\includegraphics[width=0.5\textwidth]{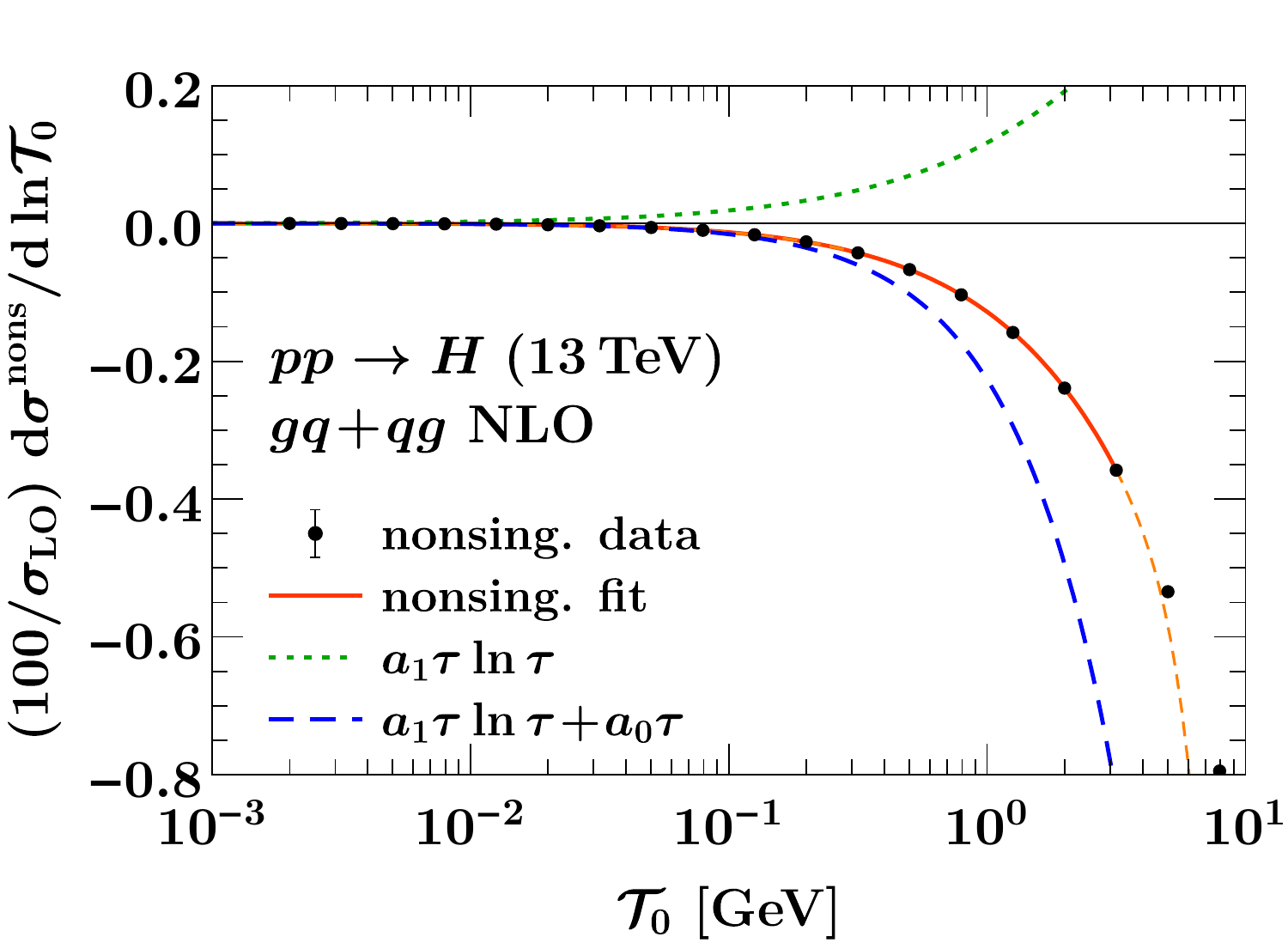}%
\hfill
\includegraphics[width=0.5\textwidth]{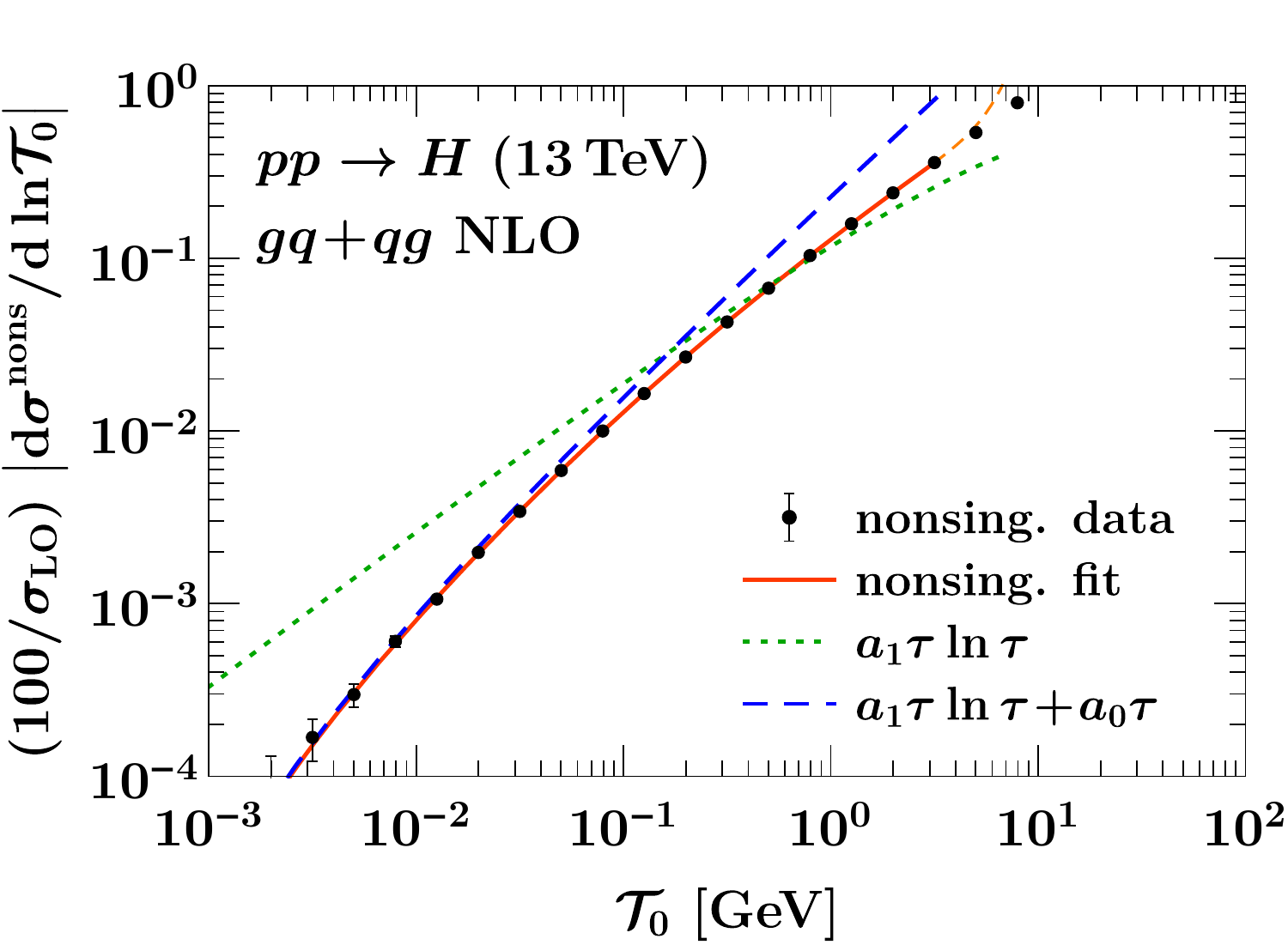}%
\\
\includegraphics[width=0.5\textwidth]{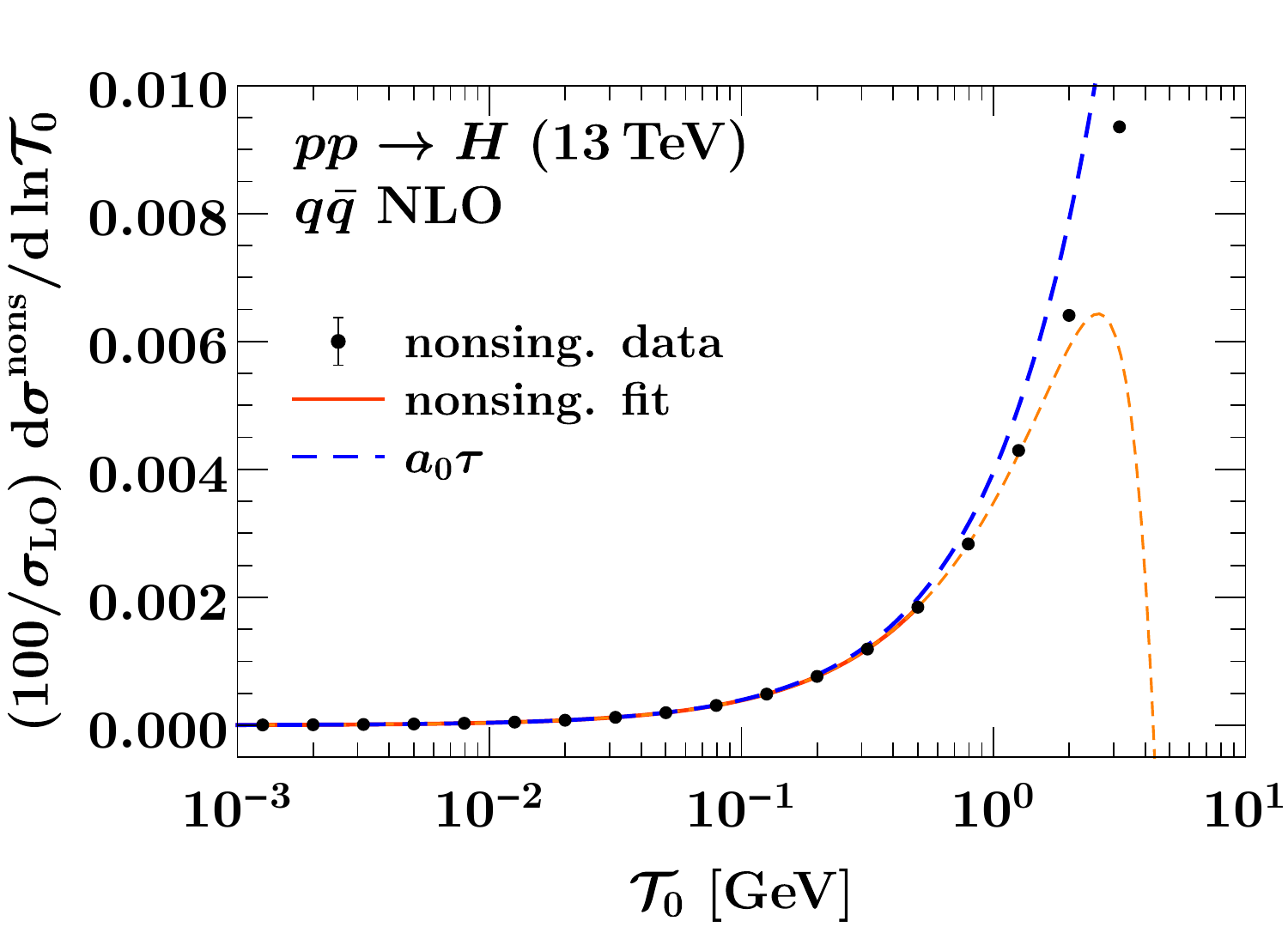}%
\hfill
\includegraphics[width=0.5\textwidth]{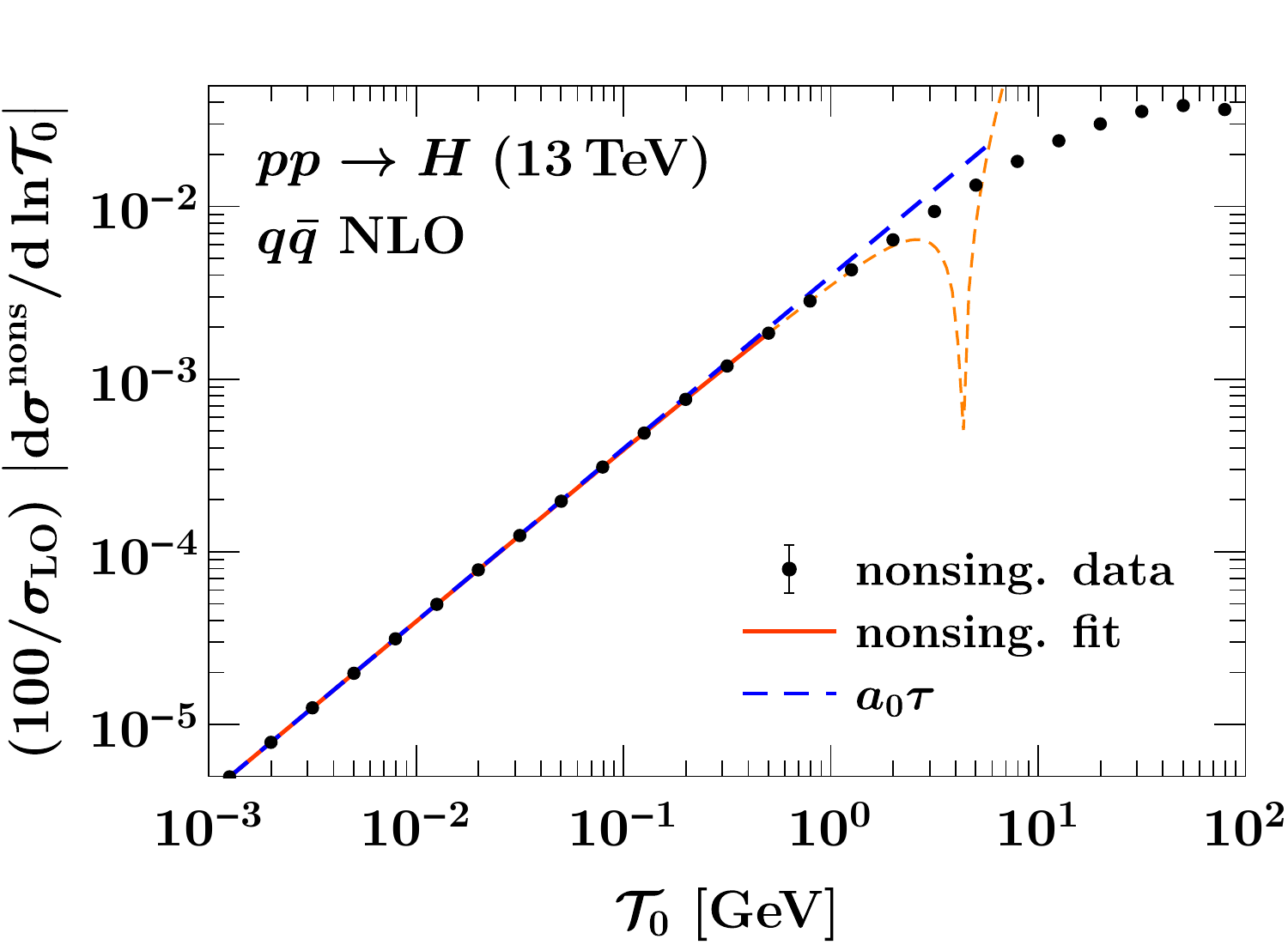}%
\caption{The $\cO(\alpha_s)$ nonsingular corrections for Higgs production for the $gg$ channel (top row), $gq+qg$ channel (middle row), and $q\bar q$ channel (bottom row). A fit to the nonsingular data (black dots) is shown by the solid red curve. The LL and NLL results are shown by green dotted and blue dashed curves, respectively. In all cases, the NLL approximation provides an excellent approximation to the complete nonsingular cross section for sufficiently small $\Tau_0$.}
\label{fig:fitNLO_gg}
\end{figure*}

\begin{figure*}[t]
\includegraphics[width=0.5\textwidth]{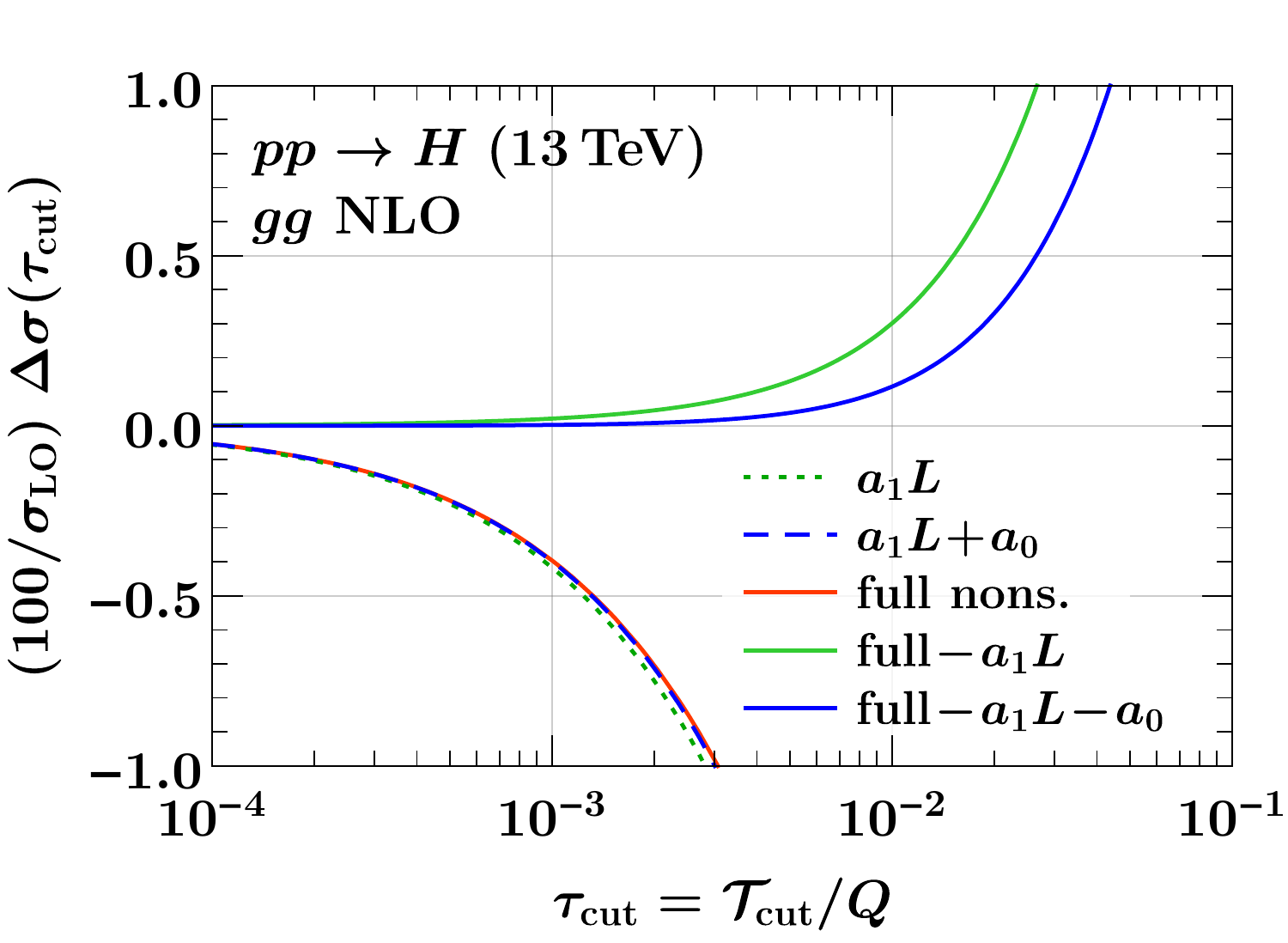}%
\hfill
\includegraphics[width=0.5\textwidth]{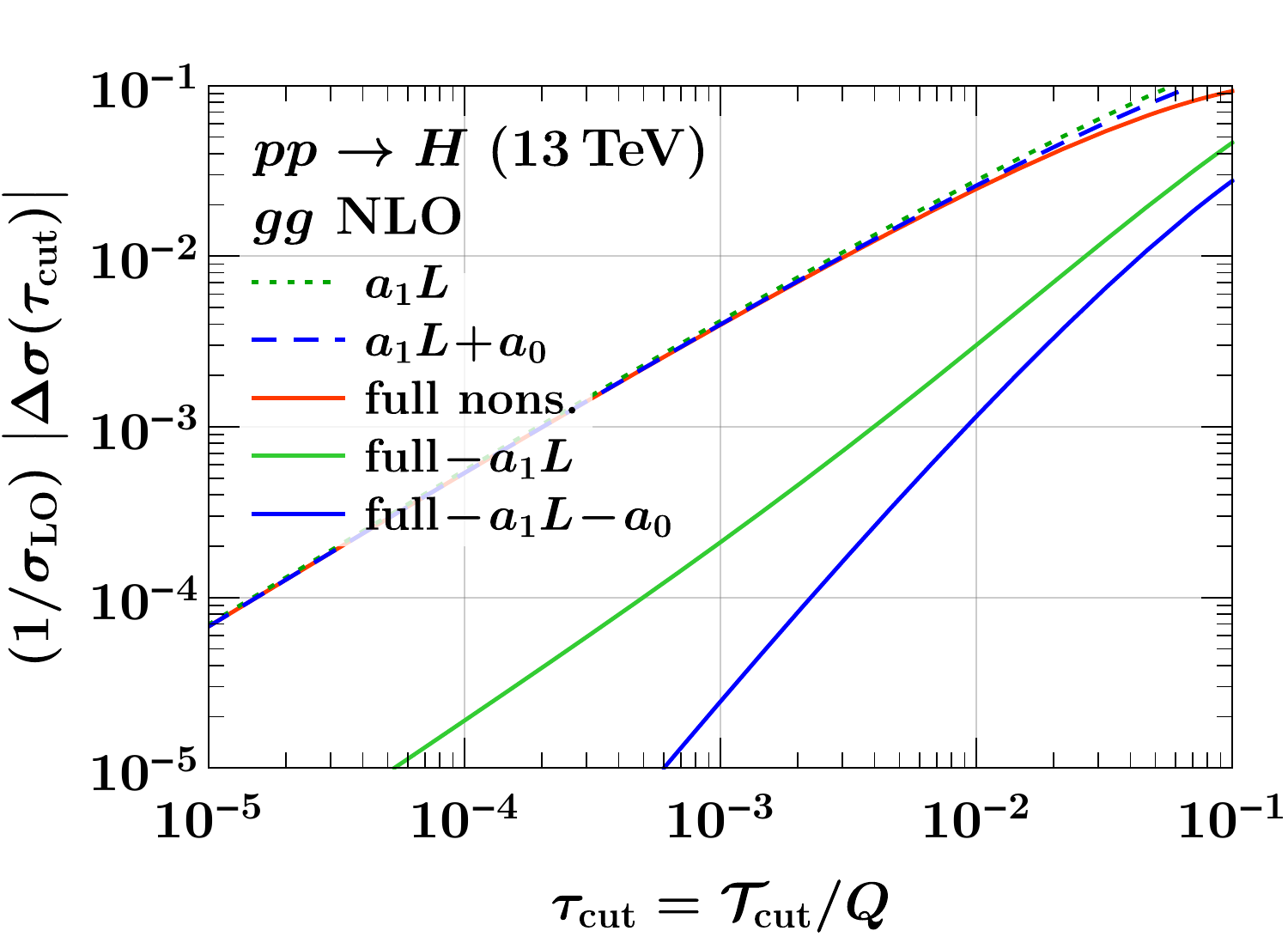}%
\\
\includegraphics[width=0.5\textwidth]{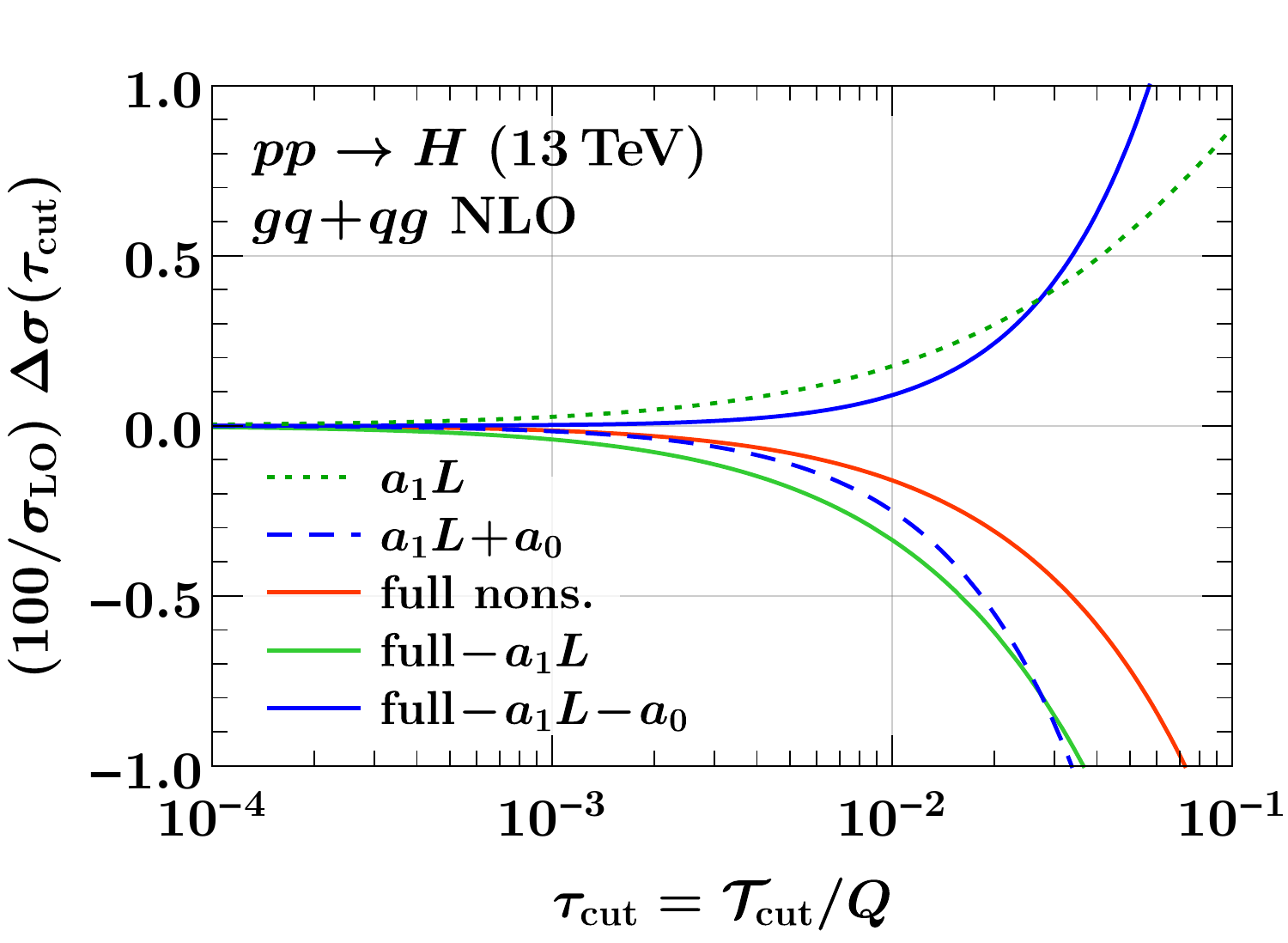}%
\hfill
\includegraphics[width=0.5\textwidth]{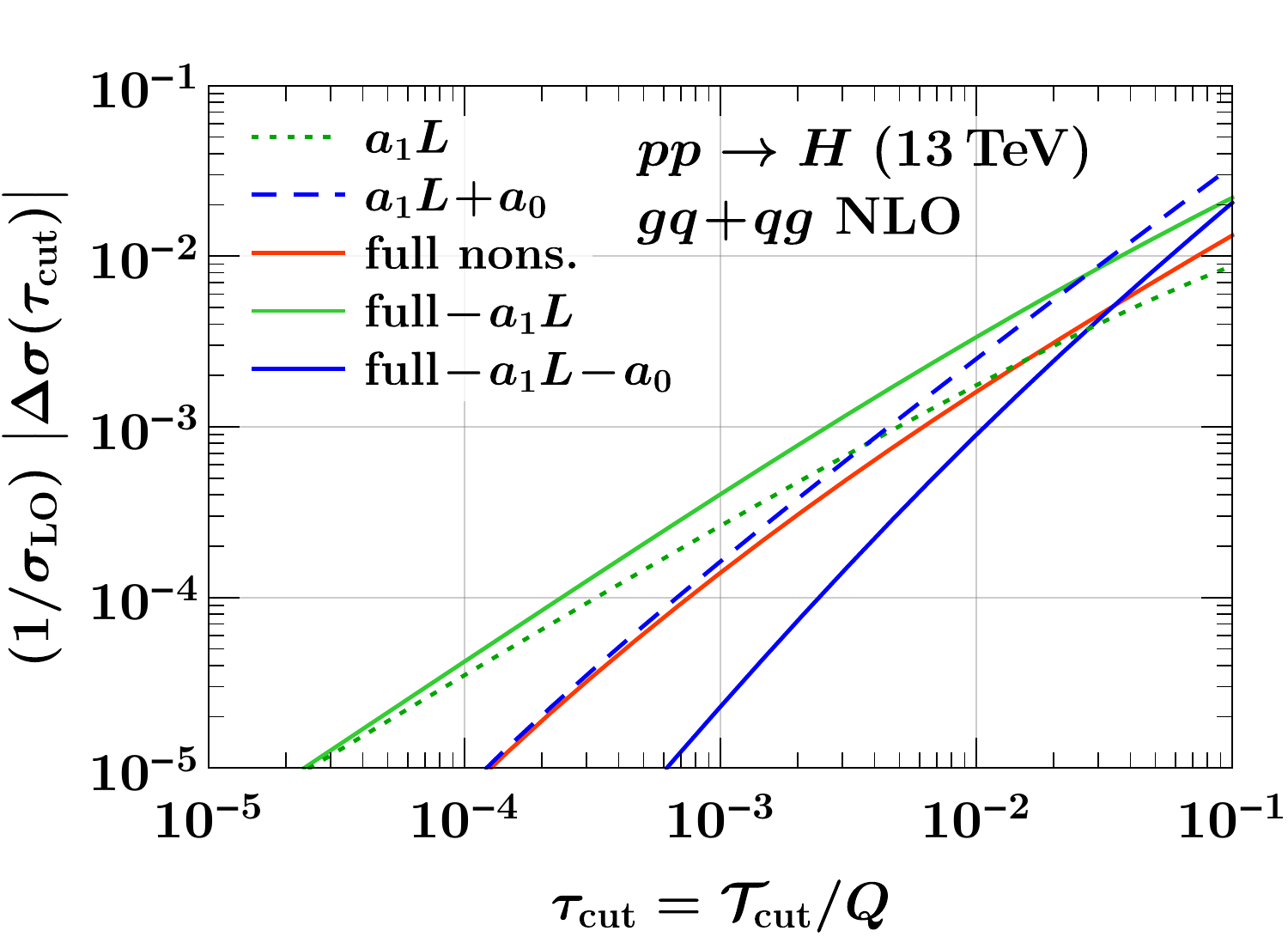}%
\\
\includegraphics[width=0.5\textwidth]{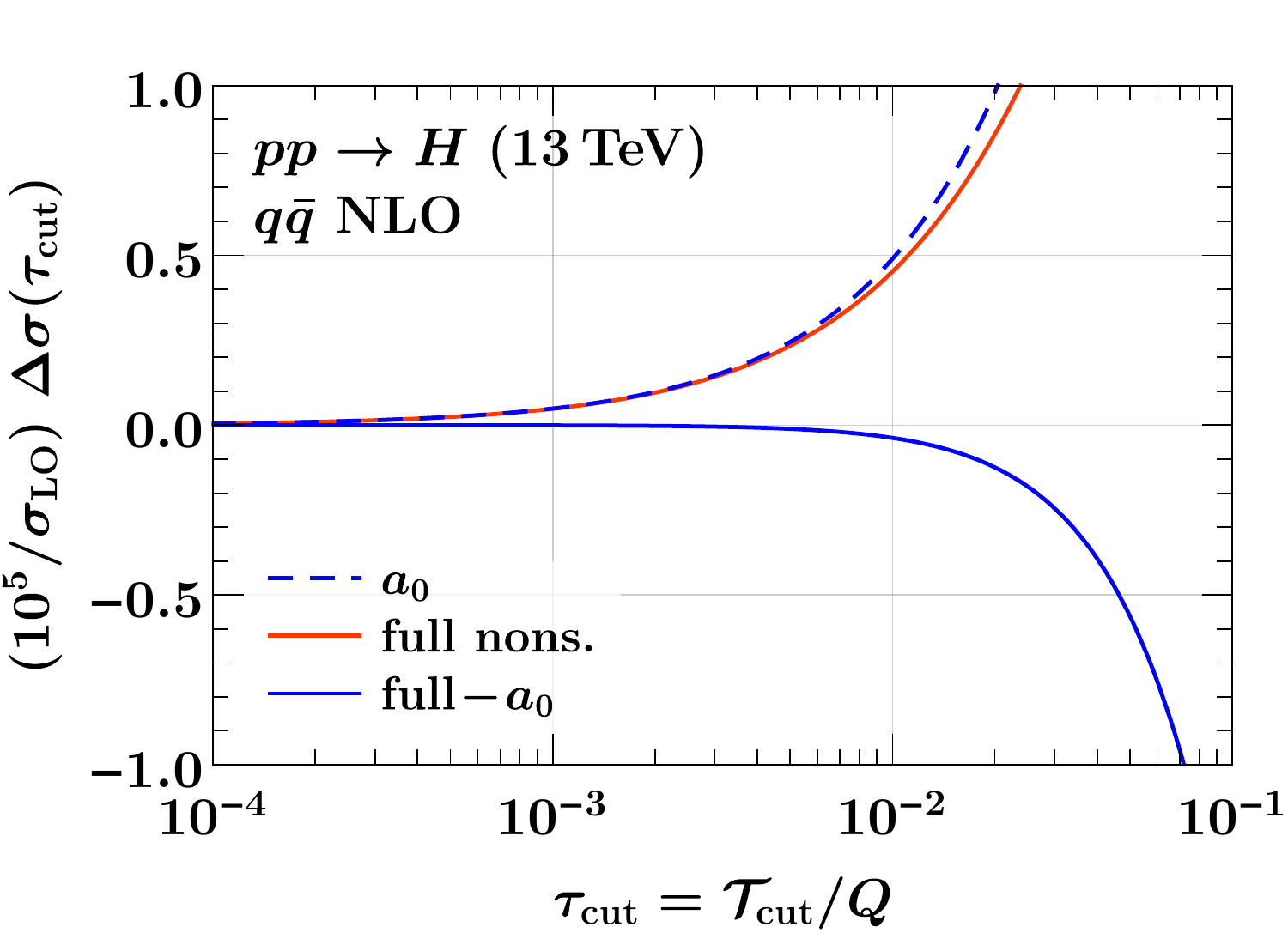}%
\hfill
\includegraphics[width=0.5\textwidth]{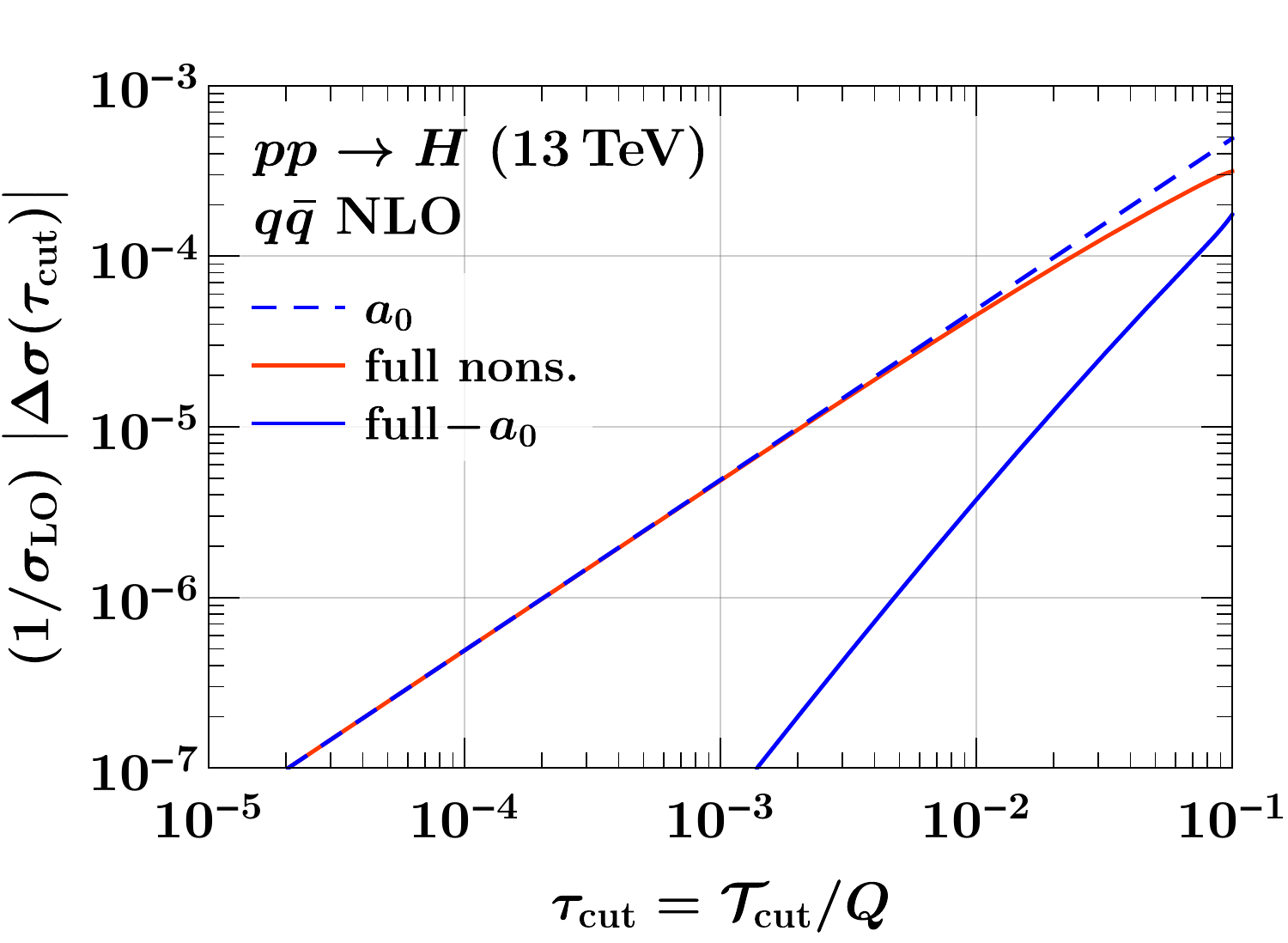}%
\caption{The power corrections for the cumulant $\Delta\sigma(\tau_\cut)$ at $\cO(\alpha_s)$ for Higgs production in the $gg$ channel (top row), $gq+qg$ channel (middle row), $q\bar q$ channel (bottom row).}
\label{fig:cumulantNLO_gg}
\end{figure*}

\begin{figure*}[t!]
\includegraphics[width=0.5\textwidth]{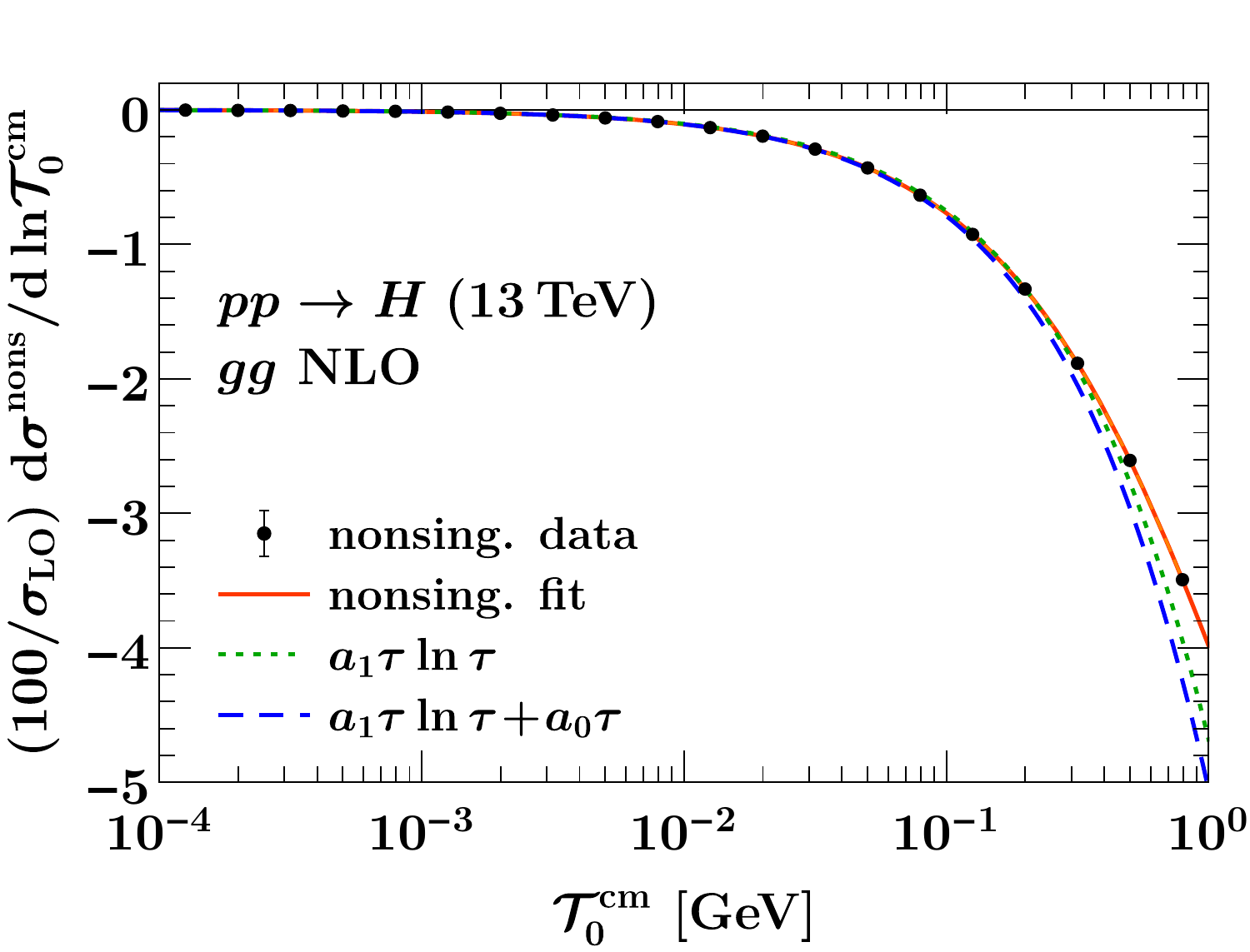}%
\hfill
\includegraphics[width=0.5\textwidth]{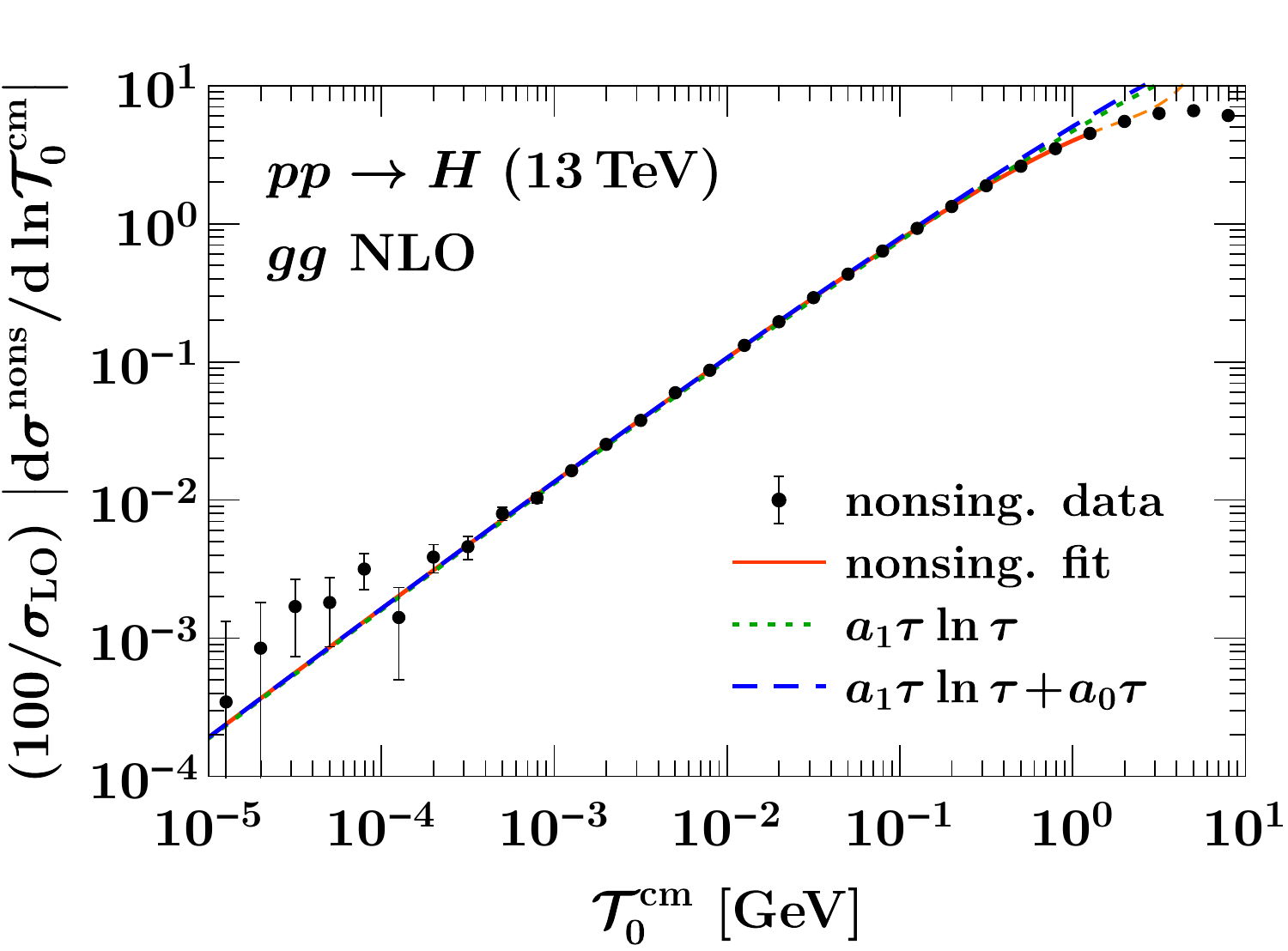}%
\\
\includegraphics[width=0.5\textwidth]{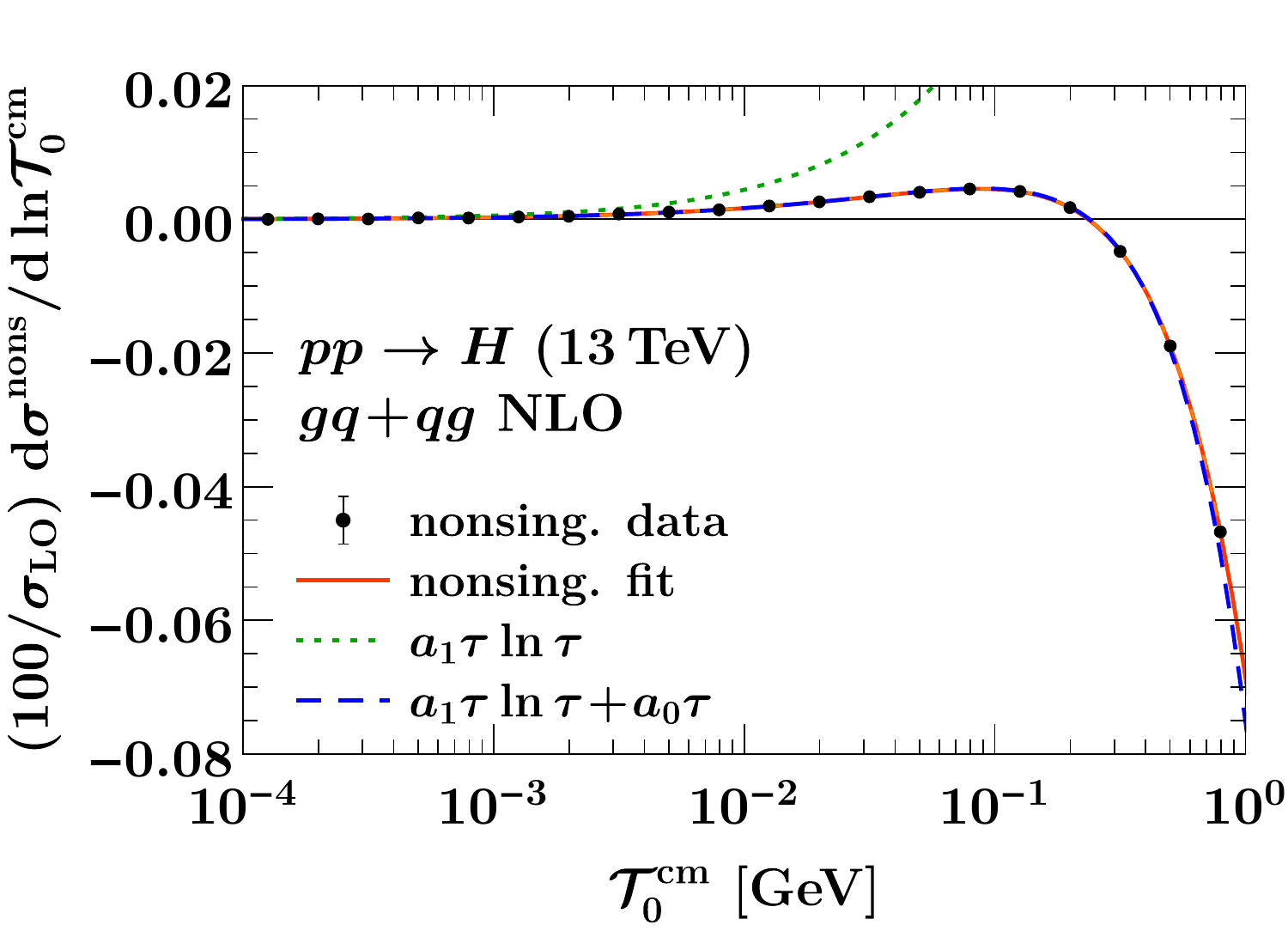}%
\hfill
\includegraphics[width=0.5\textwidth]{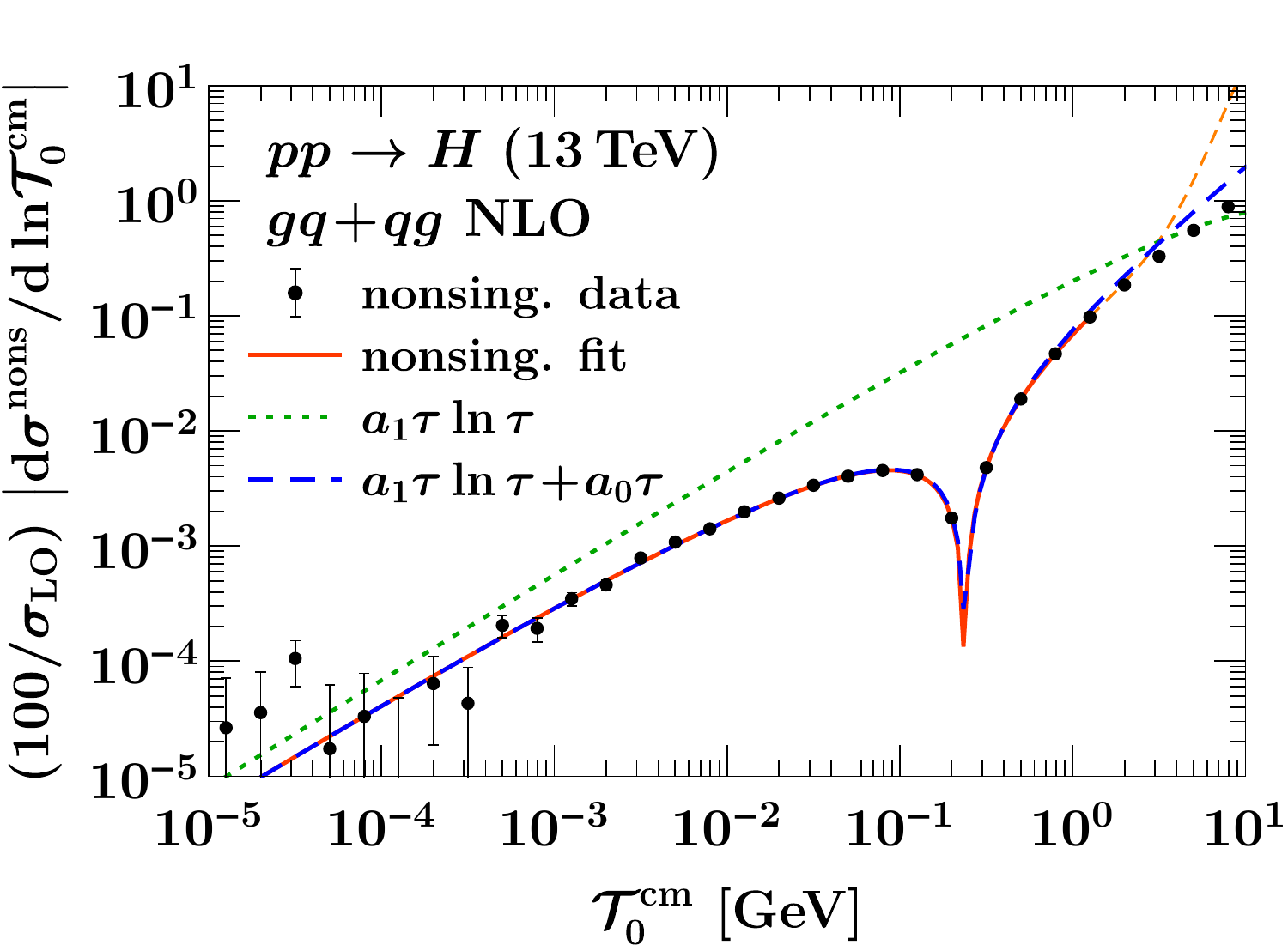}%
\\
\includegraphics[width=0.5\textwidth]{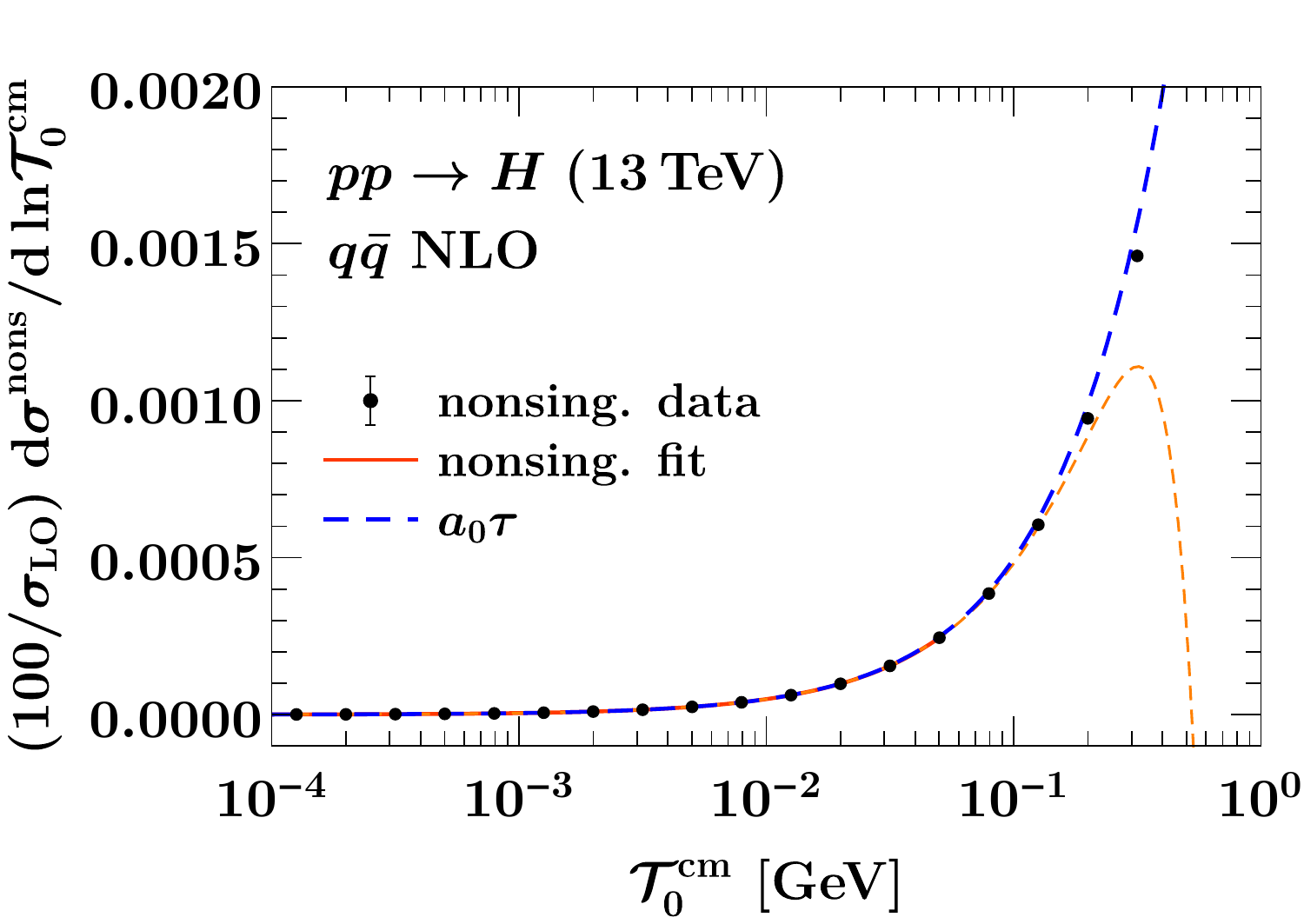}%
\hfill
\includegraphics[width=0.5\textwidth]{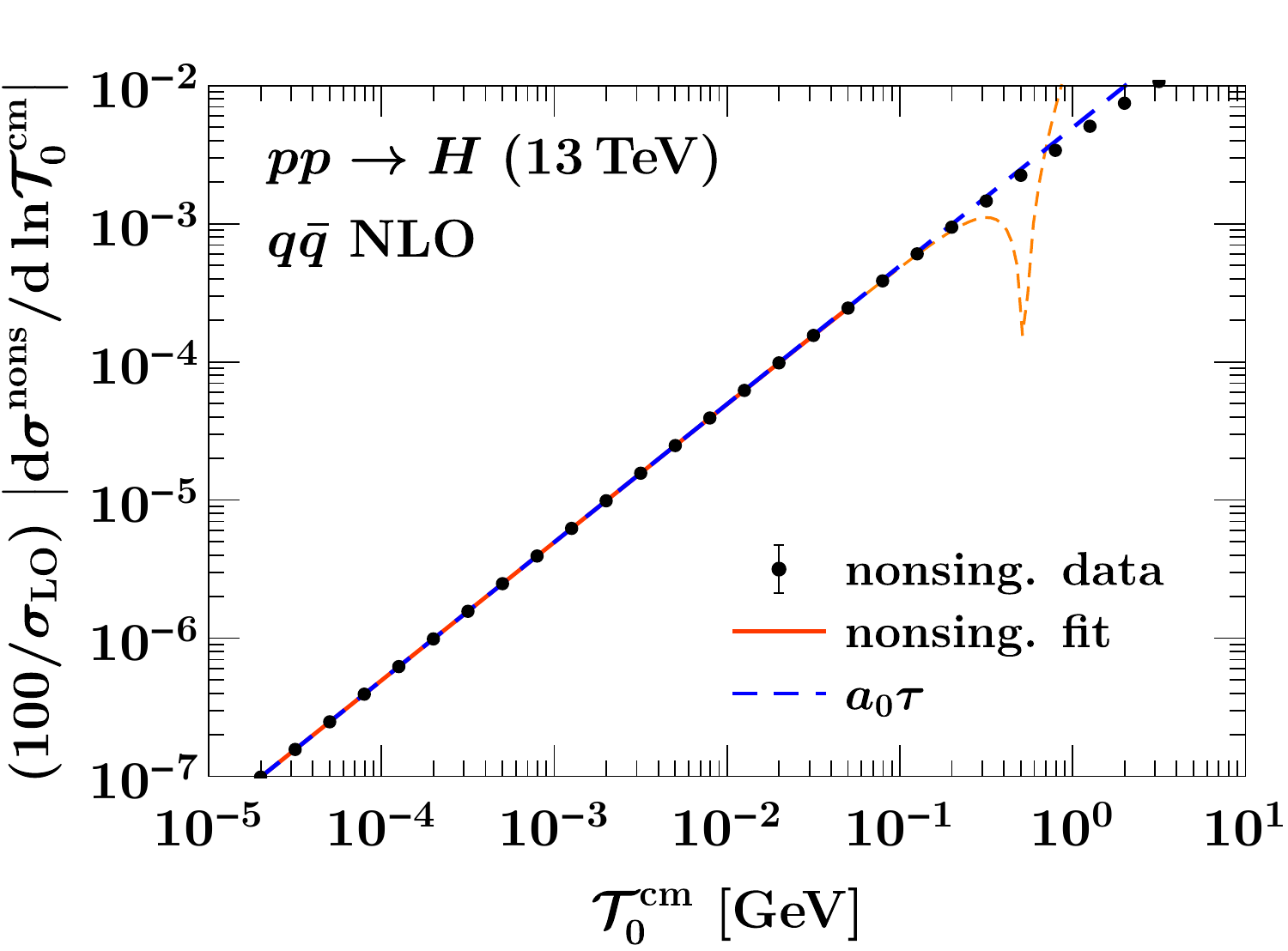}%
\caption{Same as \fig{fitNLO_gg} for the hadronic $\Tau$ definition.}
\label{fig:fitNLOcm_gg}
\end{figure*}

\begin{figure*}[t]
\includegraphics[width=0.48\textwidth]{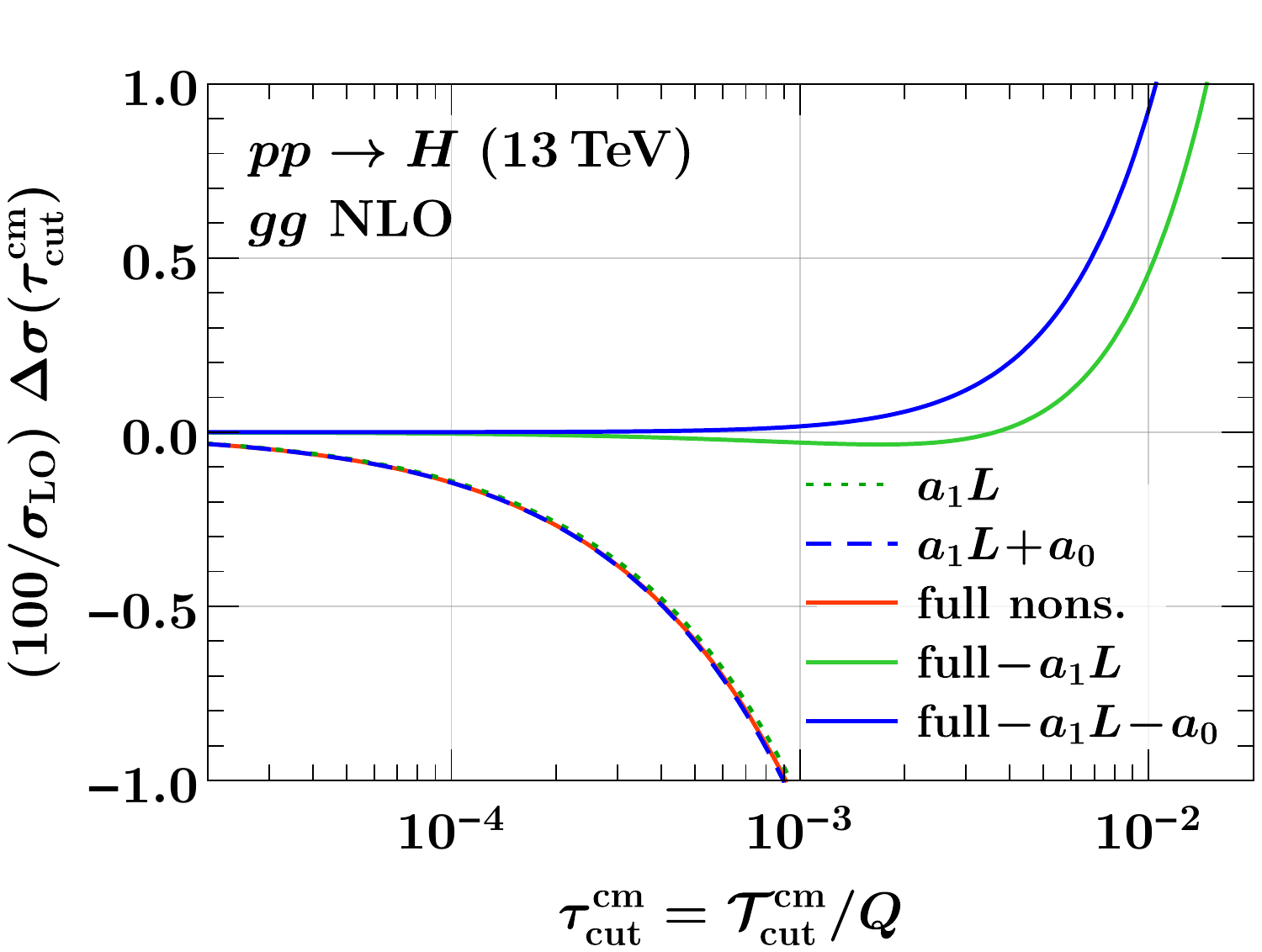}%
\hfill
\includegraphics[width=0.48\textwidth]{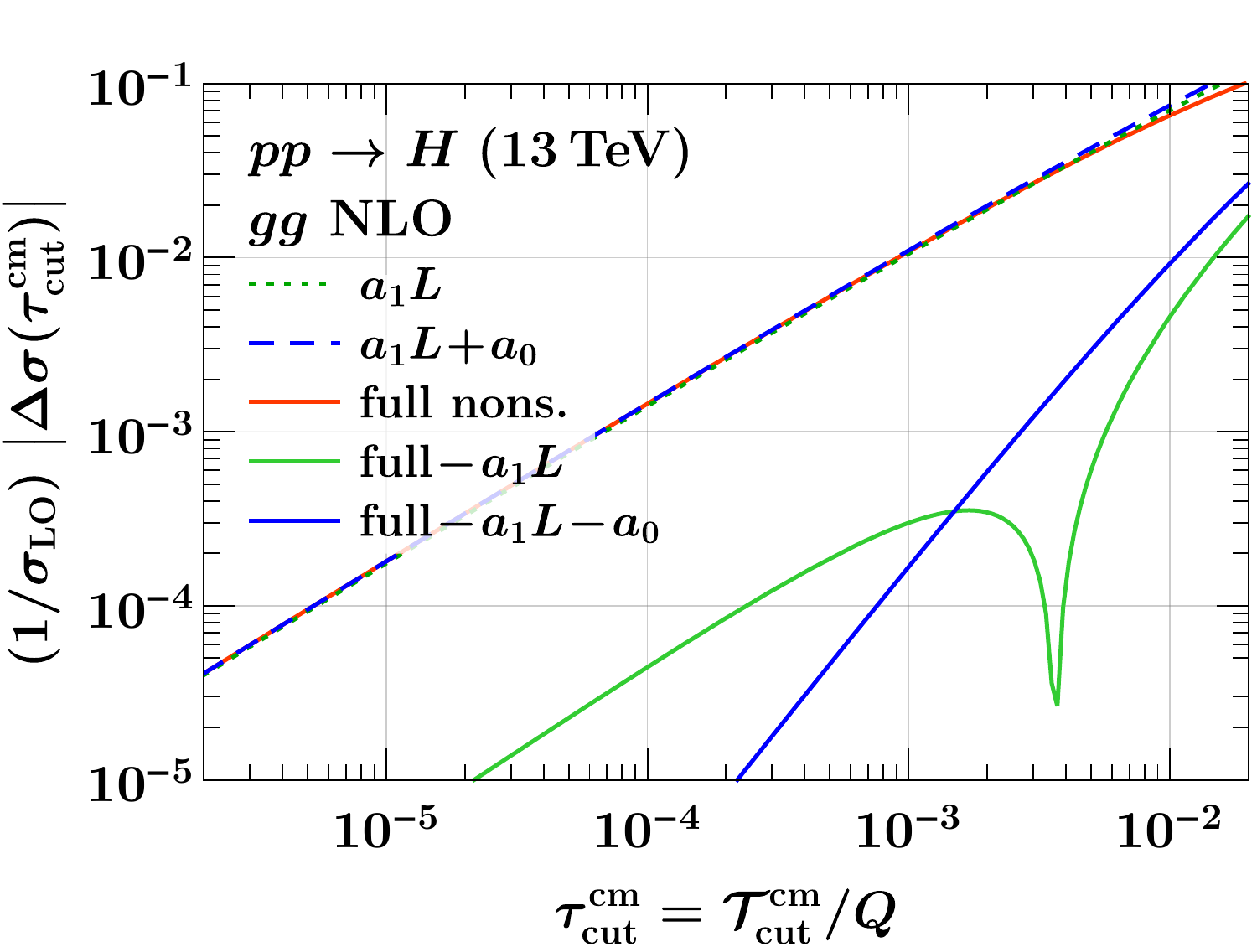}%
\hspace*{\fill}
\\
\includegraphics[width=0.5\textwidth]{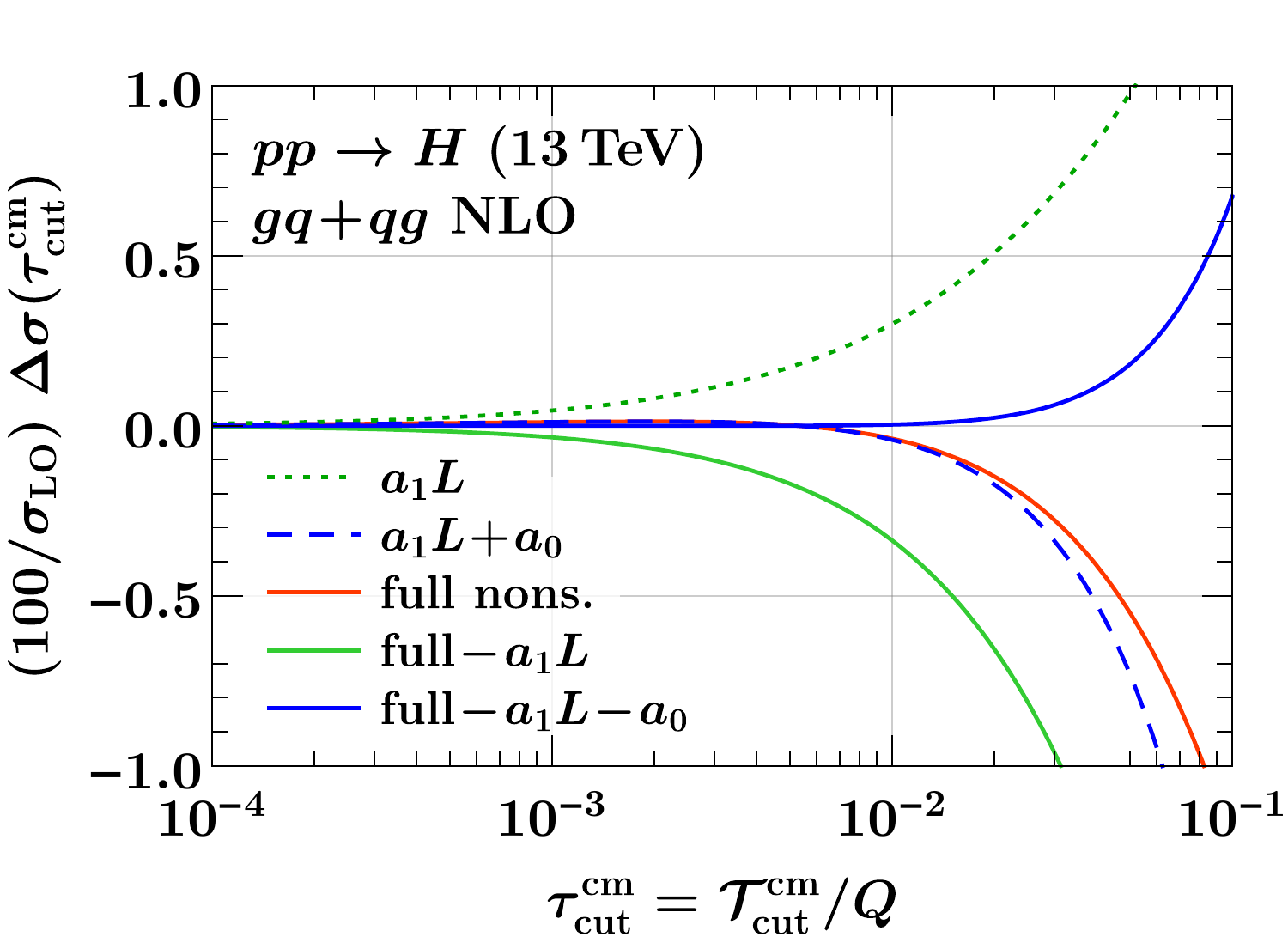}%
\hfill
\includegraphics[width=0.5\textwidth]{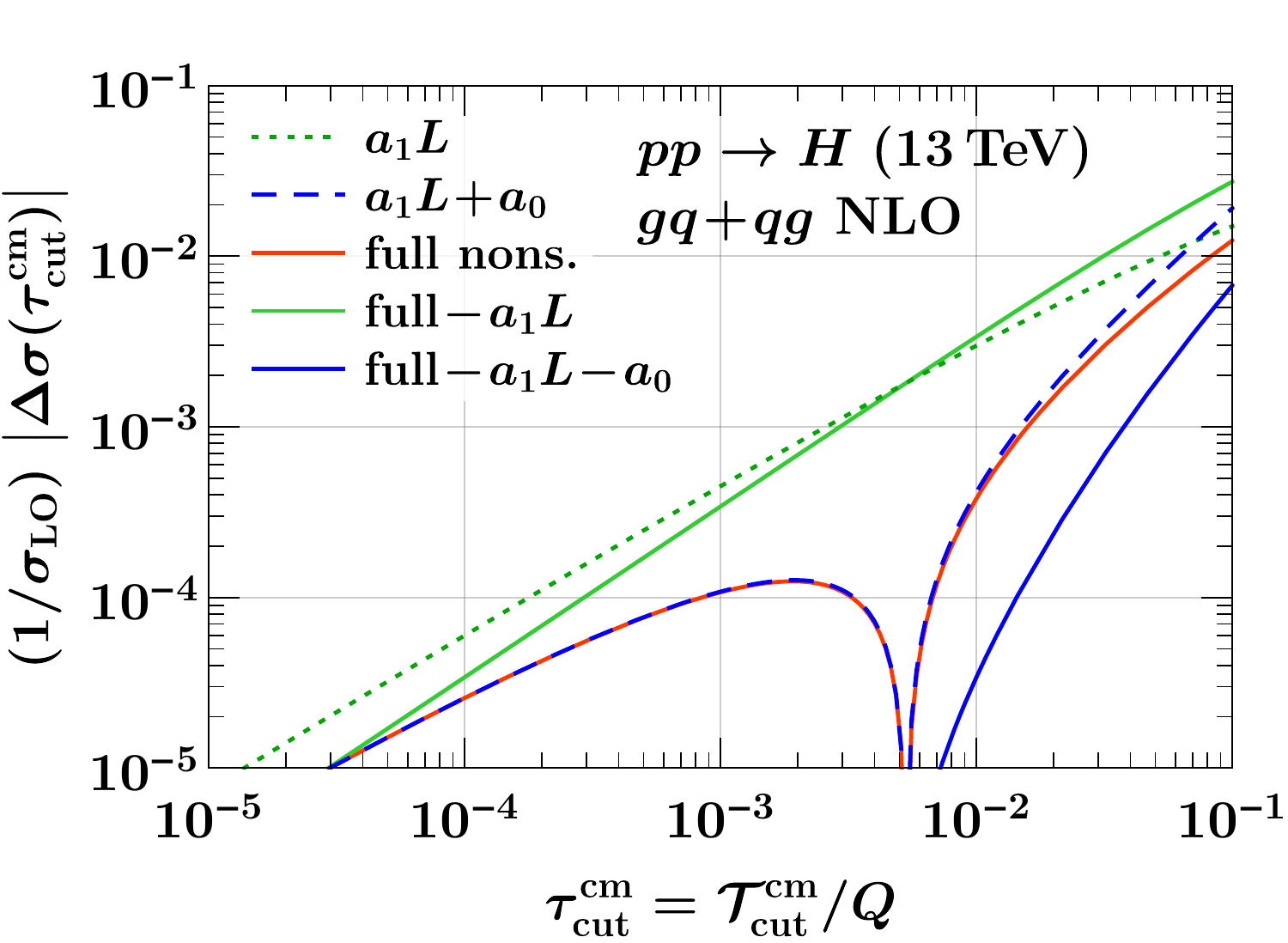}%
\\
\includegraphics[width=0.5\textwidth]{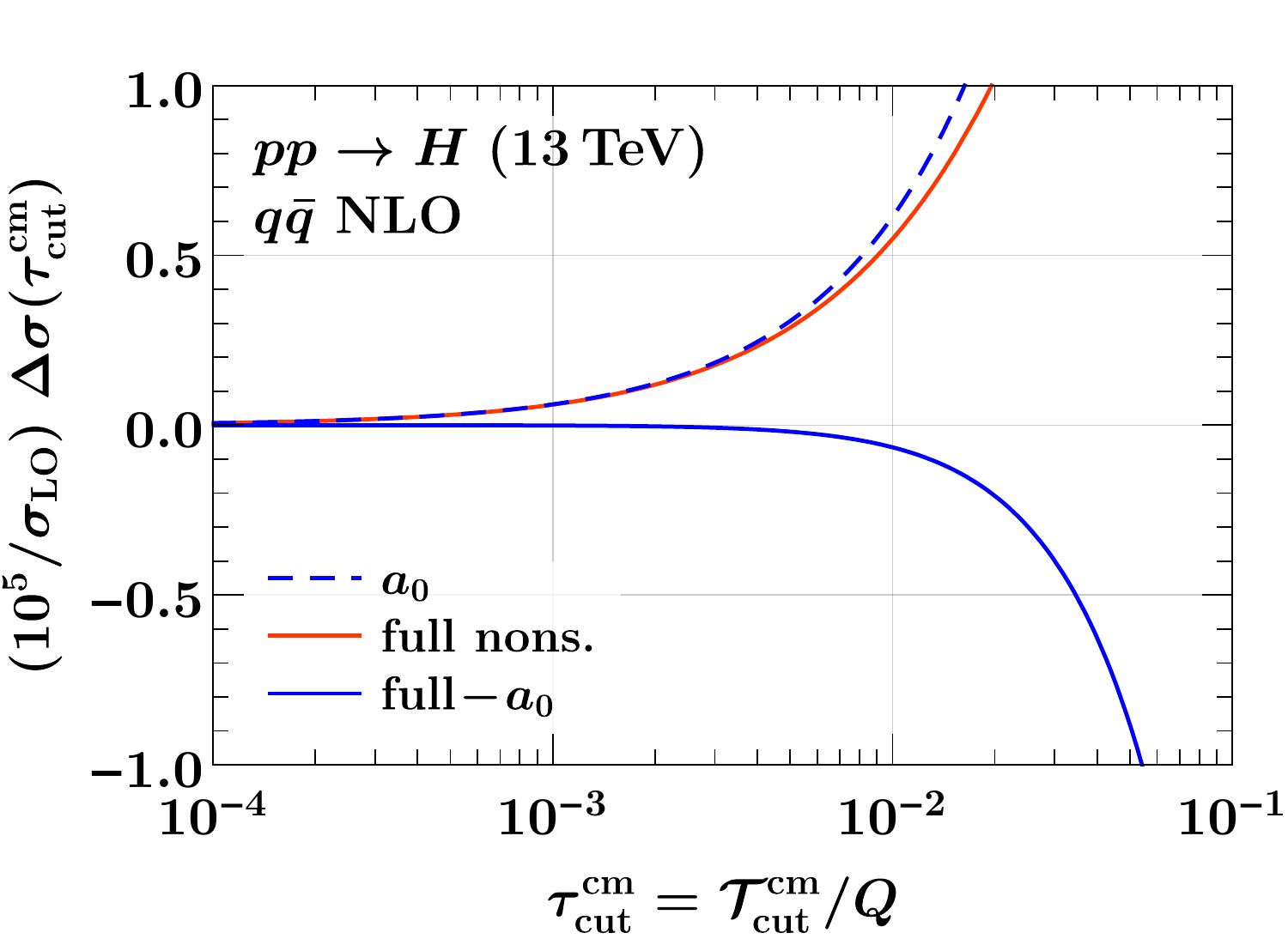}%
\hfill
\includegraphics[width=0.5\textwidth]{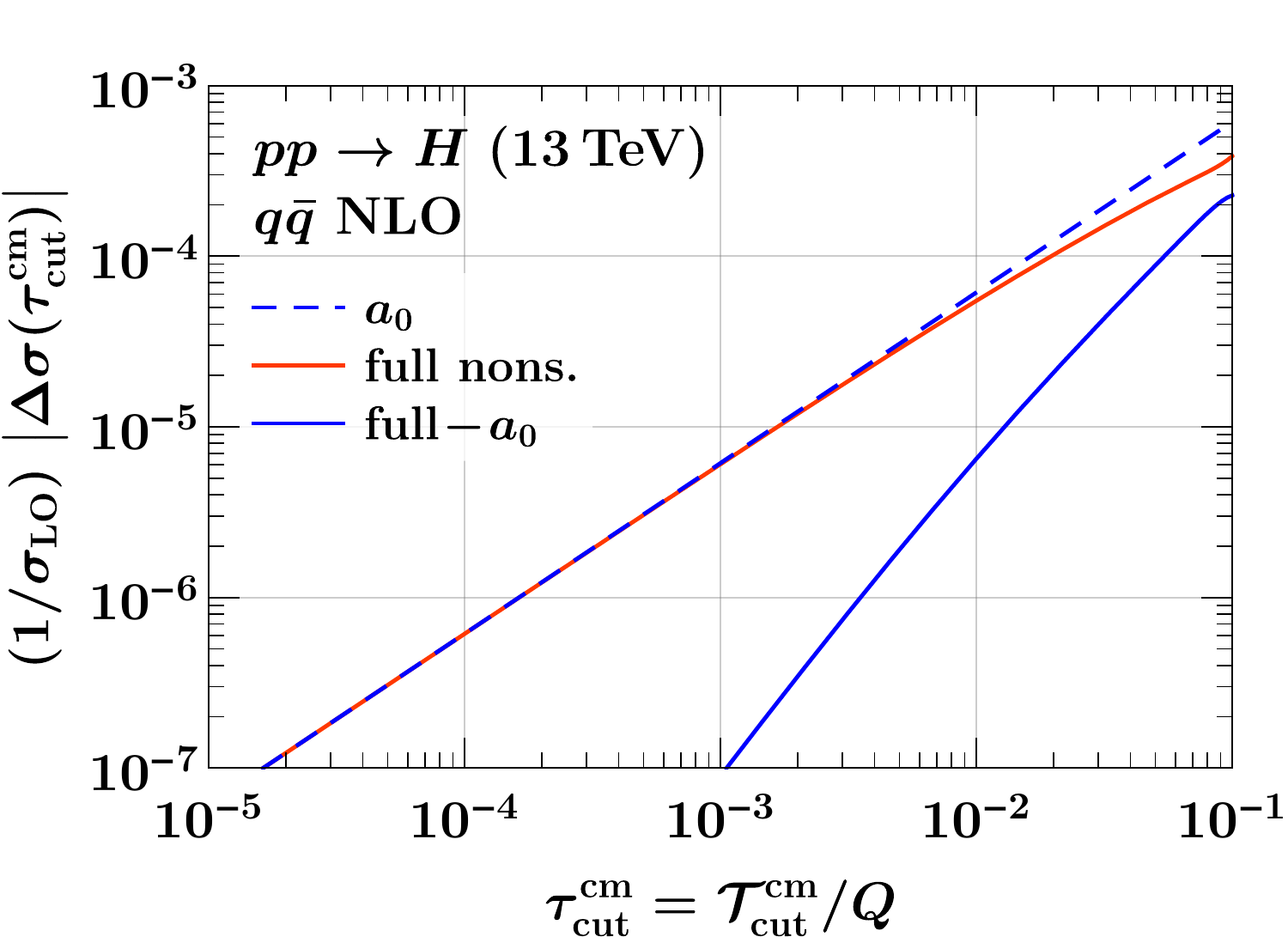}%
\caption{Same as \fig{cumulantNLO_gg} for the hadronic $\Tau$ definition.}
\label{fig:cumulantNLOcm_gg}
\end{figure*}

\begin{figure*}[t!]
\includegraphics[width=0.49\textwidth]{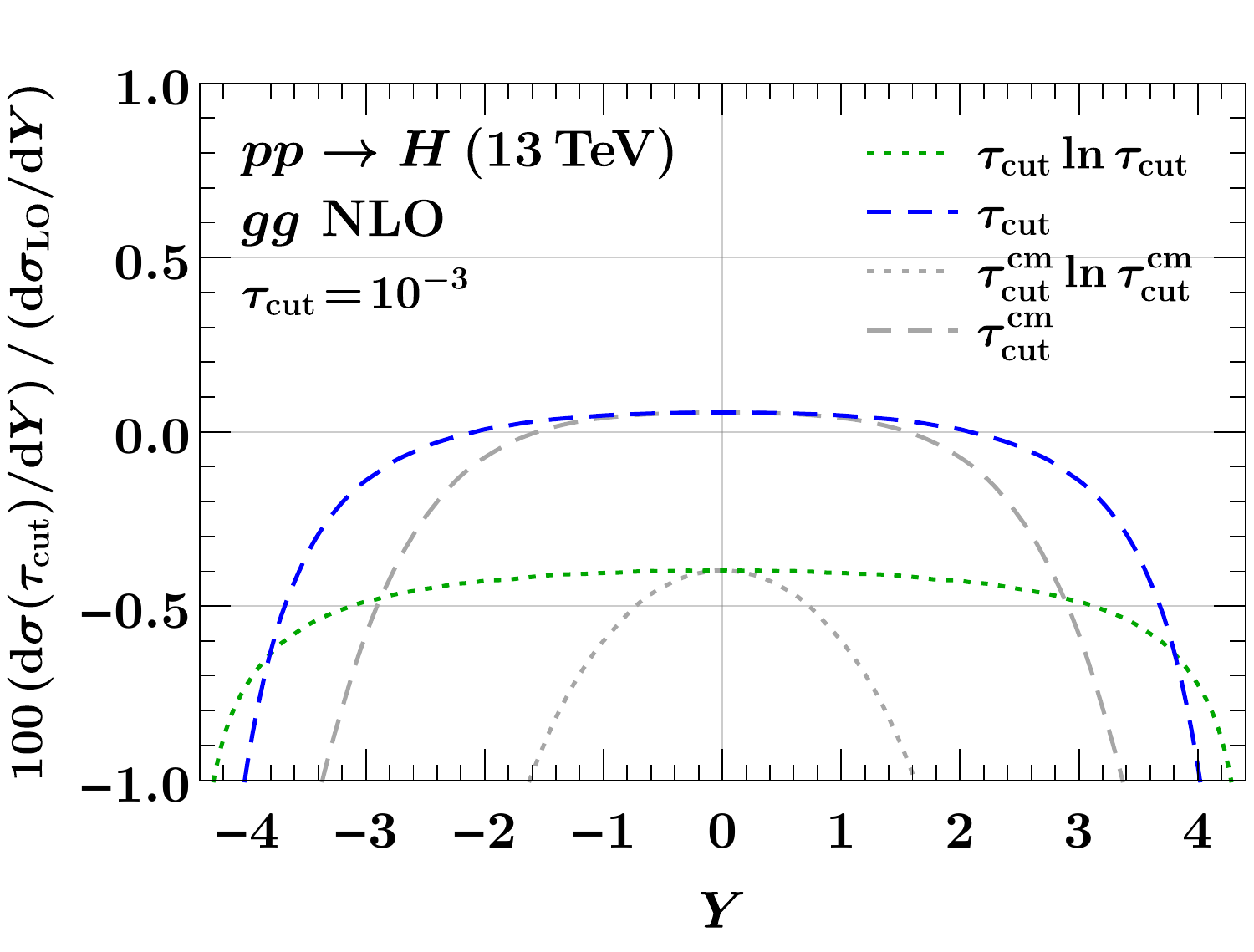}%
\hfill
\includegraphics[width=0.49\textwidth]{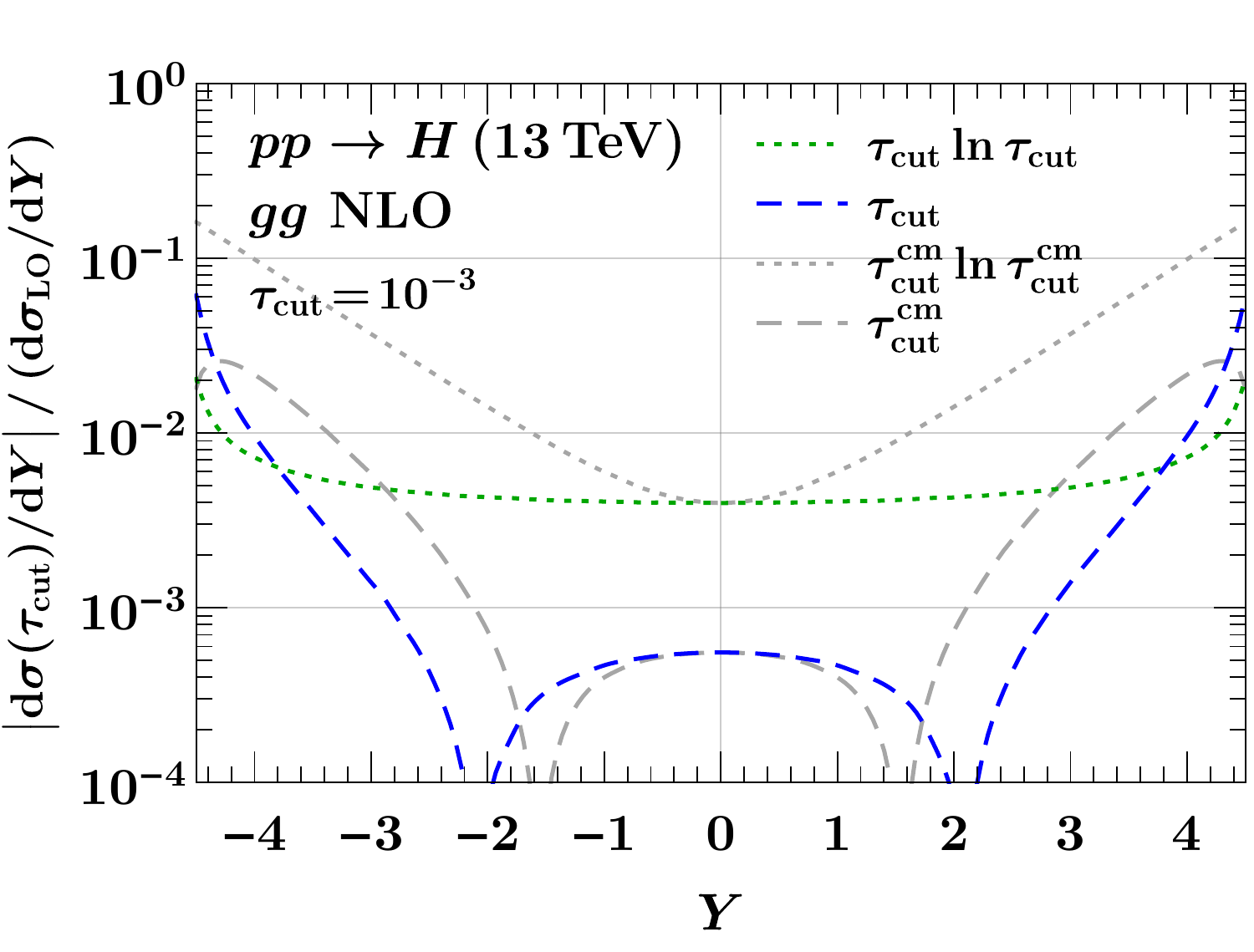}%
\\
\includegraphics[width=0.49\textwidth]{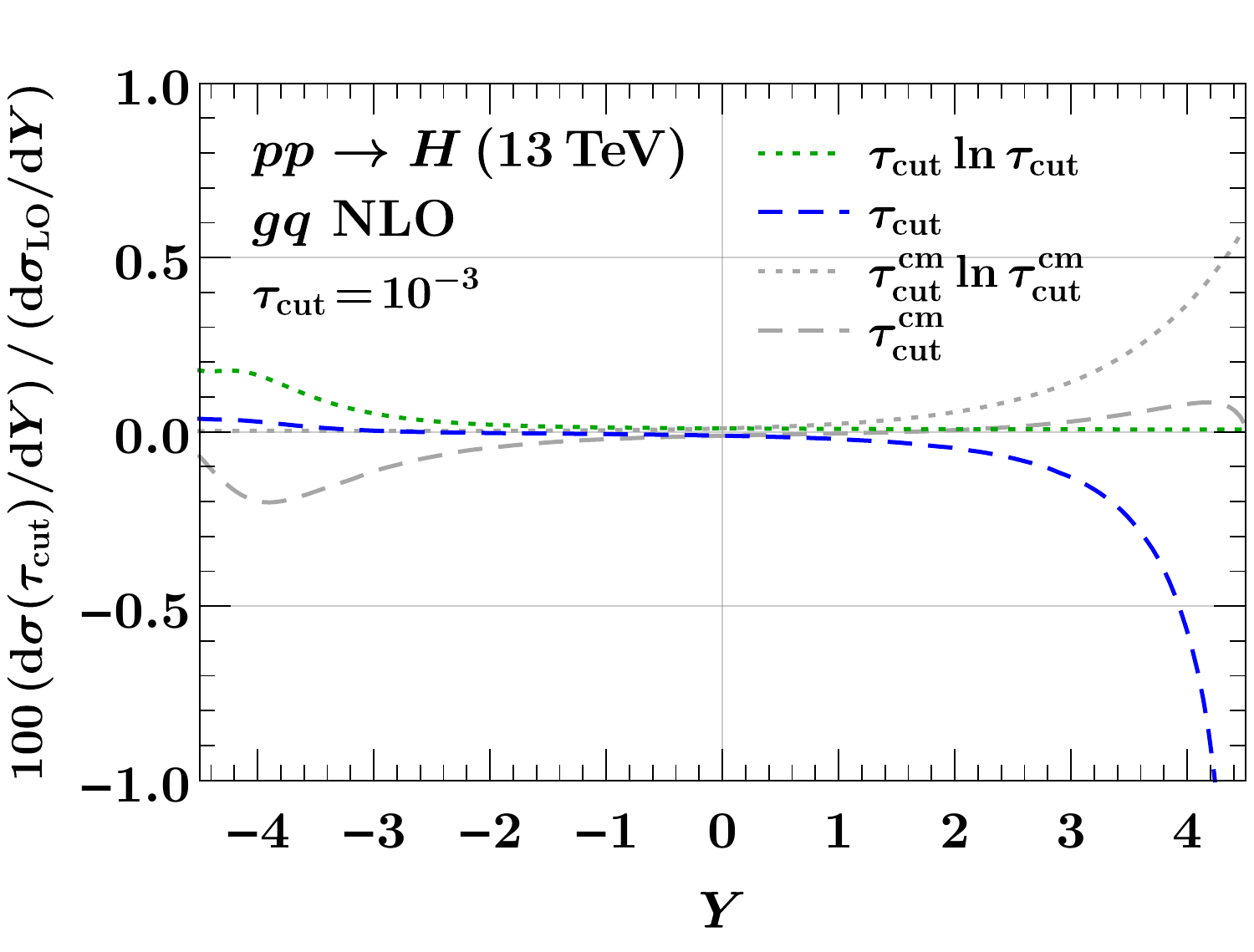}%
\hfill
\includegraphics[width=0.49\textwidth]{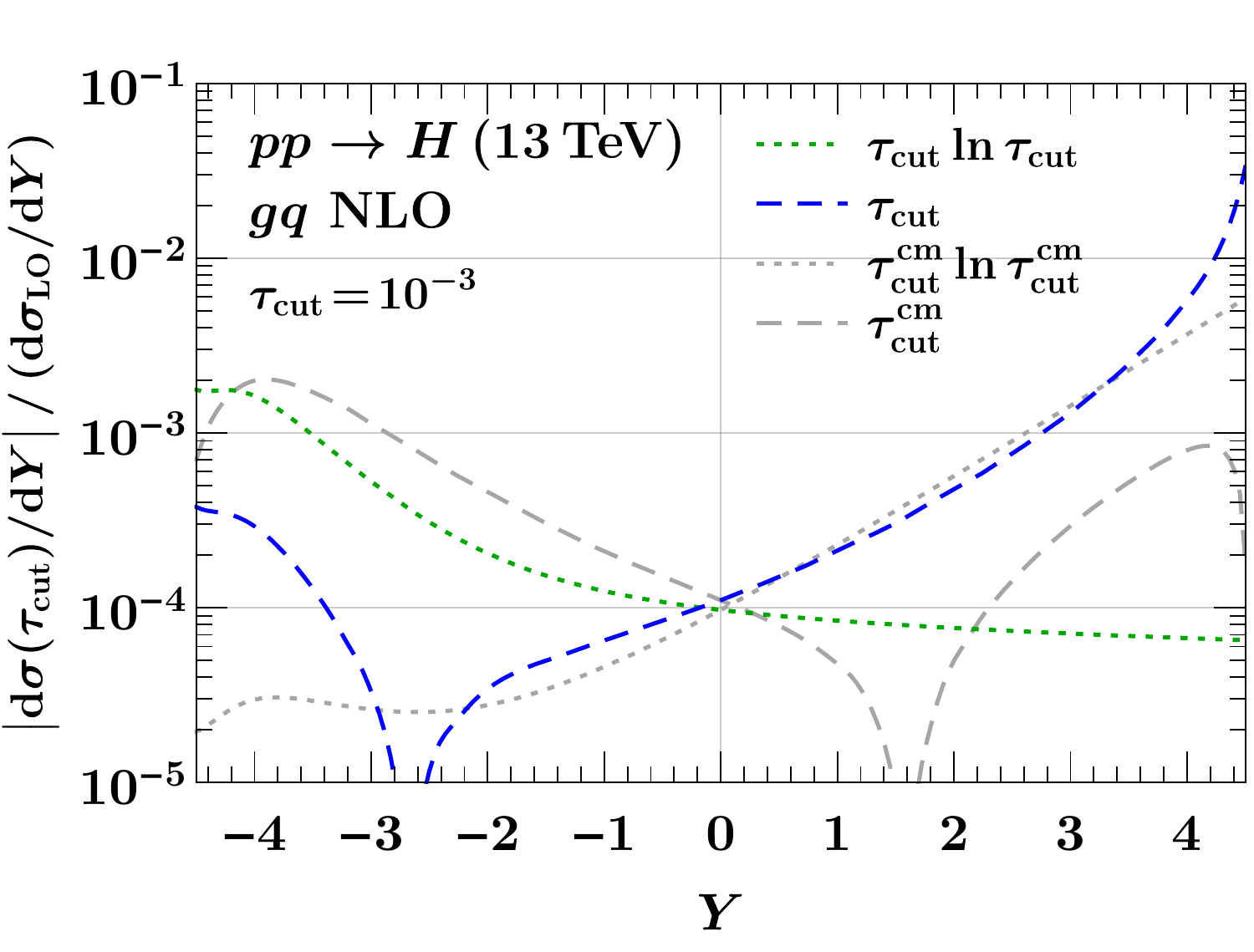}%
\caption{The NLO NLP corrections as a function of rapidity at fixed $\tau_\cut=10^{-3}$ for Higgs production for the $gg$ channel (top row) and the $gq$ channel (bottom row). The LL and NLL coefficients for leptonic $\Tau$ are shown by the green dotted and blue dashed curves and for hadronic $\Tau$ by the dotted and dashed gray curves.}
\label{fig:dsigdY_H}
\end{figure*}

In \fig{fitNLO_gg} we show as the solid red curve a fit to the full nonsingular
result  at NLO (black points), which is compared with the LL and NLL
predictions in dashed green and dashed blue, respectively. Once again this
solid red fit curve is obtained using the form in \eq{fitfun} with $a_1$ and $a_0$ fixed
by the analytic result in table~\ref{tab:NLLresults_H}, and agrees very well
with the corresponding result obtained in \refcite{Moult:2017jsg} where $a_0$ was a parameter in the fit.
In all cases, we find that the NLL result provides a good description of the full nonsingular cross section. This is expected since the NLL results includes all NLP terms in the NLO cross section.
We see, however, that particularly for the $gq+qg$ channel, the NLL result for $a_0$ is required to get a good description, and the LL power correction $a_1$ alone is not sufficient. Thus the $gq+qg$ channel provides an example where simply looking at the size of the residual nonsingular result after subtracting the $a_1$ term does not suffice to validate the value of this coefficient.

In \fig{cumulantNLO_gg}, we show a plot of the corresponding power
corrections for the cumulant, $\Delta\sigma(\tau_\cut)$, on both a linear scale (left)
and logarithm scale (right). Here we
more easily see that the inclusion of the NLL power corrections significantly
reduces the residual power corrections for the subtractions. For the dominant
$gg\to Hg$ channel at a typical value of $\tau_\cut \sim 10^{-3}$ approximately one order
of magnitude is gained at each logarithmic order that the power corrections are
computed.
From table~\ref{tab:NLLresults_H}, we see that for the $gq+qg\to Hq$ channel, the LL
coefficient is numerically suppressed, while in contrast its NLL coefficient is
quite larger. Due to this unusual behavior, the NLL result is
required to consistently reduce the power corrections as compared with the
leading-power result. In the $q\bar q$ channel there is no $a_1$ term, and
significant improvement is apparent from including $a_0$.

The analogous results for the fitted nonsingular spectrum and the residual
power corrections $\Delta\sigma(\tau^\hadcm_\cut)$ for the hadronic $\Tau$ definition
are shown in \figs{fitNLOcm_gg}{cumulantNLOcm_gg}.
The power corrections are noticeably larger, though the effect of the rapidity
enhancement is not as pronounced as for Drell-Yan, since here the PDFs suppress the
cross section contributions at larger rapidities. For the dominant $gg\to Hg$ channel
there are also numerical cancellations in the NLL coefficient. More precisely the value for $a_0$
in table~\ref{tab:NLLresults_H} arises as $a_0 = 2.356 + (-2.822) = -0.466$,
where the first term corresponds to the rapidity-enhanced version of the leptonic
$a_0$ while the second term is the NLL contribution arising from the additional
rapidity dependence in the argument of the leading logarithm discussed below \eq{kernel_Cgg_LL}.
As a result of this cancellation, including only $a_1$ in the subtractions
leads to slightly smaller power correction above $\tau_\cut > 10^{-3}$
than subtracting both $a_1$ and $a_0$
(compare the green and blue solid lines in the top row of \fig{cumulantNLOcm_gg}).
If the second NLL contribution were included as part of the LL result, the latter would
provide a much poorer approximation and including the remaining NLL contribution
would provide a substantial improvement.
Either way, the remaining power corrections after subtracting the full NLL result
shows a much steeper slope, which is as expected from its $\ord{\tau_\cut^2}$ scaling.
This provides another example where considering only the overall size of the improvement
can be potentially misleading. The $gq+qg\to Hq$ channel shows a similarly unusual
behavior as for the leptonic definition.

In \fig{dsigdY_H} we show the rapidity dependence of the NLP corrections at fixed
$\tau_\cut = 10^{-3}$ for both leptonic and hadronic $\Tau$ normalized to the LO
rapidity spectrum. The exponential enhancement for the hadronic definition at large $\abs{Y}$ is again apparent in the LL results. The NLL coefficients again exhibit an enhancement already for
the leptonic definition at large $Y$. This is again due to the different $x$ dependence of the
quark PDF and the PDF derivatives compared to the LO $gg$ luminosity to which we normalize.
The quark PDF contributions are also the main
reason why the NLL term for the $gq$ channel ($a_0$ in table~\ref{tab:NLLresults_H}) is much
larger than the LL contribution.
For the hadronic definition, the $e^{\pm Y}$ factors from the observable definition turn
out to partially compensate these PDF effects. This is best visible in the $gq$ channel,
where the PDF enhanced terms at negative (LL) or positive (NLL) $Y$ get reduced by a $e^{\pm Y}$ factor from the observable definition. The same effect is also present in the $gg$ channel at NLL.
This is the reason why the $a_0$ term for the hadronic definition in the $gq$ channel
turns out to be even slightly smaller than for the leptonic definition.

\FloatBarrier

\section{Conclusions}
\label{sec:conclusions}

In this paper, we have computed the next-to-leading power corrections in the $N$-jettiness resolution variable for Drell-Yan and gluon-fusion Higgs production at NLO. This builds on our previous work by computing the non-logarithmically enhanced terms at this order. These results enable the performance of the $N$-jettiness subtraction method to be improved, and provide important information on the structure of subleading power corrections beyond the leading logarithms.
Our calculation is based on a master formula applicable to \SCETI observables, and highlights a large degree of universality of these power corrections.

We explained in detail the issue of the treatment of Born measurements at subleading power. We have shown that an apparent disagreement in the literature arises due to the fact that the representation used to obtain the power corrections in \refscite{Boughezal:2016zws,Boughezal:2018mvf} is only valid when integrated over all rapidities, and therefore cannot be directly compared with the results of \refscite{Moult:2016fqy,Moult:2017jsg} and those in the present paper, which are differential in rapidity.
We show that after integration over rapidity the LL results agree.
Further details can be found in \sec{compare_lit}.

We find that the rapidity dependence of the NLL terms is quite sizeable and is therefore important
to know to be able to improve the subtractions.
One reason for this effect is the different $x$-dependence of parton luminosities
or derivatives of PDFs appearing in the power corrections as compared to the Born-level
parton luminosity. Hence, one can expect this to be a generic feature of subleading
power corrections.

We also compared our analytic NLL results for gluon fusion Higgs production and Drell-Yan to numerical predictions for these NLO power corrections  obtained from a fit to data from MCFM.  In all cases, excellent agreement was found. In addition we studied the extent to which the inclusion of the NLL power corrections improves the subtraction. At NLO, the inclusion of the NLL power corrections completely captures the $\cO(\tau)$ terms. Numerically, summing over production channels, the inclusion of these results reduces the size of the power corrections by two orders of magnitude in Higgs production and three orders of magnitude in Drell-Yan production.

There are a number of directions for future work. It will be interesting to extend the calculation of the NLL power corrections to NNLO. Generically we expect up to an order of magnitude improvement could also be obtained by extending the known LL power corrections at this order to NLL.  Beyond fixed order, the derivation of subleading power renormalization group evolution equations at NLL would allow for the all-orders prediction of the NLL terms. Finally, while we have focused here on color-singlet production, our results provide an important step toward the calculation of the NLP corrections at higher orders and for more complicated processes.

\begin{acknowledgments}

We thank Radja Boughezal, Andrea Isgro, Xiaohui Liu, and Frank Petriello for discussions
regarding the results of \refscite{Boughezal:2016zws, Boughezal:2018mvf}.
This work was supported in part by the Office of Nuclear Physics of the U.S.
Department of Energy under Contract No. DE-SC0011090, by the Office of High Energy Physics of the U.S. Department of Energy under Contract No. DE-AC02-05CH11231, and
the Simons Foundation Investigator Grant No. 327942. We thank the Aspen Center for Theoretical Physics, which is supported by National Science Foundation grant PHY-1607611, for hospitality while portions of this work were completed. We also thank DESY and MIT for hospitality.

\end{acknowledgments}

\appendix

\section{Derivation of NLO Leading Power Results}
\label{app:app_LP}

At leading power the singular terms for $N$-jettiness are most easily obtained from known factorization formulas \cite{Stewart:2009yx,Stewart:2010tn}, which describe the singular behavior of the observable to all orders. The fixed order approach of this paper is therefore most useful when such factorization formula are not available, or well understood, such as at subleading power. However, it can also be applied to reproduce the LP results. In this appendix we illustrate this at NLO, by reproducing the one-loop beam and soft functions for beam thrust.

To relate the beam and soft functions as defined in SCET to our calculation in this work,
recall the LP factorization formula for beam thrust \cite{Stewart:2009yx},
\begin{align}
 \frac{\df \sigma^{(0)}}{\df Q^2 \df Y \df \Tau}
 = \hat\sigma^\LO(Q,Y) \, H_{ab}(Q, \mu) \int \df t_a \df t_b \,
   B_a(t_a, x_a, \mu) B_b(t_b, x_b, \mu) S\biggr(\Tau - \frac{t_a}{Q} - \frac{t_b}{Q}, \mu \biggr)
\,,\end{align}
where the superscript $^{(0)}$ refers to LP, and as before $x_{a,b} = \frac{Q}{\Ecm} e^{\pm Y}$.
The hard function $H_{ab}$ describes virtual corrections to the hard process
$a b \to L$, $B_{a,b}$ are the two beam functions and $S$ is the soft function.
The beam functions can be further matched onto normal PDFs,
\begin{align}
 B_i(t, x, \mu) = \sum_j \int_{x}^1 \frac{\df z}{z} \cI_{ij}(t, z, \mu) f_j\biggl(\frac{x}{z}, \mu\biggr)
\,.\end{align}

All of these functions have definitions as field theory matrix elements in the EFT. 
Their fixed order definitions give rise to UV divergences, which are as usual removed
by a renormalization procedure, which in turn gives rise to RGEs that can be used
to resum large logarithms of $\Tau$.
In the approach presented in this paper, the same divergences appear as $1/\epsilon$ IR divergences in the soft and collinear limits of QCD amplitudes.

At LO, we have
\begin{align}
 H^\LO_{ij}(Q^2,\mu) = 1
\,,\quad
 \cI_{ij}^\LO(t, x, \mu) = \delta_{ij} \delta(t)
\,,\quad
 S^\LO(\Tau, \mu) = \delta(\Tau)
\,.\end{align}
At one loop, the convolution structure thus becomes trivial. Working with hard, beam, soft, and PDFs in the bare factorization theorem we have
\begin{align} \label{eq:factorization_NLO}
 \frac{\df \sigma^{(0,1)}}{\df Q^2 \df Y \df \Tau}
 &= \hat\sigma^\LO \, H^\NLO_{ab}(Q^2, \epsilon) \delta(\Tau) f_{a}(x_a,\epsilon) f_{b}(x_b,\epsilon)
 \nn\\&
   + \hat\sigma^\LO \sum_{a'} \int_{x_a}^1 \frac{\df z_a}{z_a} \, Q \, \cI^\NLO_{a a'}(Q \Tau, z_a, \epsilon) f_{a'}(x_a/z_a ,\epsilon) f_{b}(x_b,\epsilon)
 \nn\\&
   + \hat\sigma^\LO \sum_{b'} \int_{x_b}^1 \frac{\df z_b}{z_b} \, Q \, \cI^\NLO_{b b'}(Q \Tau, z_b, \epsilon) f_{a}(x_a,\epsilon) f_{b'}(x_b/z_b ,\epsilon)
 \nn\\&
   + \hat\sigma^\LO  f_{a}(x_a,\epsilon) f_{b}(x_b,\epsilon) S^\NLO(\Tau,\epsilon)
\,.\end{align}
Note the extra factor of $Q$ in the beam contributions, arising from $t_{a,b}$ and $\Tau$
having different mass dimensions.
We have written \eq{factorization_NLO} in a form similar to our master formulas,
such that we can easily read off the one-loop beam function kernels and soft function. The $\epsilon$ arguments in \eq{factorization_NLO} all refer to ultraviolet divergences and can be removed by SCET counterterms to obtain the renormalized factorization theorem. To obtain this it is important to include virtual graphs in the various sectors as well as zero-bin subtractions for the beam functions.

\subsection{Leading-Power Expansion of Matrix Elements}

The leading-power behavior of real emission matrix elements in the soft and collinear limits
is universal, see e.g.\ \cite{Catani:1996vz}, and has already been used in \sec{universality}.
Here, we briefly review the relevant formulas, and give the relevant one-loop expressions.

Given the Born process
\begin{align}
 \kappa_a(q_a) + \kappa_b(q_b) \to L(q_a + q_b)
\,,\end{align}
where the incoming momenta are given by
\begin{align}
 q_a^\mu = Q e^Y \frac{n^\mu}{2}
\,,\quad
 q_b^\mu = Q e^{-Y} \frac{\bn^\mu}{2}
\,,\end{align}
and we write the one-emission process as
\begin{align}
 \kappa'_a(q'_a) + \kappa'_b(q'_b) \to L(q'_a + q'_b - k) + \kappa_1(k)
\,.\end{align}
In the soft limit $k^\mu \ll q_a^\mu, q_b^\mu$, the squared matrix element obeys the LP relation
\begin{align} \label{eq:M2_soft_LP_app}
 \Msquared^{(0)}_{a' b' \to L k}(Q,Y; \{k\}) &
 = \frac{16 \pi \as \muMS^{2\eps} \mathbf{C}}{k^+ k^-} \times
   \delta_{a a'}\, \delta_{b b'}\, \Msquared^\LO_{a b \to L}(Q,Y)
\,,\end{align}
where we made explicit that a soft emission can not change the incoming flavors.
$\mathbf{C} = C_F, C_A$ is the Casimir constant for $ab=q\bar q, gg$.

In the LP $n$-collinear limit, the particle $k$ arises from the splitting $\kappa'_a \to \kappa_a + \kappa_1$.
If this splitting is allowed, at LP we have (in the notation of \sec{master_formula})
$q'_a = q_a / z_a$ and $q'_b = q_b$, and the LP limit of the matrix element is given by
\begin{align} \label{eq:M2_coll_LP_app}
 \Msquared^{(0)}_{a' b' \to L k}(Q, Y, \{k\})
 = \frac{8\pi \as \muMS^{2\eps}}{Q e^Y k^+} P_{a a'}(z_a, \eps) \times \delta_{b b'} \, \Msquared_{a b \to L}^\LO(Q, Y)
\,.\end{align}
Similarly, in the $\bn$-collinear limit arising from $\kappa'_b \to \kappa_b + \kappa_1$,
at LP we get $q'_a = q_a, q'_b = q_b / z_b$,
\begin{align} \label{eq:M2_bn_coll_LP_app}
 \Msquared^{(0)}_{a' b' \to L k}(Q, Y, \{k\})
 = \frac{8\pi \as \muMS^{2\eps}}{Q e^{-Y} k^-} P_{b b'}(z_b, \eps) \times \delta_{a a'} \, \Msquared_{a b \to L}^\LO(Q, Y)
\,.\end{align}
The one-loop splitting functions in $d=4-2\eps$ dimensions are given by \cite{Catani:1996vz}
\begin{align} \label{eq:Pij}
 P_{qq}(z,\eps) &= C_F \biggl[ \frac{1+z^2}{1-z} - \eps(1-z)\biggr]
\,,\nn\\
 P_{gq}(z,\eps) &= C_F \biggl[ \frac{1+(1-z)^2}{z} - \eps z\biggr]
\,,\nn\\
 P_{qg}(z,\eps) &= T_F \biggl[ 1 - \frac{2z(1-z)}{1-\eps} \biggr]
\,,\nn\\
 P_{gg}(z,\eps) &= 2 C_A \biggl[ \frac{z}{1-z} + \frac{1-z}{z} + z(1-z) \biggr]
\,.\end{align}
Note that we flipped the notation of $qg$ and $gq$ relative to \cite{Catani:1996vz},
following the standard convention.

\subsection{NLO Soft Function}
\label{app:LP_Soft}

The NLO LP soft function follows from combining \eq{sigmaLP} with \eq{M2_soft_LP_app}
using the same steps as in \sec{softmaster},
\begin{align}  \label{eq:sigLPsoft}
 \frac{\df\sigma_s^{(0,1)}}{\df Q^2 \df Y \df\Tau} &
 = \int_0^\infty \frac{\df k^+ \df k^-}{(k^+ k^-)^\eps}
   \frac{f_a(x_a) f_b(x_b)}{(4\pi)^2 2 x_a x_b \Ecm^4}
   \frac{(4\pi)^{\eps}}{\Gamma(1-\eps)} \Msquared^{(0)}(Q, Y; \{k\}) \,
   \delta\bigl[\Tau - \hat\Tau(\{k\})\bigr]
\nn\\&
 = \frac{\Msquared^\LO(Q,Y)}{2 x_a x_b \Ecm^4} \frac{\as \mathbf{C}}{\pi} f_a(x_a) f_b(x_b)
   \times
   \frac{e^{\eps \gamma_E} \mu^{2\eps}}{\Gamma(1-\eps)} \int_0^\infty \frac{\df k^+ \df k^-}{(k^+ k^-)^{\eps+1}}
   \delta\bigl[\Tau - \hat\Tau(\{k\})\bigr]
\nn\\&
 = \hat\sigma^\LO(Q,Y) \frac{\as \mathbf{C}}{\pi} f_a(x_a) f_b(x_b)
   \frac{e^{\eps \gamma_E}}{\Gamma(1-\eps)}\frac{2}{\eps} \frac{1}{\mu} (\Tau/\mu)^{-1-2\eps}
\nn\\&
 = \hat\sigma^\LO(Q,Y) \frac{\as \mathbf{C}}{4 \pi} f_a(x_a) f_b(x_b)
   \biggl[ -\frac{4}{\eps^2} \delta(\Tau) + \frac{8}{\eps} \frac{1}{\mu} \cL_0(\Tau/\mu)
   \nn\\&\hspace{5.2cm}
           - \frac{16}{\mu}  \cL_1(\Tau/\mu) + \frac{\pi^2}{3} \delta(\Tau) + \cO(\eps) \biggr]
\,.\end{align}
The $\cL_n(x)$ are the standard one-dimensional plus distributions, see e.g.\ \cite{Ligeti:2008ac} for details.
Note that there, the precise definition of the $\overline{\text{MS}}$ scheme is important.
We use
\begin{align}
 \mu^2 \equiv \mu^2_{\overline{\rm MS}} = \frac{4\pi}{e^{\gamma_E}} \muMS^2
\,.\end{align}
If one were to use $\mu^{2\eps} = \frac{(4\pi)^\eps}{\Gamma(1-\eps)} \muMS^{2\eps}$, one would miss the $\pi^2/3$ term.
For the NLP results presented in the main text, both definitions yield identical results.

Taking \eq{sigLPsoft} and adding the virtual soft diagram, and then comparing to \eq{factorization_NLO}, the one-loop bare soft function can be read off as
\begin{align} \label{eq:S_LP_NLO}
 S^\NLO(\Tau,\epsilon) =
 \frac{\as \mathbf{C}}{4 \pi} \biggl[ -\frac{4}{\eps^2} \delta(\Tau) + \frac{8}{\eps} \frac{1}{\mu} \cL_0(\Tau/\mu)
  - \frac{16}{\mu} \cL_1(\Tau/\mu) + \frac{\pi^2}{3} \delta(\Tau) \biggr]
\,.\end{align}
The finite terms precisely yield the renormalized one-loop soft function \cite{Stewart:2009yx}.

\subsection{NLO Beam Function}
\label{app:LP_Beam}

Applying the LP master formula \eq{sigmaLP} to the $n$-collinear case
and following the same steps as in \sec{sigma_NLO_NLP_coll} gives
\begin{align}
 \frac{\df\sigma_n^{(0,1)}}{\df Q^2 \df Y \df\Tau} &
 = \int_{x_a}^1 \frac{\df z_a}{z_a} \,
   \frac{f_{a'}(x_a/z_a) f_b(x_b)}{2 x_a x_b \Ecm^4}
   \frac{Q e^Y}{\rho} \biggl(\frac{Q \Tau e^Y}{\rho}\biggr)^{-\eps}
   \frac{z_a^\eps}{(1-z_a)^\eps}
   \frac{(4\pi)^{\eps}}{(4\pi)^2 \Gamma(1-\eps)}
   \nn\\&\hspace{1.5cm}\times
   \Msquared(Q,Y; \{k\})
\,.\end{align}
Using the universal $n$-collinear limit \eq{M2_coll_LP_app}, we obtain
\begin{align}
 \frac{\df\sigma_n^{(0,1)}}{\df Q^2 \df Y \df\Tau} &
 = \hat\sigma^\LO(Q,Y)
   \int_{x_a}^1 \frac{\df z_a}{z_a} \, f_{a'}\biggl(\frac{x_a}{z_a}\biggr) f_b(x_b) \,
   \frac{\as}{2\pi} \frac{e^{\eps \gamma_E}}{\Gamma(1-\eps)}
   \frac{1}{\Tau} \biggl(\frac{Q \Tau}{\mu^2} \frac{e^Y}{\rho}\biggr)^{-\eps}
   \frac{z_a^\eps P_{a a'}(z_a, \eps)}{(1-z_a)^\eps}
\,,\end{align}
where $P_{a a'}(z)$ is the standard $\epsilon$-dependent splitting kernel at NLO.
Comparing to \eq{factorization_NLO}, we can read off a result that will enable us to obtain the real radiation bare NLO beam function kernel,
\begin{align} \label{eq:I_ij}
 \cI^{(1){\rm real}}_{ij}(t, z, \epsilon) =
 \frac{\as}{2\pi} \biggl[ \frac{e^Y}{\rho}  \frac{e^{\eps \gamma_E}}{\Gamma(1-\eps)}
   \frac{1}{\mu^2} \biggl(\frac{t}{\mu^2} \frac{e^Y}{\rho}\biggr)^{-1-\eps} \frac{z^\eps P_{i j}(z, \eps)}{(1-z)^\eps}
\biggr]
\,.\end{align}
The splitting function $P_{ij}(z)$ may contain divergences as $z \to 1$,
which are regulated by the overall $(1-z)^{-\eps}$.
All divergences thus arise from the two expansions
\begin{align}
 \frac{1}{\mu_\rho^2} \left(\frac{t}{\mu_\rho^2}\right)^{-\eps-1} &=
 - \frac{\delta(t)}{\eps} + \frac{1}{\mu_\rho^2} \cL_0(t / \mu_\rho^2)
 - \eps  \frac{1}{\mu_\rho^2}  \cL_1(t/\mu_\rho^2) + \cO(\eps)^2
\,,\\
 (1-z)^{-1-\eps} &= - \frac{1}{\eps} \delta(1-z) + \cL_0(1-z) - \eps \cL_1(1-z)  + \cO(\eps)^2
\,,\end{align}
where we defined $\mu_\rho^2 = \mu^2 \frac{\rho}{e^Y}$ for ease of notation.
As written \eq{I_ij} does not yet contain the corresponding collinear virtual and zero-bin contributions.

\paragraph{\boldmath Example: $qq$ Kernel}
From \eq{Pij}, we obtain
\begin{align}
 \frac{z^\eps P_{qq}(z, \eps)}{(1-z)^\eps} &
 = C_F \frac{z^\eps}{(1-z)^\eps} \biggl[ \frac{1+z^2}{1-z} - \eps(1-z)\biggr]
\\&
 =  P_{qq}(z) + C_F \biggl\{ - \frac{2}{\eps} \delta(1-z) - \frac{3}{2} \delta(1-z)
 \nn\\\nn &\qquad
   + \eps \Bigl[ (z-1)  - (1+z^2) \cL_1(1-z) + (1+z^2) \ln(z) \cL_0(1-z) \Bigr]\biggr\}
   + \cO(\eps^2)
\,,\end{align}
where the LO quark splitting function is given by
\begin{equation}
 P_{qq}(z) = C_F \biggl[\Theta(1-z) \frac{1+z^2}{1-z} \biggr]_+ = (1+z^2) \cL_0(1-z) + \frac{3}{2} \delta(1-z)
\,.\end{equation}
Adding the corresponding virtual collinear and zero-bin contributions, \eq{I_ij} yields
\begin{align} \label{eq:I_qq}
 \cI^{(1)}_{qq}(t, z, \epsilon) =
 \frac{\as C_F}{2\pi} \biggl\{&
    \frac{2}{\eps^2} \delta(t) \delta(1-z)
  + \frac{ \delta(1-z)}{\eps} \biggl[ \frac{3}{2} \delta(t)  - 2 \frac{1}{\mu_\rho^2} \cL_0(t/\mu_\rho^2) 
 \biggr]
  - \frac{\delta(t)}{\eps} \frac{P_{qq}(z)}{C_F}  
  \nn\\&
   + 2 \delta(1-z) \frac{1}{\mu_\rho^2} \cL_1(t/\mu_\rho^2) +  \frac{1}{\mu_\rho^2} \cL_0(t/\mu_\rho^2) \biggl[P_{qq}(z) - \frac{3}{2} \delta(1-z) \biggr]
  \nn\\&
  + \delta(t) \biggl[ (1+z^2) \cL_1(1-z) - \frac{1+z^2}{1-z} \ln(z)  + (1-z) - \frac{\pi^2}{6} \delta(1-z) \biggr]
 \biggr\}
\,.\end{align}
Note that all divergences are proportional to $\delta(1-z)$, such that they cancel
after adding the soft, $\bn$ collinear and the virtual hard contribution from $H_{ab}^{\rm NLO}(Q^2,\epsilon)$,
as the latter also has the universal structure (for Drell-Yan)
\begin{align}
 \frac{\df \sigma_\text{virt}^{(0,1)}}{\df Q^2 \df Y \df \Tau}
 &= \hat\sigma^\LO(Q,Y) f_a(x_a)\, f_b(x_b)\delta(\Tau) \frac{\as C_F}{\pi} \biggl[
   -\frac{1}{\eps^2} - \frac{1}{\eps} \biggl(\frac{3}{2} - \ln\frac{Q^2}{\mu^2}\biggl) + \cO(\eps) \biggr]
\,.\end{align}
The cancellation of the $1/\eps^2$ and the remaining $3/(2\eps)$ pieces is obvious
from comparing to \eqs{I_qq}{S_LP_NLO}. The $P_{qq}(z)/\epsilon$ term cancels with the ultraviolet divergence from the bare quark PDF. 
The remaining $\ln(Q^2/\mu^2)/\epsilon$ term cancels when combining the $\cL_0(\Tau/\mu)/\epsilon$ and $\cL_0(t/\mu^2)/\epsilon$ distribution terms.
The remaining ${\cal O}(\epsilon^0)$ piece in \eq{I_qq} gives the renormalized beam function
and agrees with the result in \refcite{Stewart:2010qs}.

\paragraph{\boldmath Example: $qg$ Kernel}
For the full LP correction to Drell-Yan production, $q \bar q \to Z$,
we also require the quark-gluon kernel.
Here we only need
\begin{align}
 \frac{z^\eps P_{qg}(z, \eps)}{(1-z)^\eps} &
 = \frac{z^\eps}{(1-z)^\eps} T_F \biggl[ 1 - \frac{2z(1-z)}{1-\eps} \biggr]
\nn\\&
 = P_{qg}(z) \biggl[ 1 + \eps \biggl( \ln\frac{z}{1-z} + 1 \biggr)\biggr] - \eps  T_F + \cO(\eps^2)
\,,\end{align}
where the finite quark-gluon splitting function is defined as
\begin{align}
 P_{qg}(z) = T_F [(1-z)^2 + z^2 ]
\,.\end{align}
Equation~\eqref{eq:I_ij} thus yields
\begin{align} \label{eq:I_qg}
 \cI^{(1)}_{qg}(t, z, \mu) &=
 \frac{\as T_F}{2\pi} \biggl\{
  - \frac{\delta(t)}{\eps} \frac{P_{qg}(z)}{T_F}  
  + \frac{1}{\mu_\rho^2} \cL_0(t/\mu_\rho^2) P_{qg}(z)
  + \delta(t) \biggl[ P_{qg}(z) \biggl( \ln\frac{1-z}{z} - 1\biggr) + 1 \biggr] \biggr\}
\,.\end{align}
Again the $P_{qg}(z)/\epsilon$ divergence cancels against the same mixing term from the bare gluon PDF. 
The remaining ${\cal O}(\epsilon^0)$ terms give the mixing term in the one-loop quark beam function, agreeing with the result in \cite{Stewart:2010qs}.

\section[Comparison of NLP Contributions for \texorpdfstring{\boldmath $gg \to H g$}{gg -> Hg} at NLO]
{\boldmath Comparison of NLP Contributions for $gg \to H g$ at NLO}
\label{app:ggHg_NLPs}

Here, we give the explicit calculation to obtain our full NLP result for hadronic $\Tau$
in the $gg \to Hg$ channel in the form of \eq{sigma_NLL_comparison}.
Our result prior to integration by parts is obtained by inserting \eq{kernel_Cgg} with $\rho = 1$ into \eq{sigma_NLP_NLO_1},
\begin{align} \label{eq:sigma_NLL_comparison_1}
 \frac{\df\sigma^{(2,1)}}{\df \Tau^\hadcm} &
 = \frac{\as C_A}{\pi} \int_0^1 \df x_a \df x_b \,
   2\pi \delta(x_a x_b \Ecm^2 - m_H^2) \frac{|\cM_{gg\to H}^\LO(m_H)|^2}{2 x_a x_b \Ecm^2}
   \int_{x_a}^1 \frac{\df z_a}{z_a} \, \frac{e^Y}{m_H}
   \nn\\&\quad\times
   \biggl\{
    f_g\biggl(\frac{x_a}{z_a}\biggr) f_g(x_b) \biggl[
      \biggl(\ln\frac{\Tau^\hadcm }{m_H e^Y} + 1 \biggr) \delta(1-z_a)
      \nn\\&\hspace{4cm}
      - \cL_0(1-z_a) + \frac{1 - z_a + 9z_a^2 - 5 z_a^3 + 7 z_a^4 - 3 z_a^5}{2 z_a^2}
    \biggr]
   \nn\\&\qquad
   + \frac{x_a}{z_a} f'_g\biggl(\frac{x_a}{z_a}\biggr) f_g(x_b) \frac{(1-z_a + z_a^2)^2}{2 z_a}
   \nn\\&\qquad
   + f_g\biggl(\frac{x_a}{z_a}\biggr) \, x_b f'_g(x_b) \biggl[
      \biggl( - \ln\frac{\Tau^\hadcm }{m_H e^Y} - 1  \biggr) \delta(1-z_a)
      \nn\\&\hspace{4.5cm}
      + \frac{(1+z_a)(1-z_a+z_a^2)^2}{2z_a^2}\, \cL_0(1-z_a)
   \biggr]
   \biggr\}
\nn\\&
 + \bigl( a \leftrightarrow b \,,\, e^Y \leftrightarrow e^{-Y} \bigr)
\,.\end{align}
Here, we also separated the pure $\cL_0(1-z_a)$ term from terms regular as $z_a \to 1$.

We now apply integration by parts to the $f_g f'_g$ piece, except for its $-\delta(1-z_a)$ term.
In the notation of \eq{sigma_NLP_NLO_2}, this is achieved by choosing
\begin{align}
 D^{(2,1)}(z_a, z_b, \Tau^\hadcm) = 4 C_A \frac{e^Y}{m_H}
 \biggl[&- \ln\frac{\Tau^\hadcm}{m_H e^Y} \delta(1-z_a)
 \nn\\&
 + \frac{(1+z_a)(1-z_a+z_a^2)^2}{2z_a^2}\, \cL_0(1-z_a) \biggr]
 \delta(1-z_b)
\,.\end{align}
Here, we only consider being inclusive in $Y$, so we do not write down the boundary term.
Eq.\ \eqref{eq:sigma_NLL_comparison_1} becomes
\begin{align} \label{eq:sigma_NLL_comparison_2}
 \frac{\df\sigma^{(2,1)}}{\df \Tau^\hadcm} &
 = \frac{\as C_A}{\pi} \int_0^1 \df x_a \df x_b \,
   2\pi \delta(x_a x_b \Ecm^2 - m_H^2) \frac{|\cM_{gg\to H}^\LO(m_H)|^2}{2 x_a x_b \Ecm^2}
   \int_{x_a}^1 \frac{\df z_a}{z_a} \, \frac{e^Y}{m_H}
   \nn\\&\quad\times
   \biggl\{
    f_g\biggl(\frac{x_a}{z_a}\biggr) f_g(x_b) \biggl[ 2 \delta(1-z_a)
      + \biggl(\frac{(1+z_a)(1-z_a+z_a^2)^2}{2z_a^2} - 1 \biggr) \cL_0(1-z_a)
      \nn\\&\hspace{3.75cm}
      + \frac{1 - z_a + 9z_a^2 - 5 z_a^3 + 7 z_a^4 - 3 z_a^5}{2 z_a^2}
    \biggr]
   \nn\\&\qquad
   + \frac{x_a}{z_a} f'_g\biggl(\frac{x_a}{z_a}\biggr) f_g(x_b) \biggl[
      - \delta(1-z_a) \ln\frac{\Tau^\hadcm}{m_H e^Y}
      + \frac{(1-z_a+z_a^2)^2}{z_a^2}\, \cL_0(1-z_a)
      \nn\\&\hspace{4.3cm}
      - \frac{(1-z_a)(1-z_a+z_a^2)^2}{2z_a^2}
   \biggr]
   \nn\\&\qquad
   - f_g\biggl(\frac{x_a}{z_a}\biggr) \, x_b f'_g(x_b) \, \delta(1-z_a)
   \biggr\}
\nn\\& \quad
 + \bigl( a \leftrightarrow b \,,\, e^Y \leftrightarrow e^{-Y} \bigr)
\,,\end{align}
where as usual, $e^Y = \sqrt{x_a/x_b}$.
Next, we apply the following integration by parts:
\begin{align}
 &\int_{x_a}^1 \frac{\df z_a}{z_a} \frac{x_a}{z_a} f'_g\biggl(\frac{x_a}{z_a}\biggr) f_g(x_b)
   \biggl[ -\frac{(1-z_a)(1-z_a+z_a^2)^2}{2z_a^2}\biggr]
\nn\\
 =~&
  \int_{x_a}^1 \frac{\df z_a}{z_a} \, f_g\biggl(\frac{x_a}{z_a}\biggr) f_g(x_b)
  \frac{2 - 3 z_a + 5 z_a^3 - 6 z_a^4 + 3 z_a^5}{2 z_a^2}
\,.\end{align}
Putting this back into \eq{sigma_NLL_comparison_2}, we can rewrite it in a form close to \eq{sigma_ggHg_NLL_lit},
\begin{align} \label{eq:sigma_NLL_comparison_4}
 \frac{\df\sigma^{(2,1)}}{\df \Tau^\hadcm} &
 = \frac{\as C_A}{\pi} \int_0^1 \df x_a \df x_b \,
   2\pi \delta(x_a x_b \Ecm^2 - m_H^2) \frac{|\cM_{gg\to H}^\LO(m_H)|^2}{2 x_a x_b \Ecm^2}
   \int_{x_a}^1 \frac{\df z_a}{z_a} \, \frac{e^Y}{m_H}
   \nn\\&\quad\times
   \biggl\{
    f_g\biggl(\frac{x_a}{z_a}\biggr) f_g(x_b) \biggl[
      2 \delta(1-z_a) + \biggl( \frac{(1-z_a+z_a^2)^2}{z_a^2} - 1 \biggr) \cL_0(1-z_a)
      \nn\\&\hspace{3.7cm}
      + \frac{3 z_a^2 + 1 - z_a + z_a^3}{z_a^2}
    \biggr]
   \nn\\&\qquad
   + \frac{x_a}{z_a} f'_g\biggl(\frac{x_a}{z_a}\biggr) f_g(x_b) \biggl[
      - \delta(1-z_a) \ln\frac{\Tau^\hadcm}{m_H e^Y}
      + \frac{(1-z_a+z_a^2)^2}{z_a^2}\, \cL_0(1-z_a)
   \biggr]
   \nn\\* &\qquad
   - f_g\biggl(\frac{x_a}{z_a}\biggr) \, x_b f'_g(x_b) \, \delta(1-z_a)
   \biggr\}
\nn\\* & \quad
 + \bigl( a \leftrightarrow b \,,\, e^Y \leftrightarrow e^{-Y} \bigr)
\,.\end{align}
To compare this result to \eq{sigma_ggHg_NLL_lit}, use the relations
\begin{align}
 \frac{e^Y}{m_H} = \frac{x_a \Ecm}{m_H^2}
\,,\quad
 \ln\frac{\Tau^\hadcm}{m_H e^Y} = \ln\frac{\Tau^\hadcm}{x_a \Ecm}
\,.\end{align}

\bibliography{../subleading}
\bibliographystyle{jhep}
\end{document}